\documentclass[fleqn,usenatbib]{rasti}

\usepackage{newtxtext,newtxmath}
\usepackage[T1]{fontenc}
\usepackage{lipsum}
\usepackage{graphicx}
\usepackage{caption}
\usepackage{subcaption}
\usepackage{booktabs}

\DeclareRobustCommand{\VAN}[3]{#2}
\let\VANthebibliography\thebibliography
\def\thebibliography{\DeclareRobustCommand{\VAN}[3]{##3}\VANthebibliography}

\usepackage{graphicx}	% Including figure files
\usepackage{amsmath}	% Advanced maths commands

\title[A public Ariel simulated observations dataset]{A public dataset of Ariel simulated observations for developing exoplanetary atmosphere data reduction pipelines.}

\author[Mugnai L. V. and Yip K. H. et al.]{Lorenzo V. Mugnai$^{1,2,3,4}$\thanks{E-mail: mugnail@cardiff.ac.uk},
Kai Hou Yip$^{7}$\thanks{E-mail: kai.hou.yip@ucl.ac.uk},
Andrea Bocchieri$^{3}$, Andreas Papageorgiou$^{1}$,
\newauthor{Virginie Batista$^{5}$, Orphée Faucoz$^{6}$, Angèle Syty$^{5}$, Tara Tahseen$^{2}$,  Enzo Pascale$^{3}$,  Ingo Waldmann$^{2}$}\\
$^{1}$School of Physics and Astronomy, Cardiff University, Queens Buildings, The Parade, Cardiff, CF24 3AA, UK\\
$^{2}$Department of Physics and Astronomy, University College London, Gower Street, London, WC1E 6BT, UK\\
$^{3}$Dipartimento di Fisica, La Sapienza Universit\`a di Roma, Piazzale Aldo Moro 5, 00185 Roma, Italy\\
$^{4}$INAF – Osservatorio Astronomico di Palermo, Piazza del Parlamento 1, I-90134 Palermo, Italy\\
$^{5}$Sorbonne Université, CNRS, UMR 7095, Institut d’Astrophysique de Paris, Paris, France\\
$^{6}$CNES, Centre national d'études spatiales, Paris, France\\
$^{7}$Department of Physics, King's College London, University of London, Strand, London, WC2R 2LS, United Kingdom\\
}

\date{Accepted XXX. Received YYY; in original form ZZZ}

\pubyear{2025}

\begin{document}
\label{firstpage}
\pagerange{\pageref{firstpage}--\pageref{lastpage}}
\maketitle

% Abstract of the paper
\begin{abstract}

Detecting and characterising exoplanet atmospheres remains challenging because atmospheric signals can be comparable to residual noise and instrumental/astrophysical systematics. Spectral features span from a few ppm for small planets up to $\sim 10^3$ ppm for warm/hot giants, while high-quality JWST time-series spectroscopy typically reaches $\sim 10$--$50$ ppm (occasionally $\sim 100$--$200$ ppm in the presence of stellar variability or stronger systematics), making correlated noise across temporal and spectral dimensions a key limitation. With JWST delivering an increasing volume of high-precision transmission spectra, and Ariel set to extend this to a homogeneous survey of $\sim 10^3$ exoplanet atmospheres, robust benchmarking resources with known ground truth are essential to develop and validate data-driven (including ML-based) detrending approaches.

As a major step towards this goal, we use ExoSim2 and TauREx to generate one of the most comprehensive public datasets based on the current payload design of the ESA Ariel mission, specifically intended to benchmark detrending algorithms. We also provide a deep neural network baseline for time-series reduction, and use it to highlight the limitations of ML based detrendng methods, i.e. the risks posed by dataset shift when observed distributions diverge from those of the training set, a scenario likely to arise in real observations.

This dataset is featured in the Ariel Data Challenge 2024 on Kaggle and has been field-tested for robustness and simulation fidelity. By making these resources publicly available, we aim to support the community in developing, comparing, and stress-testing scalable and reliable methods for exoplanet transmission spectroscopy.

% Detecting and characterising exoplanetary atmospheres is a challenging task due to the presence of substantial noise in observations, often orders of magnitude higher than the planetary signals themselves. This noise is correlated across spatial, temporal, and spectral domains, making it a complex denoising problem for the scientific community. \comment{make aDC as a side task, not main thing here} The NeurIPS Ariel Data Challenge 2024 aims to encourage innovative solutions for detrending such models. In this paper, we utilise Exosim2 to create one of the most comprehensive public datasets based on the current payload design of the ESA Ariel Space Mission. \comment{This paper served as a documentation of the data generation and post processing process. As an initial benchmark on this task with Deep learning method, we implemented a CNN based pipeline , and highlight the danger of distribution shift that could result in blindly applying neural networks. }  \textbf{We have also proposed a metric and the result from our baseline solutions.} By releasing this extensive dataset and the metric, we hope to foster the development and advancement of data-detrending methods within the exoplanet characterisation pipeline, ultimately enabling breakthroughs in achieving higher signal-to-noise ratios for scientific analysis of exoplanetary atmospheres.

\end{abstract}

% Include between one and six keywords.

%%%%%%%%%%%%%%%%%%%%%%%%%%%%%%%%%%%%%%%%%%%%%%%%%%

%%%%%%%%%%%%%%%%% BODY OF PAPER %%%%%%%%%%%%%%%%%%

\section{Introduction}

Recent and ongoing space missions represent a major advancement in the atmospheric characterisation of exoplanets. The James Webb Space Telescope (JWST) mission \citep{Greene2016ApJ} is now delivering high-precision spectroscopic observations of transiting and directly imaged exoplanets, with typical noise levels of tens of ppm, enabling detailed constraints on atmospheric composition, temperature structure, and cloud properties (e.g. \citealt{Radica2023MNRAS,Carter2024NatAs,Challener2025NatAs, Espinoza2025ApJ,Taylor2025MNRAS}). These observations are opening new frontiers in the study of planetary atmospheres across a wide range of planetary regimes. The Ariel mission (Atmospheric Remote-sensing Infrared Exoplanet Large-survey) of ESA, scheduled for launch in 2029 \citep{Tinetti2018ExA}, is designed as a dedicated atmospheric survey. Its primary objective is to observe a large and statistically significant sample of exoplanets, spanning super-Earths to gas giants, in order to enable comparative planetology. Ariel is expected to characterise of order $\sim$1000 planetary atmospheres with a homogeneous observing strategy, providing a unique dataset to study population-level trends in atmospheric chemistry and physics.

The scale of exoplanet atmospheric characterization is entering a new era. While JWST provides detailed spectra of individual targets, Ariel will survey approximately 1000 exoplanet atmospheres. This transition from targeted observations to large-scale surveys requires scalable, automated analysis approaches capable of maintaining uniform quality standards across diverse planetary systems.

Public benchmark datasets with known ground truth are essential preparation for this data-intensive regime. Unlike real observations where atmospheric properties remain unknown, simulated datasets enable rigorous quantification of systematic biases, assessment of noise propagation, and direct comparison of methodological approaches under controlled conditions. Such resources accelerate technique development through transparent community-wide benchmarking.

The Ariel Data Challenge series \citep{nikolaou2023lessons, yip_esa-ariel_2022, changeat_esa-ariel_2023} exemplifies this community-driven approach, connecting planetary scientists, instrument experts, and data scientists to tackle key challenges in exoplanet characterisation. The 2024 edition \citep{yip2024ariel} addresses the need for realistic training data by introducing a large-scale public dataset of simulated raw observations from the Ariel mission, specifically designed to benchmark detrending algorithms against realistic instrumental systematics and noise sources. This dataset was generated by coupling ExoSim2 \citep{EXOSIM} and TauREx3 \citep{Taurex}, incorporating the latest Ariel mission specifications, noise maps derived from publicly available JWST detector calibrations, and diverse atmospheric compositions. The scale and complexity of this resource provide an invaluable platform for developing and testing innovative data reduction techniques essential for future observational programmes.

Developing and optimising bespoke data‑reduction pipelines for exoplanet characterisation is essential in the lead‑up to forthcoming missions. Such a pipeline comprises a suite of analysis techniques, implemented as specialised algorithms, that transform raw photometric and spectroscopic time‑series data into scientifically usable products, most notably the planetary spectrum as a function of wavelength. An ideal pipeline should behave as an unbiased estimator that minimises post‑processing noise by detrending systematic effects without introducing bias. This requirement is particularly stringent for the Ariel mission, given its ambitious objectives in comparative planetology and its commitment to a uniform observational strategy, which demands precise and accurate results for a self-consistent, homogeneous characterisation across its exoplanet sample. Currently, several data reduction pipelines with distinct methodologies exist, and the JWST pipelines serve as a notable example of this diversity, with multiple independently-developed pipelines introduced in the ERS context \citep[e.g.][]{2023G395H, 2023PRISM, 2023NIRCam, 2023NIRISS} as well as additional community pipelines released in the following years \citep[e.g.][]{2023MNRAS.524..377H, 2024MNRAS.531.2731S}.

The coexistence of multiple JWST pipelines has also enabled direct pipeline-to-pipeline comparisons on the same datasets. In several ERS studies, independent reductions and analysis frameworks have been applied to the same observations, generally finding consistent spectra within uncertainties while highlighting that specific methodological choices (e.g. systematics modelling, outlier rejection, background treatment, time-dependent instrument trends) can still induce measurable differences \citep[e.g.][]{2023G395H, 2023PRISM, 2023NIRCam, 2023NIRISS, Schmidt2025AJ, Welbanks2025NatAs}.

% The heterogeneity of analysis pipelines for instruments like \textbf{JWST and HST} has highlighted the risk of systematic biases, manifesting as discrepancies in spectra and astrophysical parameters obtained for the same object when processed with different pipelines \textbf{\citep[e.g.][]{Swain2021, Mugnai2021b, Taylor2025RNAAS, Schmidt2025AJ, Welbanks2025NatAs}.
% }
Machine learning (ML) and deep learning (DL) approaches have been successfully applied in various imagery data domains for denoising \citep{elad2023image}, reconstruction \citep{su2022survey}, and even generating new images \citep{zhang2023text, bie2023renaissance, dubey2024transformer}. Recent studies have explored the use of ML techniques to denoise astronomical data in different forms, such as time series light curve data \citep{Ingalls2016, Krick2020, Morvan2020, Morvan2022, nikolaou2023lessons} and imagery data \citep{Vojtekova21, sweere22, Park24}.

However, most of these studies have focused on either time series (light curves) or images from photometry, while actual transit observations are 3D cubes that are acquired as a time series of still images across a wide wavelength range. This combination results in an exceptionally rich information content within a single observation. The aforementioned approaches have not utilised the full extent of this information, inevitably neglecting crucial spatial and spectral information about the noise characteristics. Past studies have indicated that incorporating temporal information (light curves) from simulated HST \citep{Yip2020} and JWST data \citep{Changeat_2024} can mitigate some of the temporal aspects of noise effects (such as star spots), and could yield better constraints on downstream tasks such as retrieving the atmospheric parameters.

% The Ariel Data Challenge series connects the global data science community with planetary scientists and engineers to tackle difficult challenges in exoplanet research \citep{nikolaou2023lessons, yip_esa-ariel_2022, changeat_esa-ariel_2023}. The 2024 edition of the Ariel Data Challenge \citep{yip2024ariel} addresses the need for realistic training data by introducing a large-scale public dataset of simulated raw observations of exoplanetary transits from the Ariel mission. This dataset was generated by coupling ExoSim2 \citep{EXOSIM} and TauREx3 \citep{Taurex}, and incorporates the latest Ariel mission specifications, noise maps derived from publicly available calibration measurements of HxRG MCT (Mercury Cadmium Telluride) detectors flown on JWST, and a diverse range of atmospheric species. \textbf{This is a realistic assumption given that Ariel will employ near-IR HgCdTe/MCT detector technology of the same class as the HxRG arrays flown on JWST}.

% The scale and complexity of this dataset provide an invaluable platform for the scientific community to develop and test innovative detrending techniques. These techniques are crucial for the successful analysis of data from future generations of instruments, which will face similar challenges in terms of noise and atmospheric diversity. By making this dataset publicly available, we aim to foster collaboration and accelerate progress in the field of exoplanetary atmospheric characterisation as well as mission design.

As a first attempt at building a neural network-based data reduction and extraction pipeline, we trained a CNN-based neural network as a benchmark baseline. Since machine learning models are prone to overfitting to training data, it is important that our benchmark data account for this tendency. To thoroughly test any pipeline, whether parametric, machine learning, or hybrid, we have designed a data partition that explicitly includes out-of-distribution test examples. This approach exposes challenges that models will likely face when applied to real data, such as interpreting spectra from unknown exoplanets with unknown composition and atmospheric properties, known as ``data shift'' or ``domain shift'' in the Machine Learning and Artificial Intelligence community. This allows us to provide an estimate of the model's performance on both training data and data that are unknown to us.

The structure of this paper is organised as follows: Section \ref{sec:data} outlines the various components involved in data generation and the validation processes that occur before commissioning. Section \ref{sec:data_partition} details the data partitioning methodology, which divides the dataset into traditional training and testing subsets. In Section \ref{sec:baseline}, we present the implementation of the CNN\footnote{Convolutional Neural Network, more explanation in the Section 5.3}-based data reduction pipeline and our proposed metric to evaluate its performance, followed by an analysis of the results and performance metrics on the datasets in Section \ref{sec:results}. In Section \ref{sec:discussiom}, we discuss our methodology limitations and our results. Finally, Section \ref{sec:conclusion} concludes with a discussion of our findings, future research directions, and the challenges that lie ahead.

% Moreover, the creation of this large-scale dataset allows for a comprehensive assessment of the current state-of-the-art experimental pipeline. By subjecting the pipeline to a wide range of noise effects, atmospheric features, and jitter scenarios, we can identify its strengths and weaknesses. This assessment will provide valuable insights into the performance of the pipeline under different conditions and guide future improvements and optimisations.
The training dataset produced for this manuscript is publicly available on the Kaggle platform\footnote{\url{https://www.kaggle.com/competitions/ariel-data-challenge-2024/data}}.

% \section{Problem Statement}
% \comment{??'s section - here is where Science should be}

\section{Data description}
\label{sec:data}

The simulated data are generated using the publicly available end-to-end instrument simulator ExoSim2\footnote{in this work, we used code version \texttt{v2.0.0-rc2}, available at \url{https://github.com/arielmission-space/ExoSim2-public/releases/tag/v2.0.0-rc2}.}
 \citep{EXOSIM}. ExoSim2 models the full Ariel observing chain, including the telescope throughput, the dispersive optics, the detector response, and the noise sources, and outputs realistic time-series detector images for a given target and observing scenario. To build physically consistent simulations, ExoSim2 is coupled with (i) the atmospheric retrieval and forward-modelling framework TauREx 3 \citep{Taurex}, which provides the wavelength-dependent planetary transmission spectrum used as the astrophysical input signal, and (ii) the physical optical simulator PAOS \citep{PAOS}, which generates realistic, wavelength-dependent Ariel point-spread functions (PSFs) and optical aberrations. Using the three tools together ensures that both the astrophysical signal (planet spectrum) and the instrument response (optics/PSF and detector-level systematics and noise) are treated consistently, enabling the generation of mock images representative of Ariel observations. For a complete description of the simulation process and Ariel's specificities, please refer to the ExoSim2 documentation\footnotemark{} and published literature \citep{EXOSIM, Bocchieri2025}.

\footnotetext{\url{https://exosim2-public.readthedocs.io/}}

The following section outlines the strategies and assumptions employed in constructing the dataset. In  Section \ref{sec:payload}, we begin by discussing the model of the Ariel payload utilised for the simulated observations. In Section \ref{sec:detector_noise}, we describe both the calibration products used to realistically simulate the detector response and a comprehensive list of all simulated noise sources included in the process.  Once the telescope setup is described, we proceed to discuss the observed target star in Section \ref{sec:target} and the planetary models included in the simulation (Section \ref{sec:planets}). Finally, in Section \ref{sec:production} and Section \ref{sec:validation}, we describe the data generation process and the validation of the final product, including statistics on the dataset.

\subsection{Payload model}
\label{sec:payload}
For this dataset, we refer to the payload configuration presented in \cite{EXOSIM}, and instead of all six channels, we use one photometer (FGS1) collecting light between 0.6 and 0.8 $\mu$m and one spectrometer (AIRS-CH0) sensitive from 1.95 to 3.90 $\mu$m with spectral resolving power $R=100$ \citep{Arielrad}. The effects of Ariel's design can be summarised as a combination of Photon Conversion Efficiency (PCE) and Point Spread Functions (PSFs) as described below. Additionally, we reduced the area (in pixels) of Ariel's channels to optimise storage space in the simulations. For the FGS, we used a $32 \times 32$ pixel array instead of the original $64 \times 64$ window, and for the AIRS, we used a $32 \times 356$ pixel array instead of the original $64 \times 356$ region of interest.

Because some parameters are not yet available for the flight model, using the JWST equivalent (see Sec. \ref{sec:calibration_products}) is expected to be a realistic alternative, as Ariel relies on near-IR detector technology comparable to that adopted by JWST/NIRSpec, making CRDS-based dark, read-noise, flat-field and bad-pixel maps a physically motivated proxy for detector-level behaviour. Nevertheless, differences in operating conditions and in-flight environment (e.g. temperature, readout electronics and radiation-driven ageing) could shift the absolute noise properties and the level of correlated systematics. At the same time, we use critical elements from Ariel, such as the photon conversion efficiency and the point spread functions.

\begin{figure}
    \centering
    \includegraphics[width=\linewidth]{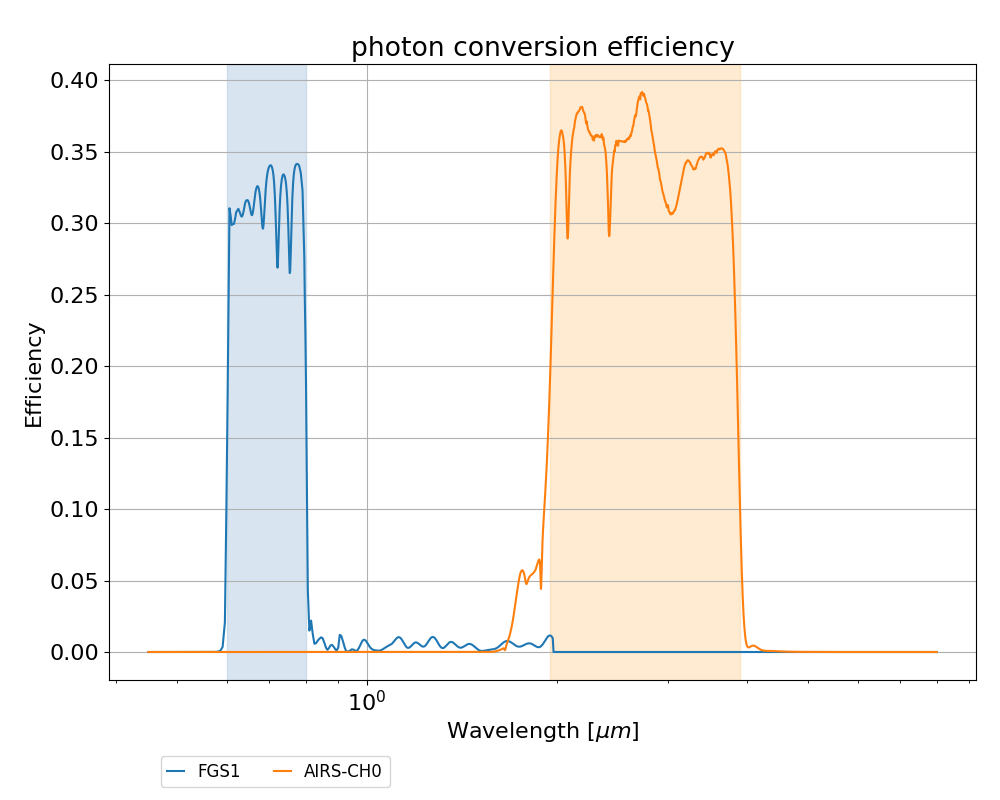}
    \caption{The figure shows the photon conversion efficiency (PCE) considered in this work for the two Ariel channels simulated: FGS1 and AIRS-CH0. The PCE is computed as the product of the telescope optical transmission efficiency and the detector quantum efficiency. The background coloured bands highlight the nominal wavelength ranges for each channel.}
    \label{fig:PCE}
\end{figure}

The photon conversion efficiency is the product of the telescope transmission efficiency and the detector quantum efficiency, and it describes the ability of the payload to convert light into electrons. For a detailed description of the conversion process, we refer the reader to the following: ExoSim2 \citep{EXOSIM}, the Ariel radiometric model ArielRad \citep{Arielrad}, and the ExoSim2 documentation\footnotemark[\value{footnote}]. The PCE used is reported in \autoref{fig:PCE}.

\begin{figure*}
    \centering
    \includegraphics[width=\textwidth]{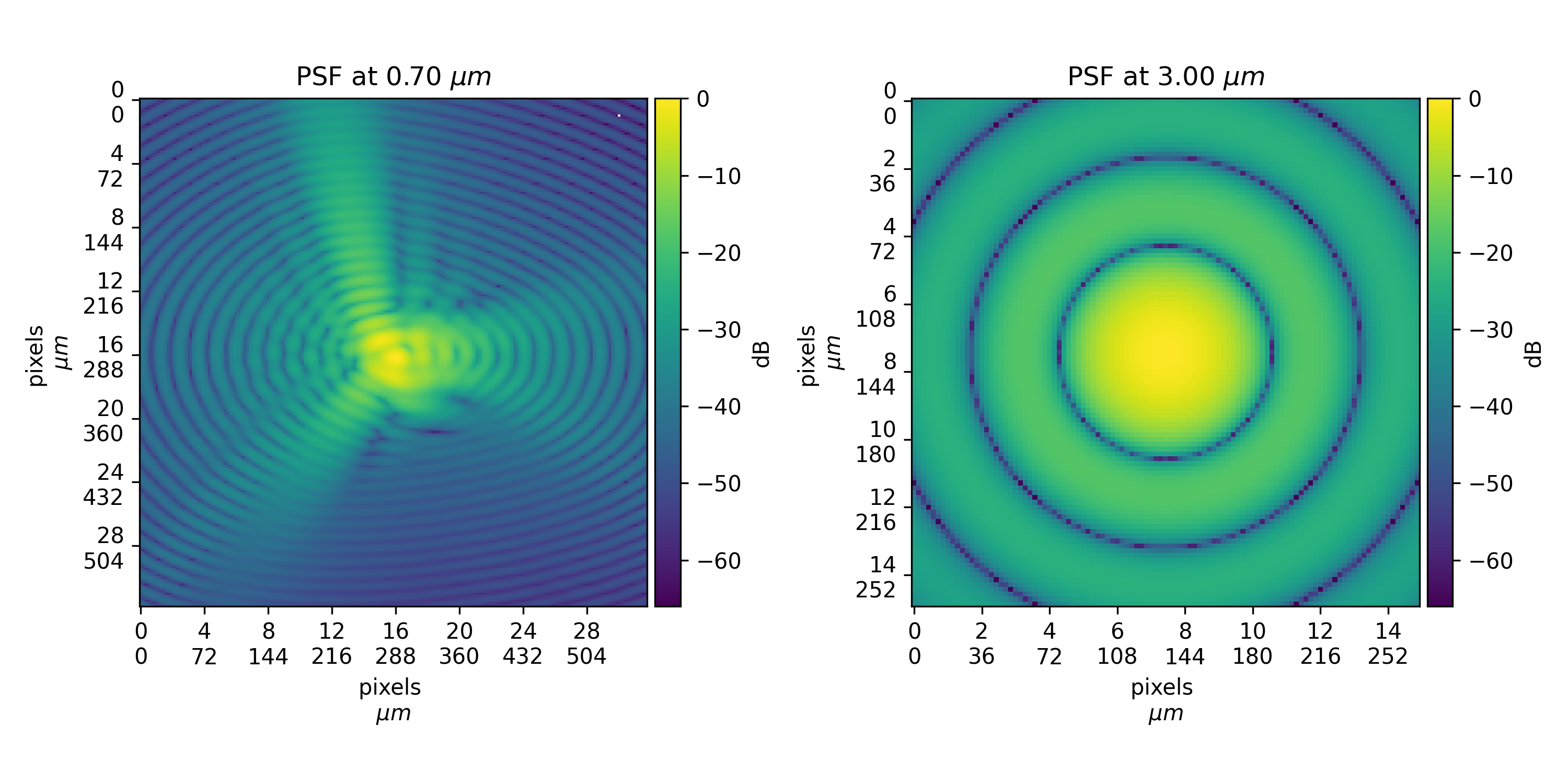}
    \caption{The figure reports two of the PSFs used to simulate FGS1 (left) and AIRS-CH0 (right). The PSFs are sampled at the sub-pixel level and reported in decibels normalised to the maximum value in the field. The axes report the grid into pixel and micron units. The reader may note that the FGS1 PSF is highly aberrated and complex, so it is sampled on a grid of the same size as the detector ($32 \times 32$), while the AIRS-CH0 PSF is an Airy function and here is reported on a smaller grid. For details, please refer to the software documentation.}
    \label{fig:psf}
\end{figure*}

For a realistic representation of the telescope's performance, we use the PSFs already presented in other Ariel-related works \citep{EXOSIM, Vinooja2025, Bocchieri2025}. The FGS1 PSFs are produced using the PAOS simulator \citep{PAOS} for a range of wavelengths sampled by the detector, as described in \cite{EXOSIM}. For the AIRS-CH0 PSFs, we use the default ExoSim2 Airy PSF model for every wavelength sampled by a pixel. The utilised PSFs are shown in \autoref{fig:psf}.

To create the final image, these PSFs are convolved with the intra-pixel response function, which models the pixel response to light, as typically, a pixel is more sensitive near the centre and less toward the edges. This model is created by a default ExoSim Task using the prescription from \citet{Barron2007}.

\subsection{Detector and noise models}
\label{sec:detector_noise}
\subsubsection{JWST calibration products} \label{sec:calibration_products}
To simulate realistic detector maps, we utilised calibration products for NIRSpec from the JWST Calibration Reference Data System (CRDS). Specifically, we collected dark frames\footnote{\url{https://jwst-crds.stsci.edu/browse/jwst_nirspec_dark_0352.fits}}, read noise\footnote{\url{https://jwst-crds.stsci.edu/browse/jwst_nirspec_readnoise_0039.fits}}, flat fields\footnote{\url{https://jwst-crds.stsci.edu/browse/jwst_nirspec_dflat_0002.fits}}, and bad pixel maps\footnote{\url{https://jwst-crds.stsci.edu/browse/jwst_nirspec_mask_0049.fits}}. These calibration files were carefully selected to ensure consistency. All CRDS reference files were taken from the NIRSpec detector \texttt{NRS2}. The selection of \texttt{NRS2} was not driven by a specific scientific preference, but was adopted for practical consistency in this dataset release.

From this data, we extracted segments of the detector to align with the dimensions of Ariel’s FGS1 and AIRS-CH0 detectors. Several regions were selected to represent various potential detectors for Ariel. To thoroughly assess the detrending performance of any proposed pipeline, we selected one detector free of bad pixels in the illuminated area and another that includes some bad pixels or defects in the FGS1 and CH0 areas, thereby simulating a worst-case scenario.

The dark current contribution is estimated for each pixel based on the expected counts for that pixel in the input map, multiplied by the exposure time of each sub-exposure. This estimate is then subject to randomisation due to Poisson noise. The read noise, on the other hand, is provided as a variance map, so for each NDR, we add a random number of counts normally distributed according to the pixel variance.

The flat fields are maps of the inter-pixel quantum efficiency variation, here used normalised to the median detector quantum efficiency used to compute the PCE. Finally, bad pixels are non-functional pixels that are no longer able to collect light.

To summarise how these effects combine to form the measured signal ($S_{mes}$) starting from the original signal ($S$), we consider the following equation:
\begin{equation}\label{eq:signal-meas}
    S_{mes} = (S \cdot \Delta t \cdot \text{Flat}) + (\text{Dark} \cdot \Delta t) + \text{Read}
\end{equation}
where $\Delta t$ is the integration time for the considered NDR, $\text{Flat}$ is the flat field, and $\text{Dark}$ is the dark current map.
In this equation, $S \cdot \Delta t$ represents the signal integrated over the exposure time, and the flat field modulates it. $\text{Dark} \cdot \Delta t$ accounts for the dark current contribution over the same integration time. $\text{Read}$ is the read noise, which is a random realisation based on the read noise variance map. Both $S \cdot \Delta t$ and $\text{Dark} \cdot \Delta t$ are subject to randomisation due to Poisson noise, reflecting the statistical nature of photon counting and dark current generation.

\begin{figure}
    \centering
    \includegraphics[width=\linewidth]{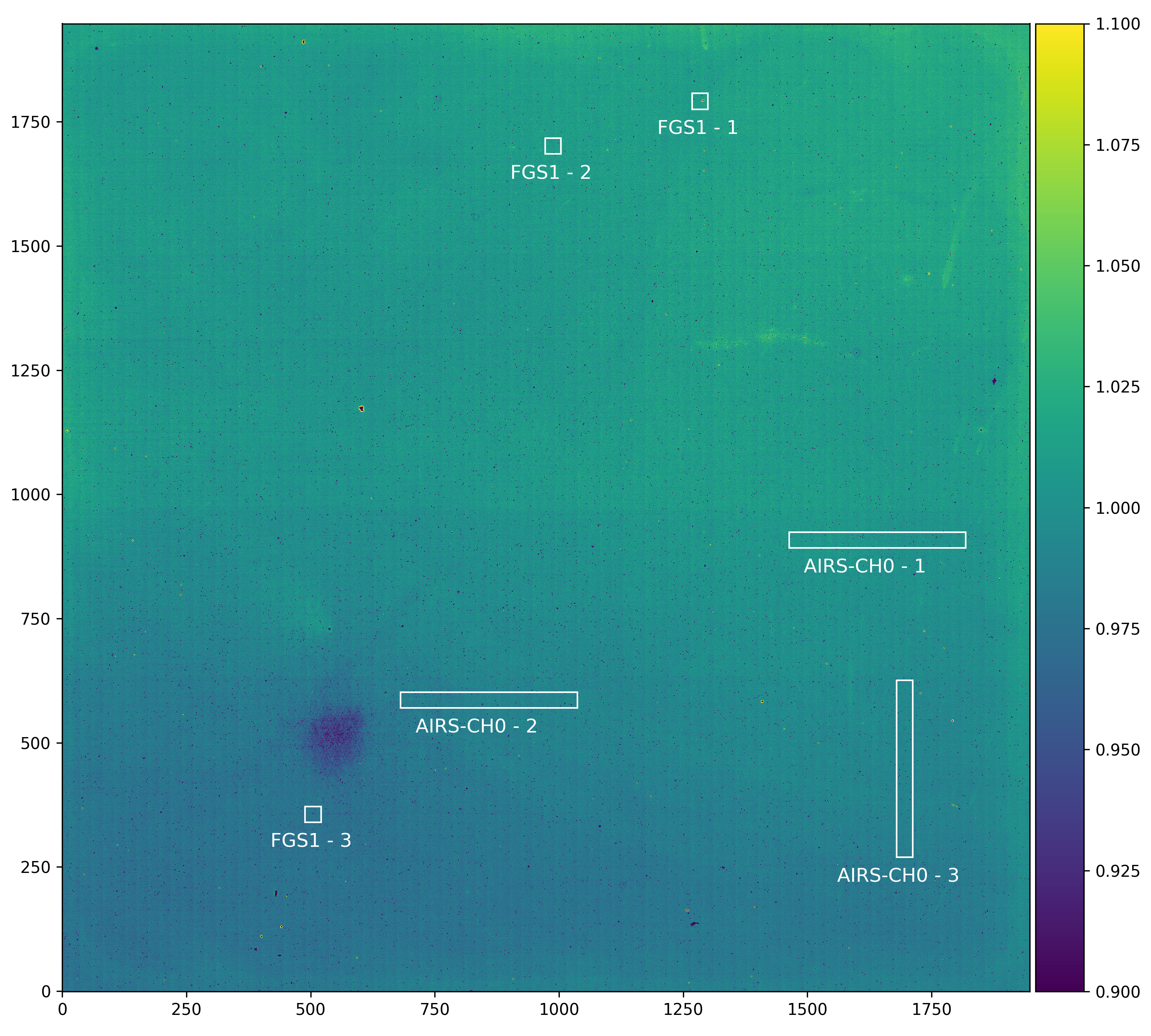}
\caption{JWST flat field variation map used for the simulation. The highlighted regions indicate the three areas selected for the each detector (FGS and AIRS-CH0) flat field maps. This flat field is presented as a variation from the median value. The colour scale is set between 0.9 and 1.1 to enhance detector features, masking the outlier values of bad pixels.}
    \label{fig:maps}
\end{figure}

\autoref{fig:maps} shows the selected regions on the JWST flat field calibration product. These regions were manually selected by visually comparing different calibration products to include a comprehensive mixture of possible detector effects. Note that this flat field is calculated as the variation from the median value, i.e., using the flat field divided by the median.

\subsubsection{Pixel non-linearity and saturation} \label{sec:non_linearity}
To simulate pixel non-linearity and saturation effects for the JWST-NIRSpec\footnote{\url{https://jwst-crds.stsci.edu/browse/jwst\_nirspec\_linearity\_0025.fits}}, we utilized the published correction coefficients. We first inverted the relationship using these coefficients, applying a fifth-order polynomial to convert ideal counts to observed counts. This relationship was then adjusted so that the mean saturation point—defined as a $5\%$ deviation from the ideal measurement—reaches fixed levels of 100000 counts for FGS1 and 85000 counts for AIRS-CH0. This results in 6 coefficients for every pixel that map observed to corrected counts:
\begin{equation} \label{eq:linearity}
    F_c = c_0 + c_1 F + c_2 F^2 + c_3 F^3 + c_4 F^4 + c_5 F^5
\end{equation}
where $F$ is the observed counts, $c_n$ are the coefficients and $F_c$ is the corrected counts.

At these thresholds, pixel counts are capped to prevent any further increase, effectively simulating the saturation effects. Additionally, we introduced a randomisation of $0.5\%$ in the recomputed fifth-order polynomial correction coefficients to mimic measured values.

\subsubsection{Pointing jitter}
ExoSim2 models the impact of pointing jitter, which refers to the inevitable vibrations and drifts of the telescope in space while observing a source, by simulating the focal plane movements during the ramp integration within a ``sub-exposure''. The timelines utilised in this simulation are provided by Airbus Defence and Space (ADS), the prime contractor for the \textit{Ariel} service module. Specifically, a 10-hour long timeline sampled at 1 kHz is employed. A more detailed analysis of the timeline used is reported in \cite{Bocchieri2025} along with a detrending algorithm developed by the Ariel Consortium. Given that only one time series is available at the time of generation, we enhance the variability of disturbances induced by this single jitter realisation by randomly phase-shifting the time series for each simulation.

The pointing jitter represents a critical systematic effect in our simulations, as it directly impacts the photometric stability of the measured signal. During the integration time, the point spread function is not stationary on the detector but continuously shifts due to the telescope line-of-sight variations. As a consequence, the incoming flux is sampled by different regions of the detector, each characterised by non-uniform intra-pixel and inter-pixel response functions. This induces time-correlated fluctuations in the measured signal, even in the absence of intrinsic astrophysical variability. Such effects are particularly challenging to correct, as they couple spatial detector inhomogeneities with temporal variations in the pointing, producing a source of correlated noise that can be comparable in amplitude to the atmospheric features. This makes pointing jitter one of the primary challenges to be addressed in this dataset, and a key driver for the development and benchmarking of robust detrending techniques.

% \begin{figure*}
%     \centering
%     \includegraphics[width=\linewidth]{jitter.png}
%     \caption{The left panel shows the jitter timeline in arcseconds sampled at 1KHz on both spectral and spatial directions. The right panel shows the Power spectral density of such timelines. \comment{If we want to publish these plots we need to ask for permission by ADS}}
%     \label{fig:jitter}
% \end{figure*}

\subsubsection{Gain drift}
The simulation of gain drift in ExoSim2 models the variation in detector gain over time and across different wavelengths. The gain noise enters the payload stability budget and needs to be at a low level for high-accuracy photometry and spectroscopy. In our simulations, the amplitude of the gain drift is randomly defined between $1\%$ and $5\%$ of the incoming signal. This range is consistent with worst-case scenarios observed in JWST data.

The gain drift is applied as a 5th-order polynomial function to model the drift over time and wavelength. The polynomial coefficients are randomly drawn from a uniform distribution between -1 and 1 in both temporal and spectral directions, ensuring a comprehensive representation of potential drift behaviours.

% This approach allows the simulation to account for significant variability in detector response, enhancing the robustness of photometric measurements by compensating for time and wavelength-dependent variations in gain.

The simulated gain drift is applied to the sub-exposures by creating a multiplicative noise map that affects each pixel's signal. The drift is modelled separately for the spectrometer (AIRS-CH0) and the photometer (FGS1). For the spectrometer, the drift is calculated as a 5th-order polynomial function of both time and wavelength, resulting in different drift values for each column (representing different wavelengths) in the detector. This polynomial trend is determined by randomly generated coefficients within specified ranges, ensuring a realistic variability in the gain drift. The absence of correlation between adjacent pixel columns, expected to be strong in the real focal plane, is assumed as a worst-case scenario to provide a more conservative test case. The resulting polynomial values are then multiplied by the original signal to simulate the drift effect. For the photometer, the drift is modelled as a polynomial function of time only, creating a uniform drift effect across all pixels. This approach ensures that the gain drift is accurately represented for both the spectrometer and photometer channels, taking into account their specific characteristics and enhancing the robustness of the photometric measurements.

\subsubsection{NDRs read out mode}
\label{sec:ndrs}
For both channels, we implemented the Correlated Double Sampling (CDS) readout mode to sample the detector light collection ramp. This method involves recording one Non-Destructive Read (NDR) at the beginning of the ramp and another at the end. This strategy allows us to reconstruct the incoming flux using the equation:
\begin{equation}\label{eq:cds}
    \text{Flux} = \frac{\text{NDR}_1 - \text{NDR}_0}{\Delta t}
\end{equation}
where $\text{NDR}_0$ is the initial NDR, $\text{NDR}_1$ is the final NDR, and $\Delta t$ is the integration time between them. While more advanced ramp sampling strategies (e.g. multi-accumulation schemes) are expected to be available for Ariel, a detailed treatment of detector readout modes is beyond the scope of this work; CDS is adopted here to provide a simple and uniform framework for dataset generation and method benchmarking, while it provides at the same time the highest SNR for Ariel's photon noise limited observations, when compared to alternative detector sampling schemes.

Each ramp cycle in both channels operates with a detector clock unit set to 10 Hz, meaning each clock cycle lasts 0.1 seconds. The sequence of a complete ramp is as follows:

\begin{enumerate}
    \item Ground State: the ramp starts with the detector in a virtually empty state for 1 clock unit (0.1 s).
    \item First NDR: the first NDR ($\text{NDR}_0$) is collected, which takes another clock unit (0.1 s).
    \item Second NDR: the detector collects photons for the integration time. For FGS1, this lasts for 1 clock unit (0.1 s) to avoid saturation. For AIRS-CH0, this lasts for 45 clock units (4.5 s) to accommodate the selected stars. The second NDR ($\text{NDR}_1$) is then collected and, because the clock unit is 10 Hz, the collection lasts an additional 0.1~s.
    \item Ramp Reset: the ramp is reset for 1 clock unit (0.1~s).
\end{enumerate}

Each ramp is thus sampled by two NDRs, and the ramps are separated by an additional 2 clock units (0.2 s) for resetting and preparation for the next ramp. Therefore, the effective integration time is 0.2~s and 4.6~s for FGS1 and CH0, respectively. Each ramp lasts 0.5~s and 4.9~s, respectively. So, the duty cycle is $0.2/0.5\sim 40\%$ and $4.6/4,9\sim 93\%$, respectively.

% \subsubsection{Cosmic rays}
% The simulation of cosmic rays in ExoSim2 accounts for the interaction of energetic particles with the detector. The cosmic ray rate parameter specifies the incidence rate of cosmic rays, expressed in units of $ct/(cm^2 \cdot s)$, with a default value of 5.5, based on measurements for JWST \citep{cosmic_rays}. The impact pattern is represented by various interaction shapes for single events, each with a specific probability. In our simulations, we set an $89\%$ probability for a cosmic ray to saturate a single pixel, a $5\%$ probability to saturate two vertically aligned pixels, a $5\%$ probability to saturate two horizontally aligned pixels and a $1\%$ probability to saturate a square of $2 \times 2$ pixels.

\subsubsection{Conversion to adu} \label{sec:adu}
The Analog-to-Digital Converter (ADConv) simulates the conversion of analog signals into digital signals, represented as Analog-to-Digital Units (ADUs). This conversion process is crucial for accurately modelling the digital readout of detectors in astrophysical instruments.

The ADC conversion operates with a 16-bit resolution, providing a high degree of precision in the digital representation of the analog signal. The gain ($ADConv_{gain}$) and offset ($ADConv_{offset}$) of the ADC are set to `auto', allowing the system to dynamically adjust these parameters based on the characteristics of the measured signal ($S_{meas}$), to make use of the full values range offered by 16-bit unsigned integers. The resulting 16-bit digital signal ($S_{16bit}$) is calculated as:

\begin{equation} \label{eq:adu}
    S_{16bit} = \left\lfloor ADConv_{gain} \left( S_{meas} - ADConv_{offset} \right) \right\rfloor
\end{equation}
where the floor function $\left\lfloor \cdot \right\rfloor$ represents the integer part of the scaled signal. This approach ensures that the conversion process can accommodate a wide range of signal intensities, capturing the variability inherent in astrophysical observations.

\begin{figure*}
    \centering
    \includegraphics[width=\linewidth]{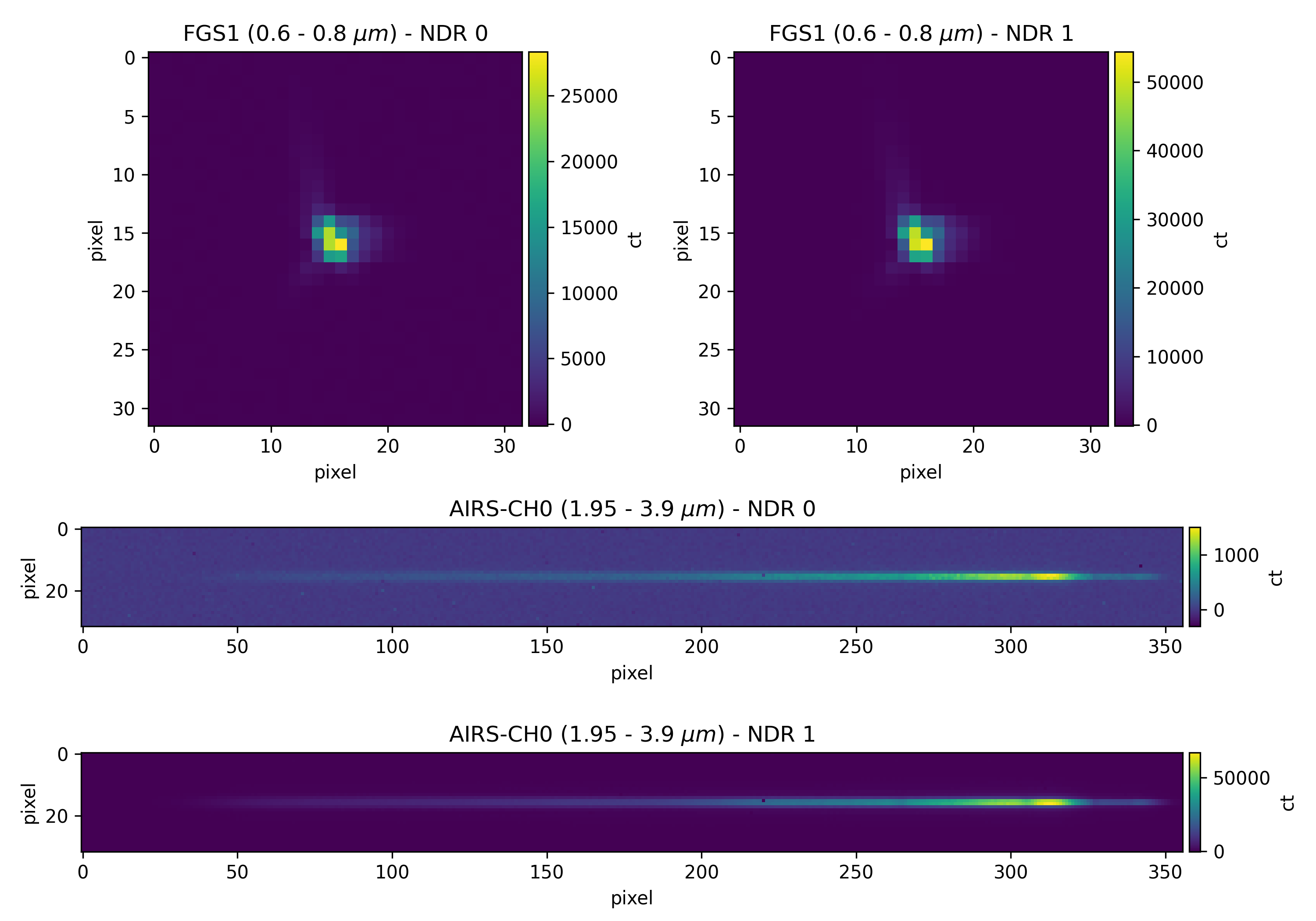}
    \caption{The figure shows the first two NDRs for every channel. NDR 0 is collected at the beginning of the photon collection ramp, while NDR 1 is collected at the end. The figures are reported in counts units as they have been converted from ADUs by inverting \autoref{eq:adu} as $S_{meas} = \frac{S_{adu}}{ADC_{gain}}+ADC_{offset}$. The top row shows the photometer FGS1 images, while the bottom row shows the spectrometer AIRS-CH0 images.}

    \label{fig:focalplanes}
\end{figure*}

Examples of the resulting raw images, converted from ADUs into counts are reported in \autoref{fig:focalplanes}.

\subsection{Target selection}
\label{sec:target}

\subsubsection{Selection of Candidate Stars}

We select our candidate stars from the Ariel candidate sample \citep{edwards2019}, which includes a diverse selection of planetary systems that meet Ariel's scientific requirements. From this pool, candidates are chosen based on their brightness levels. Here, we assumed the transit event is always captured in full, and there is only a single transit event happening each time. To adequately sample the signal from the star during the ingress and egress of the planet crossing event (transit), which typically lasts a few minutes, the exposure time must be short, typically less than 45 seconds. As this part of the challenge of this exercise is to detrend the pointing jitter, we aimed to further shorten the integration times. The jitter timeline is sampled at 1 kHz, meaning that in 45 seconds, we would combine 45,000 positions, resulting in a blurred image that is impossible to detrend as the spatial information on the focal plane has been averaged out. We set an integration time limit of an order of magnitude smaller than that: 4.5 seconds, corresponding to 450 jitter positions superimposed. To maintain consistency and comparability within the dataset, a fixed exposure time is set for all stellar targets, guiding the selection towards stars with brightness levels compatible with this observing strategy.

To ensure a high Signal-to-Noise Ratio (SNR), the stars must be sufficiently bright to fill the detector with photons in such a short time, ensuring a photon-noise dominated observation, but not so bright as to saturate the detector.

To satisfy the stringent requirement for integration time while maintaining sufficient SNR, we ended up basing our simulation around 4 host stars, as reported in \autoref{tab:stars}.

\begin{table*}
\parbox{.7\linewidth}{
    \centering
    \begin{tabular}{c|c|c|c|c|c|c}
         Name & $R_{\star}$ [$R_{\odot}$] & $M_{\star}$ [$M_{\odot}$] & $T_{\text{eff}}$ [K] & $D$ [pc] & Type & Phoenix model \\
         \hline
         KELT-11 & 2.705 & 1.439 & 5372.5 & 98 & G8/K0 &lte054.0-3.5-0.0a+0.0.BT-Settl.spec.fits \\
         HD\,17156 & 1.522 & 1.32 & 6066 & 77.16 & G0 &lte061.0-4.0-0.0a+0.0.BT-Settl.spec.fits \\
         HD\,209458 & 1.17967 & 1.1753 & 6086 & 47.45 & F9 &lte061.0-4.5-0.0a+0.0.BT-Settl.spec.fits \\
         HD\,149026 & 1.468 & 1.31 & 6158 & 78 & G0 &lte062.0-4.0-0.0a+0.0.BT-Settl.spec.fits \\
         \hline
    \end{tabular}
    \caption{Parameters of the selected sources and models used as input for ExoSim2. Parameters are selected to match \citet{edwards2019}.}
    \label{tab:stars}
}
\hfill
\parbox{.25\linewidth}{
    \centering
    \begin{tabular}{c|c}
         Parameter  &  value \\
         \hline
         mid-transit time ($t_0$)& 3.75 h \\
         period ($p$) & 3.525 d \\
         semi-major-axis ($a$) & 8.81 \\
         inclination ($i$) & 86.71 deg \\
    \end{tabular}
    \caption{Orbital parameters used for HD209458\,b as reported in \citet{Stassun2017AJ}. The mid-transit time has been set in this work to align the transit time with the simulated observation.}
    \label{tab:HD209458b}
    }
\end{table*}

Assuming detector saturation and not accounting for the duty cycle, we can estimate the relative uncertainties on the star signal given the saturation times by inverting the Signal-to-Noise Ratio ($\frac{S}{N}$) equation:
\begin{equation}
    \frac{S}{N} = \sqrt{\frac{T_{\text{obs} \ \cdot \ n_{e^-}}}{\Delta t}}
\end{equation}
where $T_{\text{obs}}$ is the observing time,  $\Delta t$ is the integration time, defined in \autoref{eq:cds}, and $n_{e^-}$ is the number of electrons. Considering the detector saturation levels, we assume $n_{e^-} = 100 \,\text{k}e^{-}$ for FGS1 and $n_{e^-} = 85 \,\text{k}e^{-}$ for AIRS-CH0. Under the assumption of saturation in the previously described CDS observation, we adopt a frame time of $\Delta t = 0.2 \,\text{s}$ for FGS1, and $\Delta t = 4.6 \,\text{s}$ for AIRS-CH0. For a total observing time of $T_{\text{obs}} = 7.5 \,\text{h}$, the relative noise can then be estimated as

\begin{equation}\label{eq:n/s_fgs1}
    \left( \frac{N}{S} \right)_{\text{FGS1}} = \sqrt{\frac{0.2 \, \text{s}}{7.5 \, \text{h} \cdot 100 \, \text{k}e^-} } = 8 \, \text{ppm}
\end{equation}
For FGS1 the duty cycle plays a significant role. Since only $40\%$ of the total $7.5 \, \text{h}$ is effectively used for integration, the relative noise increases by a factor of $1/\sqrt{0.4}$. Therefore, the expected value is
\begin{equation}
    \left( \frac{N}{S} \right)_{\text{FGS1, \, DC}} = \frac{8}{\sqrt{0.4}} = 12 \, \text{ppm}.
\end{equation}
In contrast, for AIRS the duty cycle can be neglected, and we estimate
\begin{equation}\label{eq:n/s_airs0}
    \left( \frac{N}{S} \right)_{\text{AIRS}} = \sqrt{\frac{4.6 \, \text{s}}{7.5 \, \text{h} \cdot 85 \, \text{k}e^-} } = 45 \, \text{ppm}
\end{equation}

These values represent the theoretical lower limit for the expected uncertainties for the selected stars, assuming they saturate the detector and utilise the full instrument capability.

In real observations, each star would have a different exposure time, leading to variations in dataset sizes. This tailored approach ensures that the observations are optimised for both the instrument and the target star characteristics. However, for this exercise, we standardised the integration time and the observing time to 7.5 hours. We ensured that every planet has the same transit time, and the mid-time of the simulated observation coincides with or is close to the transit event mid-time. While we realise this standardised integration and observing time setup is not necessarily representative of the normal operating conditions, the homogeneity in the datasets allows for easier introduction to ML-related approaches.

Nevertheless, the chosen observing duration remains physically realistic within the context of the Ariel mission. Ariel is expected to observe each target for a total duration of approximately 2.5 times the transit duration, in order to adequately sample both in-transit and out-of-transit baselines. Therefore, a total observing time of 7.5 hours corresponds to a transit duration of approximately 3 hours, which is representative of typical hot Jupiter systems and consistent with the class of targets considered in this work. This value is also broadly consistent with the average transit duration of the exoplanet population targeted by Ariel, ensuring that the simulated observations reflect realistic temporal scales. This choice ensures that, despite the imposed uniformity, the simulated observations retain a realistic temporal structure and signal-to-noise regime.
% \begin{itemize}
%     \item KELT-11b, 47ppm, 10s
%     \item HD 209458, 77ppm, 11s
%     \item HD 1756, 63ppm, 17s
%     \item HD 145026, 93ppm, 19s
% \end{itemize}

\subsubsection{Stellar models}
\begin{figure}
    \centering
    \includegraphics[width=\linewidth]{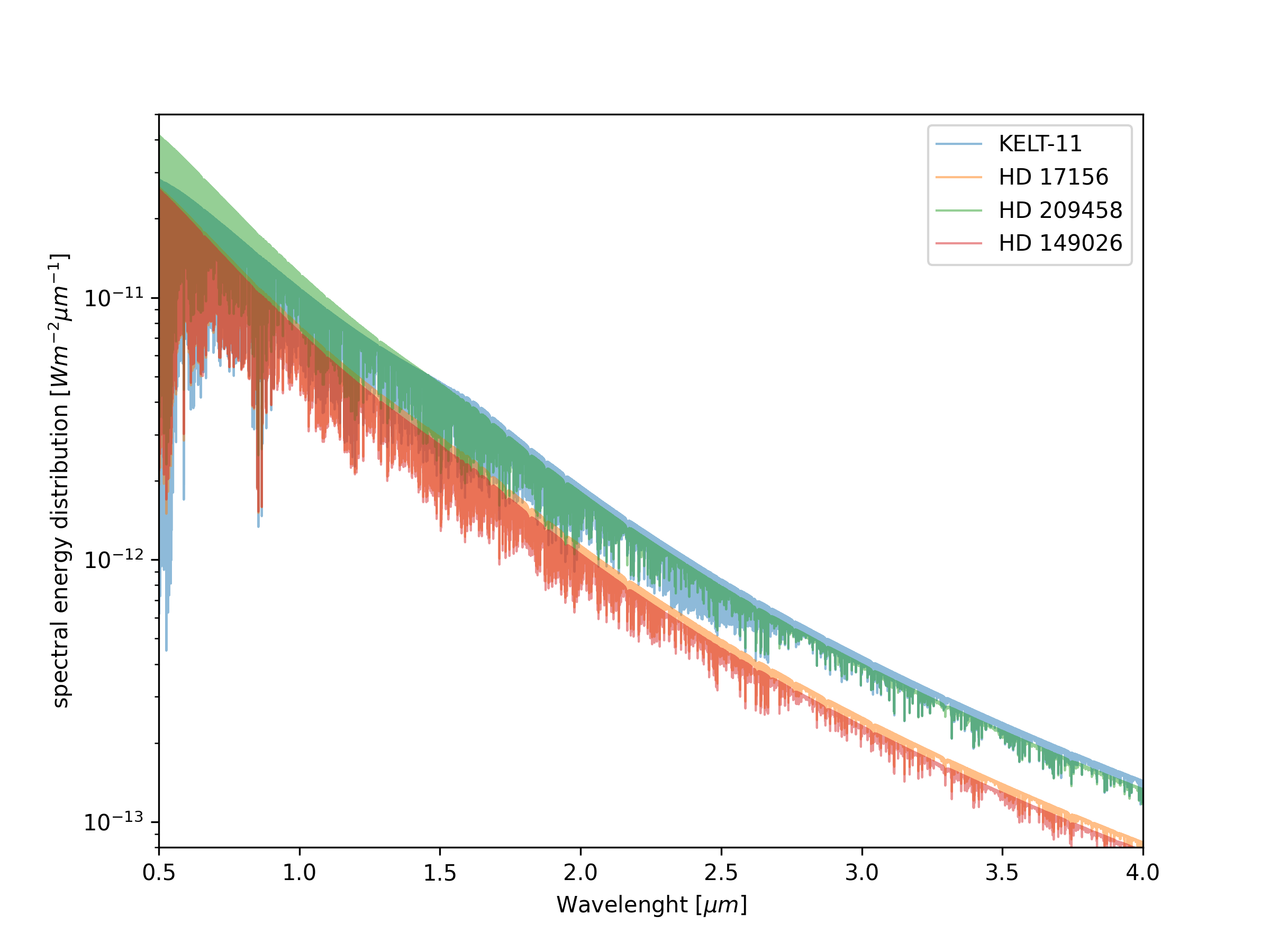}
    \caption{Stellar Spectral Energy Distribution (SED) used in this work for each star used as target and reported in \autoref{tab:stars}.}
    \label{fig:stars}
\end{figure}

To simulate the incoming light from the star, we use the stellar Spectral Energy Distribution (SED) obtained from a grid of synthetic Phoenix spectra\footnote{\url{https://phoenix.ens-lyon.fr/Grids/BT-Settl/CIFIST2011_2015/FITS/}} \citep{Baraffe2015}. The best matching model is selected based on stellar temperature, surface gravity (logg), and metallicity. The specific model for each star is chosen from the indicated SED grid according to these parameters, scaled by $(R_{\star}/D)^2$ to account for the star distance and is reported in \autoref{tab:stars}. The resulting SED has units of $W / (m^2 \cdot \mu m)$ and it is reported in \autoref{fig:stars}.

\subsection{Planetary model}
\label{sec:planets}

This section describes the planetary models used in dataset generation. To ensure diverse representation of the planetary population, we varied the following categories of parameters: planetary configurations ($R_p$, inclination, semi-major axis, mid-transit time) and atmospheric features (equilibrium temperature, trace gases detailed in \autoref{tab:linelist_and_prior}, and clouds) during the simulation process. We note that these variations cover hotter and larger exoplanets to enrich the diversity. \autoref{tab:scens} provides a detailed breakdown of varied parameters; below we describe the planetary configurations and atmospheric models in detail.

\subsubsection{Planetary Configuration}
The planetary configuration concerns with the specifies of the planetary system.

The configuration of the host star follows the targets chosen in \autoref{tab:stars}. We opt to build examples using synthetic datasets, i.e. synthetic planets orbiting around our selected host stars. We sampled the radii of different synthetic planets from a uniform distribution and used a Splined-based mass radius predictor to infer the corresponding mass (Yip et al. in prep) from the uniformly sampled radius. We have set a lower bound in the planet's mass as 10 Earth Masses in this investigation. As part of the effort to maintain a homogeneous dataset, we have assumed these planets to resemble very similar orbital configurations as HD209458\,b, as described in \autoref{tab:HD209458b}. This assumption is relaxed in some cases, as indicated in \autoref{tab:scens}. Fixed settings mean the configuration for HD209458\,b is used, otherwise, they are varied according to a specified uniform distribution. This is to allow more extensive tests on the algorithm based on possible scenarios that could occur during an observation.

% \begin{table}[]
%     \centering

% \end{table}
%In some test cases, actual planet configurations are used. The exact details of the simulation is shown in Table \ref{tab:scens}.

\subsubsection{Atmospheric Model}

To generate realistic simulations of observations from distant exoplanets and their atmospheres, we employed the forward modelling capabilities of TauREx3. This tool allowed us to create hypothetical atmospheric modulations for these observations.

Within reasonable assumptions for our synthetic planets, and with the goal of providing a benchmark dataset for data detrending, we set up a number of ``global'' assumptions for our atmospheric model: we created all the atmospheres with a 1D plane-parallel primary atmosphere composed of H/He at a solar abundance ratio of 0.17, along with additional atmospheric absorption such as Collision-Induced-Absorption (C.I.A.) and scattering effects such as Rayleigh Scattering. Furthermore, we assumed an isothermal Temperature-Pressure profile. We used the equilibrium temperature of the planet as the temperature of the planet's atmosphere. The equilibrium temperature is calculated via the following relationship from \cite{Tinetti2013}:
\begin{align}
    T_{eqm} = T_{\star} \times \sqrt{\frac{R_{\star}}{2a}} \times (1-A_B)^{1/4}
\end{align}

Where we have assumed the Bond albedo ($A_B$) to be 0, and that the host stars' radiation follows a blackbody. The exact values of $T_{\star}$ and $R_{\star}$ are varied according to their measured uncertainty (see \autoref{tab:stars}). The extent of the atmosphere is set from  $10^{-6}$ Pa to $10^6$ Pa. All observation is conducted in transmission mode and is generated in native resolution as provided from the ExoMol linelist data with no binning applied (see relevant references in \autoref{tab:linelist_and_prior}). In some cases, we implemented a grey cloud deck with pressures varying from 3 to 6 bar. We note here that the level of variation does not fully capture current findings for hot Jupiters, where high-altitude clouds could exist at lower pressures; this limitation will be addressed in future editions of the dataset.

\begin{figure}
    \centering
    \includegraphics[width=1.\linewidth]{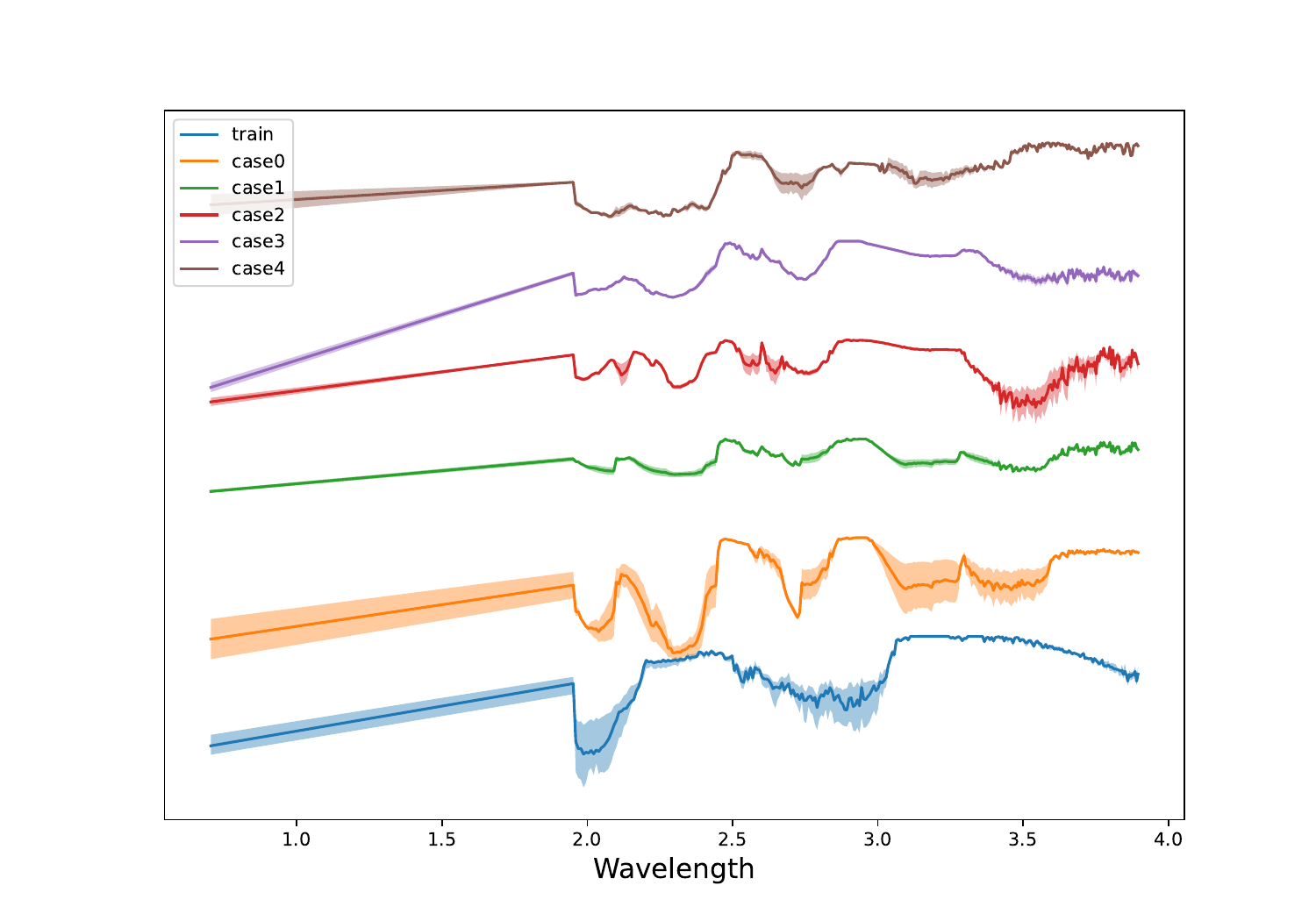}
    \caption{Visualisation in spectral space of the different training and test cases presented in \autoref{tab:scens}. For each spectrum, we show the mean spectrum for that training or test case and the spread of the spectra. Offsets are applied to individual spectra for better visualisation.}
    \label{fig:mean_distribution}
\end{figure}

Aside from these global assumptions, trace gases of various abundances are added to the atmosphere depending on the different scenarios. All of them are assumed to have constant chemical abundance throughout the extent of the atmosphere. Please see \autoref{tab:scens} for the exact species used in each case and \autoref{fig:mean_distribution} for a comparison of each case in spectral space. We note that some of these combinations and prior ranges may not be physically motivated and do not fully capture the diversity of targets that Ariel is going to observe. These settings are set up to prevent any algorithms from learning to detrend by having prior assumptions on the spectral feature of particular molecules or combinations of molecules (in other words, having prior knowledge before the extraction), and to stress test any potential models with their ability to detrend accurately and precisely, even in unknown and unseen scenarios. On top of that, we have also discarded spectra with NaN values and with transit depth larger than 2\%.

\begin{table*}
\resizebox{1.0\textwidth}{!}{
\begin{tabular}{|l|l|l|l|l|l|l|l|l|l|}
\hline
\multicolumn{1}{|c|}{\textbf{}} & \multicolumn{1}{c|}{\textbf{Training}} & \multicolumn{1}{c|}{\textbf{Training-Ext}} & \multicolumn{1}{c|}{\textbf{Test case 0}} & \multicolumn{1}{c|}{\textbf{Test case 1}} & \multicolumn{1}{c|}{\textbf{Test case 2}} & \multicolumn{1}{c|}{\textbf{Test case 3}} & \multicolumn{1}{c|}{\textbf{Test case 4}} & \multicolumn{1}{c|}{\textbf{Test case 5}}  & \multicolumn{1}{c|}{\textbf{Calib case}} \\ \hline
\multicolumn{1}{|l|}{\textbf{Number of data}} & \multicolumn{1}{l|}{576} & \multicolumn{1}{l|}{97} & \multicolumn{1}{l|}{104} & \multicolumn{1}{l|}{102} & \multicolumn{1}{l|}{103} & \multicolumn{1}{l|}{105} & \multicolumn{1}{l|}{105} & \multicolumn{1}{l|}{231} & \multicolumn{1}{l|}{4} \\ \hline
\multicolumn{1}{|l|}{\textbf{$R_{\rm p}$ range ($R_{\rm Jup}$)}} & \multicolumn{1}{l|}{0.5 - 1.2} & \multicolumn{1}{l|}{0.5 - 1.2} & \multicolumn{1}{l|}{0.5 - 1.2} & \multicolumn{1}{l|}{0.5 - 1.2} & \multicolumn{1}{l|}{0.5 - 1.2} & \multicolumn{1}{l|}{0.2 - 1.5} & \multicolumn{1}{l|}{0.5 - 1.2} & \multicolumn{1}{l|}{0.5 - 1.2} & \multicolumn{1}{l|}{Actual planet} \\ \hline
\multicolumn{1}{|l|}{\textbf{Host Stars}} & \multicolumn{1}{l|}{KELT-11, HD 17194} & \multicolumn{1}{l|}{KELT-11, HD 17194} & \multicolumn{1}{l|}{KELT 11, HD 17194} & \multicolumn{1}{l|}{KELT-11, HD 17194} & \multicolumn{1}{l|}{KELT-11, HD 17194} & \multicolumn{1}{l|}{HD 209458, HD 149026} & \multicolumn{1}{l|}{KELT-11, HD 17194} & \multicolumn{1}{l|}{KELT-11, HD 17194} & \multicolumn{1}{l|}{All four} \\ \hline
\multicolumn{1}{|l|}{\textbf{Trace gases}} & \multicolumn{1}{l|}{H$_2$O, CH$_4$, CO$_2$, Cloud} & \multicolumn{1}{l|}{H$_2$O, CH$_4$, CO$_2$, Cloud} & \multicolumn{1}{l|}{NH$_3$, C$_2$H$_2$, SO$_2$, SiO} & \multicolumn{1}{l|}{NH$_3$, C$_2$H$_2$, SO$_2$, SiO} & \multicolumn{1}{l|}{NH$_3$, C$_2$H$_2$, SO$_2$, SiO} & \multicolumn{1}{l|}{NH$_3$, C$_2$H$_2$, SO$_2$, SiO} & \multicolumn{1}{l|}{HCN, H$_2$S, AlH and CO} & \multicolumn{1}{l|}{HCN, H$_2$S, AlH and CO} & \multicolumn{1}{l|}{flat} \\ \hline
\multicolumn{1}{|l|}{\textbf{Detector type}} & \multicolumn{1}{l|}{Model A} & \multicolumn{1}{l|}{Model A} & \multicolumn{1}{l|}{Model A} & \multicolumn{1}{l|}{Model B} & \multicolumn{1}{l|}{Model C} & \multicolumn{1}{l|}{Model A} & \multicolumn{1}{l|}{Model A} & \multicolumn{1}{l|}{Model A} & \multicolumn{1}{l|}{Model A}  \\ \hline
\multicolumn{1}{|l|}{\textbf{Semi-Major Axis}} & \multicolumn{1}{l|}{Fixed} & \multicolumn{1}{l|}{$\mathcal{U}$(8.81, 11)} & \multicolumn{1}{l|}{Fixed} & \multicolumn{1}{l|}{Fixed} & \multicolumn{1}{l|}{Fixed} & \multicolumn{1}{l|}{Fixed} & \multicolumn{1}{l|}{Fixed}  & \multicolumn{1}{l|}{$\mathcal{U}$(8.81, 15.86)} & \multicolumn{1}{l|}{Fixed} \\ \hline
\multicolumn{1}{|l|}{\textbf{Inclination}} & \multicolumn{1}{l|}{Fixed} & \multicolumn{1}{l|}{$\mathcal{U}$(86.71,88)}  & \multicolumn{1}{l|}{Fixed} & \multicolumn{1}{l|}{Fixed} & \multicolumn{1}{l|}{Fixed} & \multicolumn{1}{l|}{Fixed} & \multicolumn{1}{l|}{Fixed} & \multicolumn{1}{l|}{$\mathcal{U}$(87,90)} & \multicolumn{1}{l|}{Fixed} \\ \hline
\multicolumn{1}{|l|}{\textbf{Mid Transit Time}} & \multicolumn{1}{l|}{Fixed} & \multicolumn{1}{l|}{$\mathcal{U}$($t_0$-0.1,$t_0$+0.1)} & \multicolumn{1}{l|}{Fixed} & \multicolumn{1}{l|}{Fixed} & \multicolumn{1}{l|}{Fixed} & \multicolumn{1}{l|}{Fixed} & \multicolumn{1}{l|}{Fixed} & \multicolumn{1}{l|}{$\mathcal{U}$($t_0$-0.5,$t_0$+0.5)} & \multicolumn{1}{l|}{Fixed} \\ \hline
\end{tabular}
}
\caption{Specification of different train test scenarios, including the $R_p$ range used to uniformly sample planet radii, host stars used as the sources of flux, trace gases included in the atmospheres, and different detector types and noise maps for the instrument. $\mathcal{U}(x_1, x_2)$ represents a uniform distribution between $x_1$ and $x_2$.}
\label{tab:scens}
\end{table*}

\begin{table}
\centering
\resizebox{0.5\textwidth}{!}{%
\begin{tabular}{lll}
\hline
Parameters & Abundance Range (log$_{10}$) & Line list reference \\ \hline
H2O        & -8 - -4       & \citet{Polyansky2018}                 \\
CO2        & -8 - -4       & \citet{Rothman2010}                 \\
CH4        & -8 - -4       & \citet{Yurchenko2017}                 \\
NH3        &    -8 - -4     & \citet{Coles2019}                 \\
C$_2$H$_2$ &   -8 - -4     & \citet{Chubb2020}                 \\
SO$_2$     &  -8 - -4     & \citet{Underwood2016}                 \\
SiO        &  -8 - -4         & \citet{SiO}                 \\
HCN        &   -8 - -4        & \citet{Barber2014}                 \\
H$_2$S     &  -8 - -4         & \citet{Azzam2016}                 \\
AlH        &  -8 - -4        & \citet{Yurchenko_AlH}                 \\
CO         &   -8 - -4       & \citet{Li2015}                 \\
grey cloud (Pa) &   3 - 6        & N/A
\end{tabular}%
}
\caption{Abundance range for sampling different physical parameters and their corresponding line list references.}
\label{tab:linelist_and_prior}
\end{table}

%% write down the consideration
%% description the forward model being used, basic assumptions and

\begin{figure}
    \centering
    \includegraphics[width=1\linewidth]{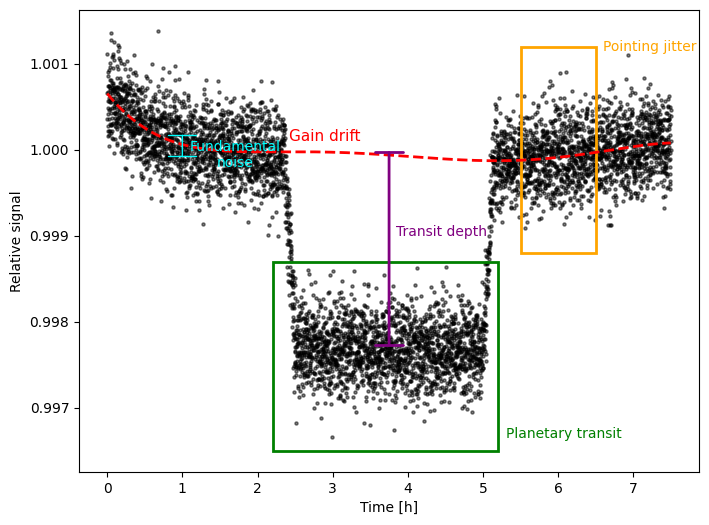}
    \caption{White light curve (i.e., the resultant light curve from using the spectrometer as a single wavelength channel) for one of the simulated observations with Ariel AIRS-CH0. The signal dip from the transit event is highlighted in the green box, with the depth of the transit also marked in purple. We highlighted other significant features: the spread of the measurements due to high-frequency and quantum noise (cyan), the polynomial drift due to simulated gain variations (red), and the photometric noise caused by the jitter disturbance affecting the measurements (orange), mostly embedded in the overall noise in this case.}
    \label{fig:enter-label}
\end{figure}

\subsection{Data production and sanitation}
\label{sec:production}
For each simulated observation, we produced three versions of the same dataset. (a) The first version includes all noise sources, representing a complex dataset on which the pipelines can be trained and tested under realistic conditions. (b) A second version excludes time-correlated noise components, such as jitter and gain drift, which require dedicated analysis strategies. Finally, (c) a third version is generated without any noise or uncertainty, serving as a reference dataset.

To generate these datasets, we developed a distributed computing system that operates across multiple machines within the Department of Physics and Astronomy at Cardiff University. This distributed approach allowed us to produce the entire dataset (over 7TB, as detailed later) in less than a day of computing time.

Since ExoSim2 is designed to model the effects of various instrumental subsystems in detail, its output includes all intermediate products required to reconstruct the simulated observation. It was therefore necessary to sanitise the data to provide only the essential versions. Specifically, we removed all auxiliary data, retaining only the data cube with the simulated images and the critical information necessary for their interpretation (wavelength solution, time axis, and integration times for each image). In addition, we supplied a separate set of calibration data with parameters uniformly altered by 5\% to simulate realistic uncertainties in calibration products, rather than providing the exact inputs used in the data generation process.

\subsubsection{Dataset statistics}

The complete dataset generated for this exercise comprises 7.0 TB and is stored on the Cardiff University and DiRAC CSD3 server, accessible upon request. This dataset includes 1435 simulated observations, as described in Section \ref{sec:production}. Each simulated observation weighs approximately 4.5 GB, as it contains all intermediate products of the simulation.

A sanitised version of the training dataset, publicly available, consists of 673 observations with a total size of approximately 174.33 GB (with each observation in the sanitised version reduced to about 500 MB). This public dataset includes images, calibration data, and ground truth spectra, as detailed in Section \ref{sec:validation}. All observations were simulated using the same integration and observation times to ensure uniformity. Each dataset contains 135000 images for FGS1 and 11250 images for AIRS-CH0. The test dataset, while hidden on Kaggle, will be made available on Zenodo, it has a total size of 184GB with 758 files

Each pair of images represents the extremes Non-Destructive Readouts (NDRs) in a Correlated Double Sampling (CDS) readout mode for the detector ramp. These images are separated by 0.2 seconds for FGS1 and 4.6 seconds for AIRS-CH0, resulting in 67500 CDS pairs for FGS1 and 5625 CDS pairs for AIRS-CH0. The total observing time across all observations is 7.5 hr.

A summary of this information, including the number of observations, images, and the total observing time, is provided in Table \ref{tab:stats}.

\begin{table}
    \centering
    \begin{tabular}{r|c|c}
        \hline
         \textbf{Parameter}& \textbf{FGS 1} & \textbf{AIRS-CH0} \\
        \hline
        Observing time & 7.5 hr& 7.5 hr\\
        Integration time & 0.2 s& 4.6 s \\
        Number of images per obs. & 135000 & 11250\\
        Number of CDS couples & 67500& 5625\\
        Size of image & $32 \times 32$& $32 \times 356$\\
        \hline
    \end{tabular}
    \caption{Dataset information for the two detectors.}
    \label{tab:stats}
\end{table}

\subsection{Data validation and overview}
\label{sec:validation}

ExoSim2 provides a detailed time-domain model of both the payload and astrophysical scene, enabling accurate simulation of Ariel's observational data with realistic noise and systematic effects. The simulator has undergone rigorous validation, with each module tested to achieve over 90\% test coverage, ensuring robust and reliable outputs.
Given the substantial scale of the generated dataset, comprehensive end-to-end validation becomes critical for confirming input accuracy and minimizing potential errors. This section examines the validation methodology and presents an overview of the resulting dataset, demonstrating its value for the Data Challenge while highlighting opportunities for broader scientific exploration.

\subsubsection{Data processing pipeline}
The data processing pipeline was developed to accurately extract transit signals, evaluate noise properties, and validate the dataset against theoretical models. The pipeline proposed here is a basic implementation that does not reflect the official Ariel pipeline, which is still under development at the time of writing. The methodology can be divided into three main phases: preprocessing, image construction, and spectral analysis.

In the preprocessing phase, the quality of the raw data was ensured by identifying and masking defective pixels, such as dead or hot pixels. This was achieved using calibration data and the \texttt{sigma\_clip} method provided by the \texttt{astropy} package. Following this, the raw 16-bit digital data were converted into 64-bit floating-point values. During this conversion, corrections for the analog-to-digital gain (\(ADConv_{gain}\)) and offset (\(ADConv_{offset}\)) were applied using the calibration files (see Section~\ref{sec:adu} and Equation~\ref{eq:adu}). Pixel non-linearity, a common feature of detector response, was then corrected using a 5th-order polynomial fit derived from calibration coefficients (see Section~\ref{sec:non_linearity} and Equation~\ref{eq:linearity}). Finally, the contribution of dark current was subtracted for each pixel, scaled appropriately by the integration time, using detector-specific calibration maps (see Equation~\ref{eq:signal-meas}).

The second phase, image construction, involved creating science frames optimised for analysis. First, Correlated Double Sampling (CDS) frames were computed by subtracting the signal from the first and last non-destructive reads (NDRs) within each integration ramp (see Equation~\ref{eq:cds}). Subsequently, pixel-to-pixel response variations were corrected by normalising the CDS frames with flat-field coefficients provided in the calibration data. This step ensured a uniform response across the detector array. The signals were then integrated across the images to obtain transit light curves. For photometric observations (FGS1), the total flux was summed over the defined aperture, while for spectroscopic observations (AIRS-CH0), the signal was integrated along pixel columns and associated with specific wavelengths. Any data outside the target wavelength range of 1.95–3.9 $\mu$m was excluded to maintain spectral consistency.

The final phase of the data processing pipeline focused on extracting transit depths and estimating noise properties from the transit light curves obtained in the last step. To isolate the transit signal, the timeline was divided into transit and out-of-transit regions. This separation leveraged the symmetry of the transit event or, when necessary, an estimation of the timeline mid-point. Margins were applied around the ingress (\(T_1\)) and egress (\(T_4\)) to account for potential edge effects. The transit depth was then calculated using the relationship:
\begin{equation} \label{eq:transit-depth}
    d(t, \lambda) = 1 - \frac{F_{in}}{F_{out}},
\end{equation}
where \(F_{in}\) and \(F_{out}\) represent the stellar flux rates (F($\lambda$), in e$^-$ s$^{-1}$) measured during and outside the transit, respectively. To first order, the transit depth can be approximated as \(d(t, \lambda) \simeq (R_p/R_*)^2\) \citep[see, e.g.,][]{Sing2018}, linking the depth to the relative sizes of the planet and the star.
The out-of-transit timelines were then detrended by fitting a 5th-order polynomial, removing systematic gain drifts.

The expected uncertainty on the transit depth, derived from Equation~\ref{eq:transit-depth}, is expressed as:
\begin{equation} \label{eq:transit-depth-sigma}
    \sigma_d^2 = \left(\frac{1}{t_{in} F_{out}}\right)^2 v_{in} + \left(\frac{F_{in}}{t_{out} F_{out}^2}\right)^2 v_{out},
\end{equation}
where \(t_{in}\) and \(t_{out}\) are the times spent in and out of transit and \(v_{in}\) and \(v_{out}\) represent the time-integrated variances in and out of transit. These variances are estimated from the residuals between the measured light curves and the ground truth transit models.

% These uncertainties are calculated as the sum of photon noise and read noise, the latter being derived from the detector inputs (see Section~\ref{sec:calibration_products}). In the case of the simulated observation affected only by fundamental noise (see Section~\ref{sec:production}) we could directly compare the residual scatter to the expected theoretical noise to assess the precision of the simulations.

This simple prototype pipeline, although not designed for comprehensive data reduction, successfully isolates planetary transit signals and quantifies measurement uncertainties, validating the reliability of the dataset.

\subsubsection{Sample validation}
This section details the validation performed on a subset of simulated observations to evaluate signal and noise levels. For this analysis, we selected a subsample of the dataset comprising 100 examples. These examples have the largest peak-to-valley measured across their ground truth spectra than all the other examples in the dataset. This initial sample was further pruned to include only those observations where the transit mid-point was centred within the timeline. The final subsample consisted of 46 simulated observations.

\begin{figure}
    \centering
    \includegraphics[width=1\linewidth]{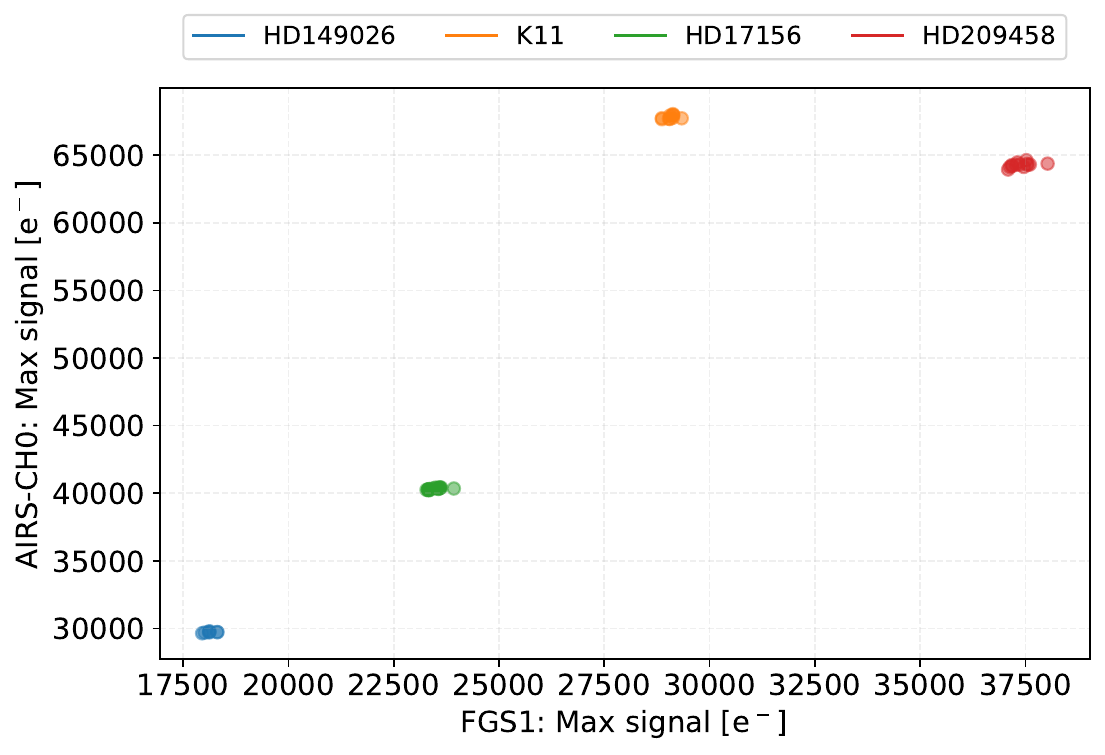}
    \caption{Maximum signal detected out-of-transit for the photometer (FGS1) and spectrometer (AIRS-CH0). The colour code indicates the stellar targets, illustrating how differences in stellar spectral energy distributions influence the observed signals.}
    \label{fig:max_signal_star_colors}
\end{figure}

The first step of this validation was to compute the maximum signal in each channel. This was done by taking the median value along the time axis of the out-of-transit CDS science frames and identifying the pixel with the highest signal. The results, presented in Figure~\ref{fig:max_signal_star_colors}, show that the signal spans from 17.6\,\(\text{k}e^-\) to 37.5\,\(\text{k}e^-\) in FGS1 and from 30\,\(\text{k}e^-\) to 68\,\(\text{k}e^-\) in AIRS-CH0. These ranges probe different depths of the detector well in both channels while avoiding saturation. Differences in the stellar spectral energy distributions (SEDs) explain why KELT-11 exhibits the highest signal in AIRS-CH0 with respect to the signal measured in AIRS-CH0 for the other stars. Similarly, the HD\,209458 signal measured in FGS1 dominates with respect to the FGS1 signal obtained for the rest of the stellar sample. Overall, we observe that for each of these stars, the AIRS-CH0 signal is stronger than that in FGS1, as expected.

% Using the median of the maximum signal for each star in each channel, we calculated the theoretical noise levels based on Equations~\ref{eq:n/s_fgs1} and~\ref{eq:n/s_airs0}. The results are summarized in Table~\ref{tab:min_n/s}, representing the theoretical lower limits for uncertainties under the given observing conditions.

% \begin{table}
% \centering
% \caption{Theoretical minimum noise-to-signal ratio (N/S) in parts per million (ppm) for each stellar target and channel, computed using the median maximum signals from Figure~\ref{fig:max_signal_star_colors}.}
% \label{tab:min_n/s}
% \begin{tabular}{@{}c|c|c@{}}
% \toprule
% \textbf{\begin{tabular}[c]{@{}c@{}}Channel/\\ Star\end{tabular}} & \textbf{\begin{tabular}[c]{@{}c@{}}FGS1\\ {[}ppm{]}\end{tabular}} & \textbf{\begin{tabular}[c]{@{}c@{}}AIRS-CH0\\ {[}ppm{]}\end{tabular}} \\ \midrule
% \multicolumn{1}{|c|}{HD\,149026} & 14.3 & 74.9 \\
% \multicolumn{1}{|c|}{KELT-11} & 11.3 & 49.6 \\
% \multicolumn{1}{|c|}{HD\,17156} & 12.6 & 64.3 \\
% \multicolumn{1}{|c|}{HD\,209458} & 10.0 & 50.9 \\ \bottomrule
% \end{tabular}
% \end{table}

From here, the validation strategy is to use two different versions of the dataset (introduced in~\ref{sec:production}) for a self-consistent and homogeneous comparison of their corresponding results, with utmost control on the simulation inputs.
Specifically, we compare the dataset version including all noise sources (version \textbf{a}; hereafter, ``test data'') with that containing only fundamental noise sources (version \textbf{c}; hereafter, ``reference'').

\begin{figure}
    \centering
    \includegraphics[width=1\linewidth]{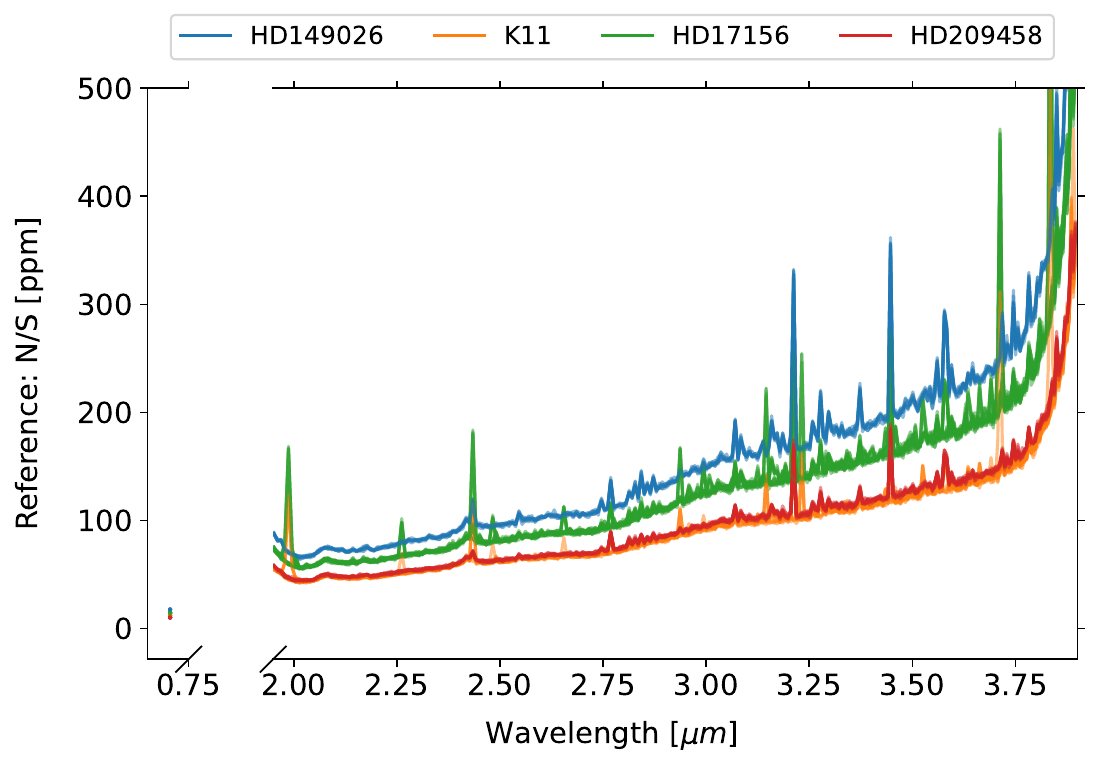}
    \caption{Relative noise (ppm) as a function of wavelength for the reference timelines. The colour coding represents different stellar targets, reflecting how their SEDs influence the noise levels across the spectrum. The large spikes come from detector imperfections such as bad pixels that are left uncorrected.}
    \label{fig:reference_N/S_star_colors}
\end{figure}

Figure~\ref{fig:reference_N/S_star_colors} illustrates the relative noise as a function of wavelength, calculated from the out-of-transit variance in the timelines and rescaled to the observation duration. Here, we use the reference timelines.
Differences between stars arise from their distinct SEDs. Differences in the curves with the same star arise from variations in the simulated detector, as we show in the following.
Spectral features below the noise curves in this figure are effectively undetectable, even assuming ideal data reduction.

Then, we evaluated the excess relative noise in the test data, by comparing it to the reference timelines. In this way, any excess noise measured in the test data with respect to the reference data can be traced to noise sources that are not corrected or insufficiently corrected, e.g., from correlated and time-dependent systematics.
The noise excess, when present, can translate to biases in the measured transit spectra, hindering our interpretation of atmospheric results.

\begin{figure}
    \centering
    \includegraphics[width=1\linewidth]{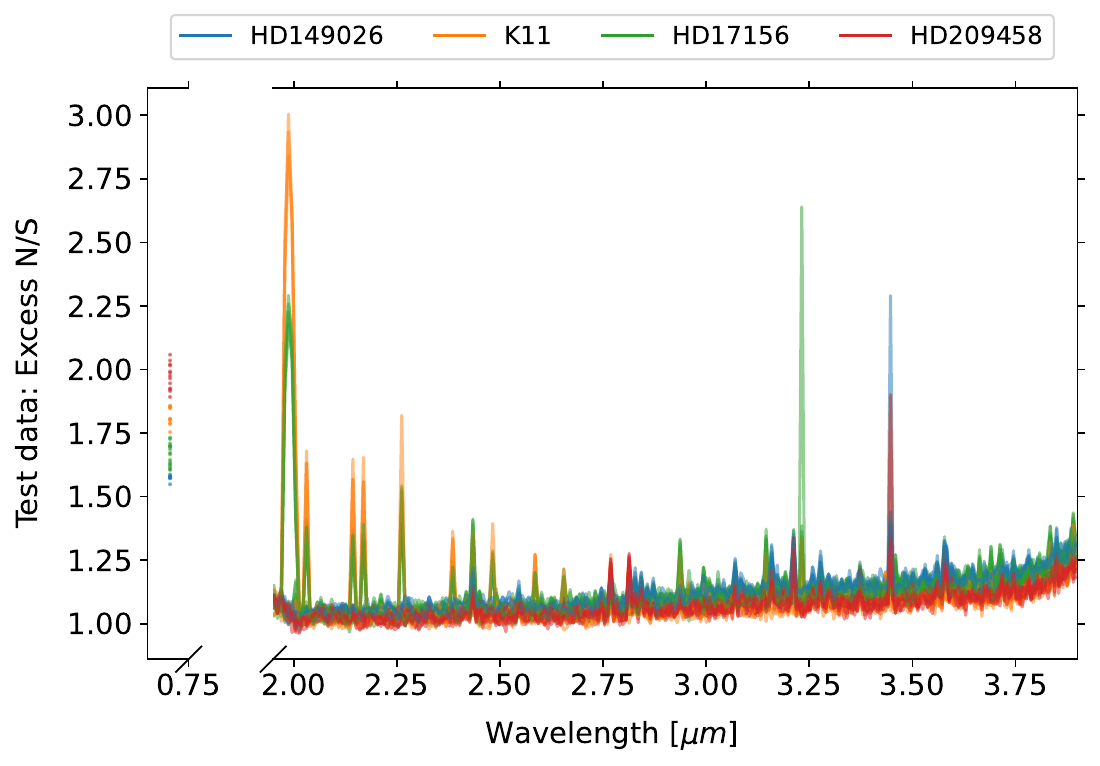}
    \caption{Excess relative noise (ppm) in the test data compared to the reference timelines, plotted as a function of wavelength. Colour coding corresponds to stellar targets, highlighting trends due to line-of-sight jitter and noise dominance.}
    \label{fig:excess_N/S_star_colors}
\end{figure}

\begin{figure}
    \centering
    \includegraphics[width=1\linewidth]{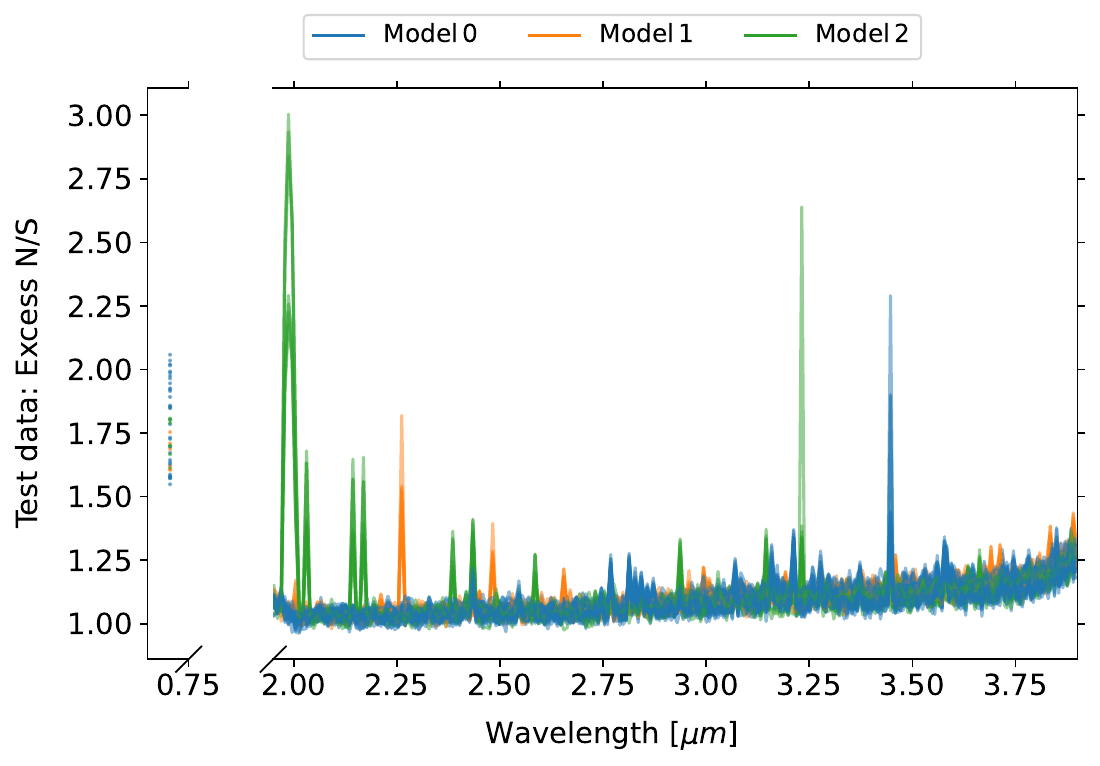}
    \caption{Excess relative noise (ppm) as in Figure~\ref{fig:excess_N/S_star_colors}, but with colour coding representing each of the different detector models used in the data creation.
    }
    \label{fig:excess_N/S_detector_colors}
\end{figure}

Figure~\ref{fig:excess_N/S_star_colors} shows that in FGS1, excess noise increases with the signal level due to uncorrected line-of-sight jitter dominating over fundamental noise sources. This trend is also visible in the brightest spectral bins of AIRS-CH0 at the blue edge of the band but reverses for wavelengths longer than 2.2 \(\mu\)m, where photon noise and readout noise dominate. Detector-specific effects, such as bad pixels, result in the additional noise spikes shown in the figure. This is evidenced in Figure~\ref{fig:excess_N/S_detector_colors}, where the color coding represents the different detector models used in the simulations. Each detector model has bad pixels appearing at different positions along the spectral trace, which cause an increase in the measured noise in the corresponding wavelength bins when no bad pixel interpolation is applied.

\begin{figure*}
    \centering
    \includegraphics[width=0.8\textwidth]{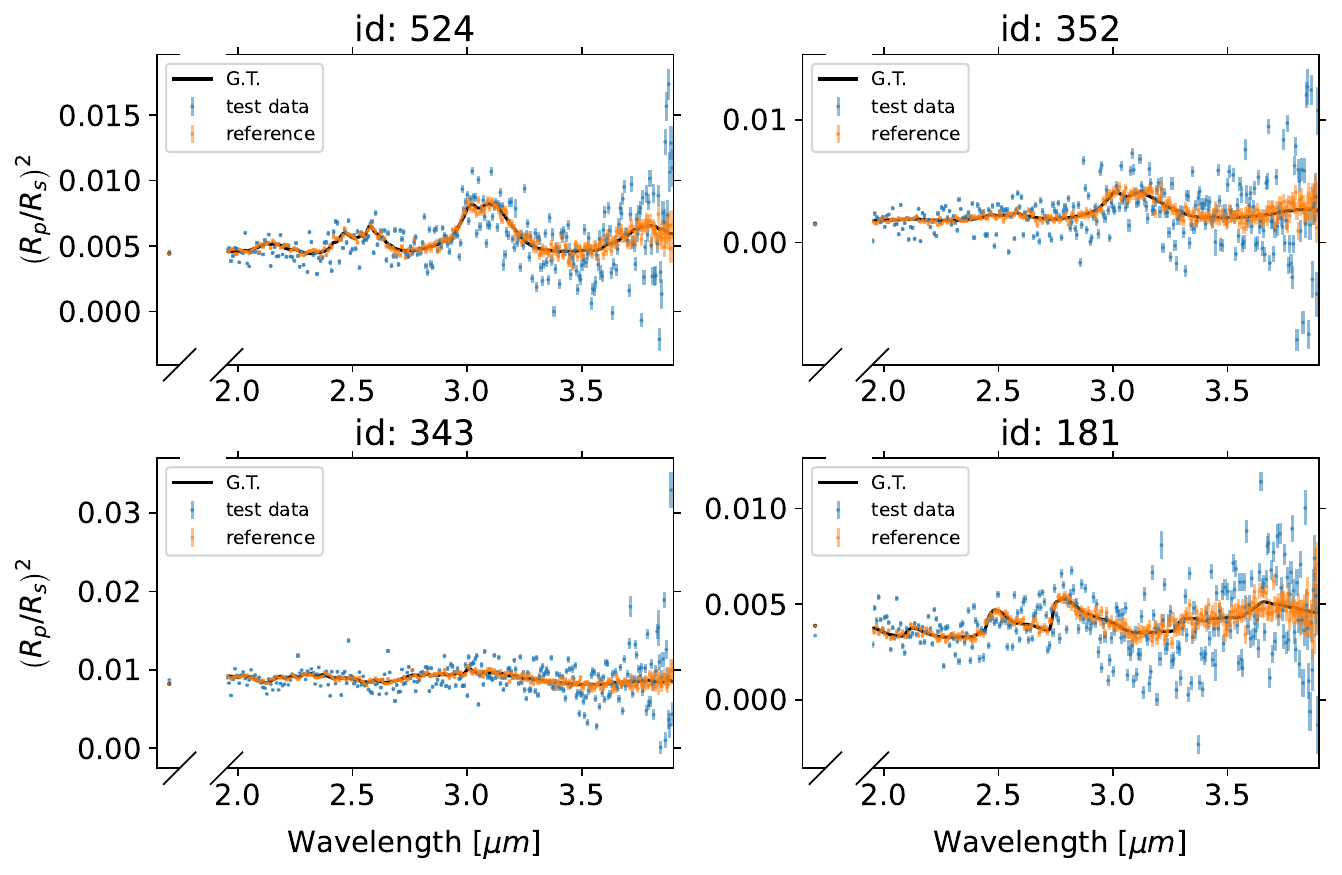}
    \caption{Comparison of transit spectra for selected examples from the subsample. Black: ground truth spectra; orange: reference data; blue: test data. Error bars are included.}
    \label{fig:rprs2_AB}
\end{figure*}

\begin{figure}
    \centering
    \includegraphics[width=1\linewidth]{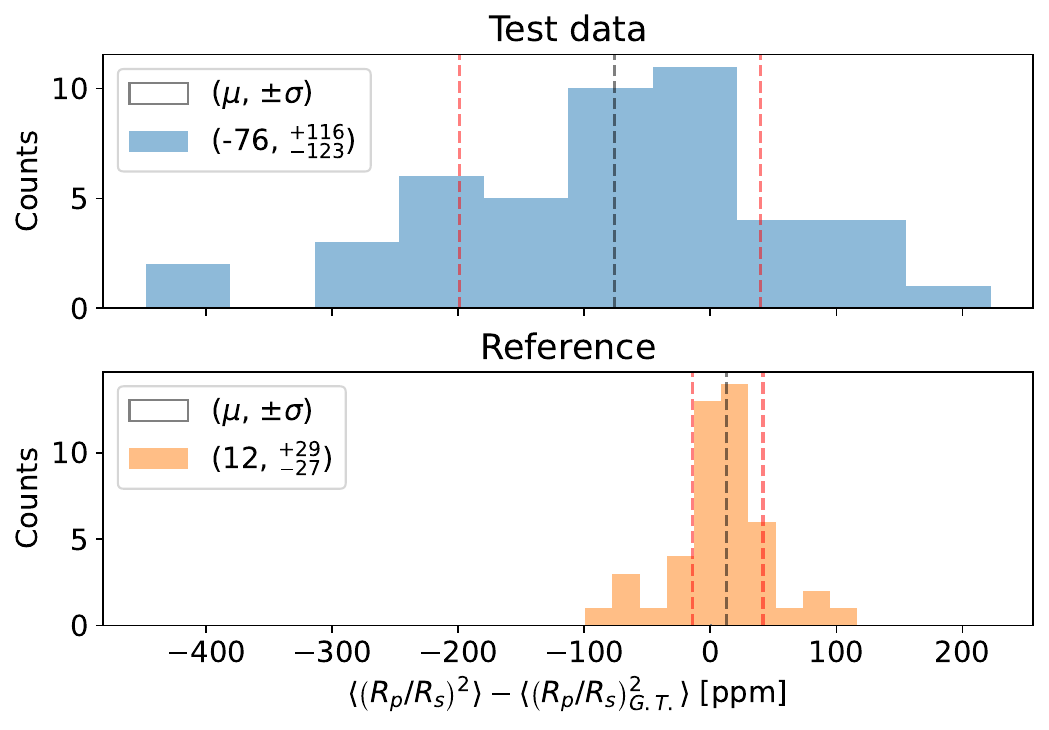}
    \caption{Histogram of residuals between median extracted transit depths and ground truth. The black line shows the mean and the red dash line shows the standard deviation. \textit{Top panel}: test data, showing larger deviations and a negative bias. \textit{Bottom panel}: reference data, with deviations within \(\pm 100\) ppm.}
    \label{fig:hist_bias_AB}
\end{figure}

Next, we assessed the accuracy of the extracted transit depths by comparing test and reference spectra to the ground truth. Figure~\ref{fig:rprs2_AB} shows examples of extracted spectra, where the reference timelines align well with the ground truth, with 68.6\(^{+3.0}_{-2.7}\)\% of points within the 1-\(\sigma\) interval. For the test data, only 17.0\(^{+3.5}_{-2.1}\)\% fall within this range, indicating the limitations of the simplistic reduction applied. Figure~\ref{fig:hist_bias_AB} highlights a bias toward smaller transit depths in the test data ($\sim 100$ ppm), with residuals between extracted and ground truth spectra showing larger deviations.

While the data reduction used here does not fully address systematic uncertainties, it provides a useful baseline for participants in the Data Challenge and other follow up studies. The validation confirms that the signals and noise levels in the dataset align with input parameters, demonstrating its suitability for benchmarking future analysis pipelines.

\section{data partition}
\label{sec:data_partition}

A unique challenge in exoplanet characterization lies in the absence of ground truth—for most cases, we do not know a priori the kind of atmosphere the telescope will observe. An ideal data reduction pipeline for exoplanetary data, or space-based data by extension, should be able to retrieve the signal without bias within the permitted noise level (above or close to the photon noise level).

To thoroughly test any pipeline—whether parametric, machine learning, or hybrid—we have designed a data partition that explicitly includes out-of-distribution test examples, as detailed in Table~\ref{tab:scens}. Unless otherwise stated, the global atmospheric setup follows the discussion in Section~\ref{sec:planets}. This partitioning strategy divides the dataset into training (including validation) and multiple test scenarios that systematically introduce domain shifts across key observational parameters. Each test case is designed to probe a specific type of distribution shift that algorithms will encounter when deployed on real Ariel observations, where target properties will inevitably differ from any training set.

The training set establishes a baseline with simulated planets having radii between $0.5$--$1.2$ $R_{\rm Jupiter}$ orbiting two host stars (KELT-11 and HD~17194) with atmospheric compositions including H$_2$O, CH$_4$, CO$_2$, and clouds, based on currently discovered molecular species. The \texttt{Training-Ext} set is an extended training set with uniform distributions for semi-major axis, inclination, and mid-transit time to introduce temporal and geometric variability in the light curves, consistent with likely scenarios in actual observations.

The test scenarios progressively introduce distribution shifts to explicitly test whether models overfit to the training set. \texttt{Test cases 0--2} maintain the same atmospheric species and host stars as the training set but employ different detector models (A, B, C), which differ in their bad pixel distributions, flat field characteristics, and noise maps (see Section~\ref{sec:calibration_products}). This configuration isolates instrumental robustness: can algorithms handle detector variations while the astrophysical signal remains constant? \texttt{Test case 3} expands to additional host stars (HD~209458 and HD~149026) with different effective temperatures and spectral energy distributions compared to the training stars, with a broader planetary radius range ($0.2$--$1.5$ $R_{\rm Jup}$) that includes smaller planets absent from training. This tests generalisation to different stellar irradiation environments and planetary bulk properties. To implement thorough screening of proposed pipelines, \texttt{Test cases 4--5} feature entirely different atmospheric compositions (HCN, H$_2$S, AlH, and CO) from those in the training set and \texttt{Test cases 0--2}, producing distinct spectral morphologies. \texttt{Test case 5} extends \texttt{Test case 4} with additional orbital parameter variations (semi-major axis, inclination, mid-transit time) similar to \texttt{Training-Ext}, introducing temporal and geometric diversity alongside compositional shifts. This stringent combination tests whether algorithms have learned general denoising principles or have simply memorised training set spectral shapes. The calibration case uses flat (featureless) spectra with no atmospheric features for all 4 stars, representing the null hypothesis and providing a baseline for assessing systematic biases in the extraction process. Note that these 4 calibration cases are not included in the actual test data in the Kaggle competition, but are provided for independent validation.

We emphasise that test case design prioritises algorithmic stress-testing over strict astrophysical plausibility. While some atmospheric combinations may not represent the most probable Ariel targets, they serve the essential function of exposing overfitting and validating that pipelines can handle diverse, unexpected scenarios—a critical requirement for survey missions. This partitioning approach deliberately exposes models to the types of challenges they will encounter when analysing real Ariel observations, where target exoplanets may have atmospheric compositions, stellar hosts, and orbital configurations that differ significantly from any training examples.

\section{ML-based Data Reduction Pipeline}
\label{sec:baseline}
To explore the possibility of having a scalable data reduction pipeline for large datasets, and to demonstrate how to manipulate the dataset for data science applications, we developed a ML-based pipeline to predict the combined transmission spectra derived from FGS1 and AIRS-CH0 instruments, or in another words, a 283 long vector. In this section we will outline its implementation. We begin our pipeline with data preprocessing, including calibrating the science frames using the calibration files, time-binning and normalising the dataset. The next stage is to feed the calibrated and normalised data into the neural network based pipeline. Here we present two different architectures, each leveraging different perspective of the data. The final stage is using the trained networks to predict planetary spectra and associated uncertainties for the test set, which will be compared against the ground truth for scoring.

% This section outlines the baseline machine learning model implemented as a baseline for the NeurIPS - Ariel Data Challenge 2024. The primary task is to predict the combined transmission spectra derived from FGS1 and AIRS-CH0 instruments. While our description focuses on the challenge format, we note that the dataset is versatile and can be adapted for various research purposes beyond this specific application.

\subsection{Data preprocessing and binning}
\label{sec:data_preprocess}
We begin the data preprocessing by calibrating the observation. We performed Analog to Digital Conversion (ADConv), flat fielding, dark current subtraction, dead pixel corrections and non-linearity correction using the calibration files provided for each observation. Each science image is corrected using the following formula:
\begin{align}
    \frac{\text{image - dark}}{\text{flat}}
\end{align}
In all these frames the dead pixels are masked using Python masked arrays. The bad pixels are thus masked but left uncorrected. Other methods can be used to correct bad-pixels, but we leave this for future exploration. The non-linearity of the pixel-wise’ response function is corrected using the inverse polynomial coefficients provided in the calibration files as well, to perform non-linearity correction.

Following the assumptions set out in Section \ref{sec:ndrs}, we computed the Correlated Double Sampling (CDS) using the difference between the end of exposure and the start of exposure. This resulted in a dataset containing 135,000 CDS frames for FGS1 and 5,610 CDS frames for AIRS-CH0.

We subsequently performed time binning on the science frames from both instruments to save storage space and homogenise the input to the neural network. The binned AIRS-CH0 frames were created by grouping 30 CDS frames together, corresponding to 187 frames with a time exposure of 2'24'' each. To match the input AIRS-CH0 data,  we applied a binning of 30$\times$12 (360) on the input FGS1 data. This consisted of 30 frames to match the AIRS-CH0 binning, and 12 frames to match the number of FGS1 images taken for each AIRS-CH0 frame.

The resultant datacubes of frames have dimensions of (187, 32, 32) for FGS1 and (187, 32, 282) for AIRS-CH0. We assume that this binning will have a small impact on the relevant information (noise + signal) contained in the data, as the dominant effect is the non-linearity correction (shown in Appendix in \autoref{sec:calibration}), which is proportional to the signal. We note, however, that the jitter motion may be blurred as a result of the binning step. Additionally, we combined the AIRS and FGS images by summing up the 32 pixels of the FGS x-axis, transforming each image into a (32, 1) format. We then concatenated this with the AIRS-CH0 image, yielding an input dimension of (187, 32, 283) for each combined image. The final step also aligns with the output target spectra, which have 283 points, one for FGS1 and 282 for AIRS-CH0.

\begin{figure}
    \centering
    \includegraphics[width=\linewidth]{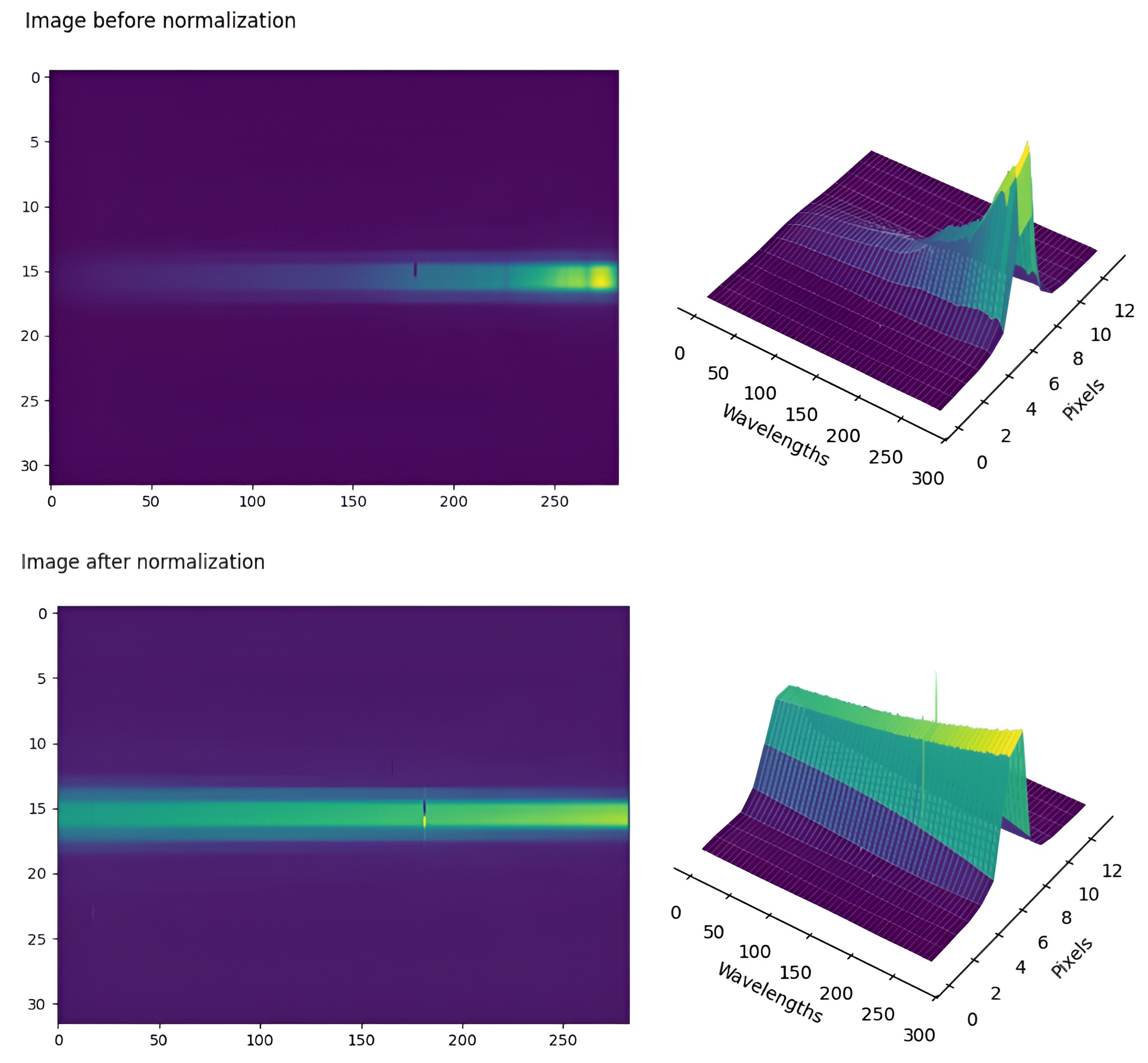}
    \caption{Stellar normalisation of the detector images. Top: raw flux before normalisation, showing the strongly wavelength-dependent stellar continuum concentrated along the spatial trace. Bottom: after dividing by the out-of-transit stellar spectrum, equalising flux across wavelengths and revealing the spatial PSF profile. Left: 2D detector images (spatial pixels × wavelength bins); right: corresponding 3D surface representations.}
    \label{fig:star_subraction}
\end{figure}

Given the predominance of stellar flux in the signal, we implemented a normalisation procedure using the stellar spectrum extracted from the images. \autoref{fig:star_subraction} shows an example of the signal before and after the normalisation has been applied.

% The dimensions of FGS input data are (135000, 32, 32), respectively for time, and two x and y spatial dimensions. To match them with the input AIRS data injected in the CNN model, we first compute the CDS just like for AIRS data, dividing by two the time dimension, to which we apply a binning of 360 = 30x12, 30 to match the AIRS binning and 12 as 12 images of FGS are taken for each AIRS image.
% Additionally, we combined AIRS and FGS images by summing up the 32 pixels of the FGS x-axis, transforming each image into a  (32, 1) format, that we assume to be an additional wavelength concatenated with the AIRS image, yielding to a final dimension of (187, 32, 283) for each image.

\subsection{Learning Strategy Overview}
\label{sec:ml_strategy}
\begin{table}
\centering
\caption{Architecture of the 1D convolutional neural network (CNN) used to retrieve
         the mean RpRs from the white
         light curve. The network takes a fixed-length time-series input of 187
         time steps and outputs a single scalar estimate of the mean transit depth.}
\label{tab:1dcnn_rprs}
\renewcommand{\arraystretch}{1.3}
\begin{tabular}{llll}
\hline\hline
\textbf{Layer} & \textbf{Type} & \textbf{Output Shape} & \textbf{Kernel Size} \\
\hline
\multicolumn{4}{l}{\textit{Feature Extraction --- 1D Convolutional Block}} \\
\hline
Input          & InputLayer         & $(N,\ 187,\ 1)$   & ---  \\
Conv1D-1       & Conv1D (ReLU)      & $(N,\ 183,\ 32)$  & 5    \\
MaxPool-1      & MaxPooling1D       & $(N,\ 91,\ 32)$   & 2    \\
BatchNorm      & BatchNormalisation & $(N,\ 91,\ 32)$   & ---  \\
Conv1D-2       & Conv1D (ReLU)      & $(N,\ 89,\ 64)$   & 3    \\
MaxPool-2      & MaxPooling1D       & $(N,\ 44,\ 64)$   & 2    \\
Conv1D-3       & Conv1D (ReLU)      & $(N,\ 42,\ 64)$   & 3    \\
MaxPool-3      & MaxPooling1D       & $(N,\ 21,\ 64)$   & 2    \\
Conv1D-4       & Conv1D (ReLU)      & $(N,\ 19,\ 128)$  & 3    \\
MaxPool-4      & MaxPooling1D       & $(N,\ 9,\ 128)$   & 2    \\
\hline
\multicolumn{4}{l}{\textit{Regression Head --- Fully Connected Block}} \\
\hline
Flatten        & Flatten            & $(N,\ 1152)$      & ---  \\
Dense-1        & Dense (ReLU)       & $(N,\ 500)$       & ---  \\
Dropout-1      & Dropout            & $(N,\ 500)$       & ---  \\
Dense-2        & Dense (ReLU)       & $(N,\ 100)$       & ---  \\
Dropout-2      & Dropout            & $(N,\ 100)$       & ---  \\
Output         & Dense (Linear)     & $(N,\ 1)$         & ---  \\
\hline\hline
\end{tabular}
\end{table}
A key feature of the dataset is its rich volume in information content, but exceedingly limited amount of training examples, standing from a machine learning point of view. Keeping data close to its original form preserves information like instrumental noise and bias, but increases computation costs and may hinder neural network convergence.

% A key challenge in this competition is handling large volumes of data, which demands substantial computational resources and time. Keeping data close to its original form preserves information like instrumental noise and bias, but increases computation costs and may hinder neural network convergence. Moreover, the training dataset size is small for machine learning.

There are two broad learning strategies:
\begin{itemize}
    \item Training on complete spatial-spectral time series to generate full spectra.
    \item Column-wise analysis along the spectral direction to determine transit depth per column.
\end{itemize}
The former strategy is straight forward but computationally intensive, while the second strategy facilitates the construction of a substantial dataset ($N_{data} \times$ 282 ). For this investigation, we have chosen to prioritise the first strategy for the following reason: The individual light curves extracted were excessively noisy, affecting our ability to accurately determine the mean spectrum value. Although we could improve the signal-to-noise ratio (SNR) by summing the flux across all wavelengths in each image to generate a ``white'' light curve for extracting the mean transit depth, this method presented certain limitations: when concentrating on a single column of the mean-subtracted image, or a specific wavelength, it is not possible to determine the level of variation from the average value without comparative information from other channels. While incorporating adjacent channels may mitigate this limitation, we will reserve that exploration for future work. For the current investigation, we will therefore focus on the first strategy.

We developed two baseline approaches for the first strategy, each leveraging different features of the dataset, without separating the image columns. Both baselines utilise a CNN-based architecture, but for different views of the data. A CNN is a neural network architecture that gained widespread popularity since its successful application to the ImageNet competition in 2012. The algorithm's design allows it to recognise patterns from structured data (in this case, spatial-temporal data) through hierarchical layers of learnable filters.

Both approaches employ a two-stage training process. The first stage is identical for both, utilising a 1D-Convolutional Neural Networks (1D-CNN) to fit the mean value of the transmission spectra (see Table \ref{tab:1dcnn_rprs}.

This CNN takes as input the transit white light curve, which represents the total flux of each image as a function of time. The second stage, however, diverges into two different approaches, each designed to capture different aspects of the data:
\begin{enumerate}
    \item \textit{First approach}: train on the 3D data cubes (N x N\_times x N\_spatial x N\_wavelengths), this approach uses all the spatial information but as a consequence requires a lot of computing resources. The model is trained on a 3D CNN to search for atmospheric features.
   \item \textit{Second approach}: preprocess the data by summing up the fluxes along the pixel y-axis, for each wavelength, resulting in 2D images of dimension (N\_times, N\_wavelengths), and transforming the images in order to enhance transit depth variations between wavelengths.
\end{enumerate}

\begin{figure*}
    \centering
    \includegraphics[width=\linewidth]{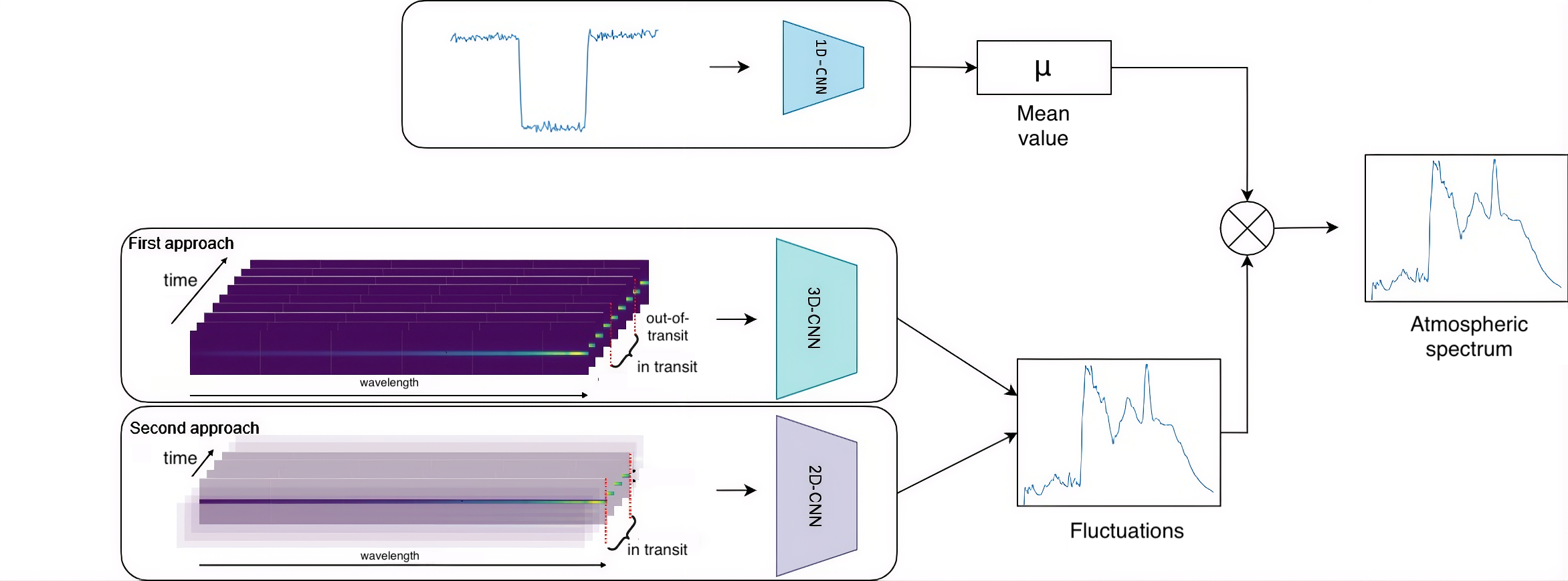}
    \caption{illustrate the two different ML strategies deployed in this investigation. The two approaches uses the same initial step (fit with a 1D CNN top row) followed by two different architectures (3D-CNN and 2D-CNN) to tackle the problem using different inputs.}
    \label{fig:ARIEL_ML_approaches_diagram}
\end{figure*}

\subsection{First baseline}
\begin{table}
\centering
\caption{Architecture of the first baseline. The network takes a 3D input corresponding to wavelength bins, time
         steps, and spectral channels, and outputs a per-wavelength fluctuation
         vector.}
\label{tab:3dcnn_fluctuation}
\renewcommand{\arraystretch}{1.3}
\begin{tabular}{llll}
\hline\hline
\textbf{Layer} & \textbf{Type} & \textbf{Output Shape} & \textbf{Kernel Size} \\
\hline
\multicolumn{4}{l}{\textit{Feature Extraction --- 3D Convolutional Block}} \\
\hline
Input      & InputLayer    & $(N,\ 283,\ 187,\ 32,\ 1)$  & ---           \\
Conv3D-1   & Conv3D (ReLU) & $(N,\ 142,\ 94,\ 16,\ 32)$  & $3\times3\times3$ \\
MaxPool-1  & MaxPooling3D  & $(N,\ 47,\ 47,\ 8,\ 32)$    & $3\times2\times2$ \\
Conv3D-2   & Conv3D (ReLU) & $(N,\ 24,\ 24,\ 4,\ 64)$    & $3\times3\times3$ \\
MaxPool-2  & MaxPooling3D  & $(N,\ 8,\ 12,\ 2,\ 64)$     & $3\times2\times2$ \\
Conv3D-3   & Conv3D (ReLU) & $(N,\ 8,\ 12,\ 2,\ 64)$     & $3\times3\times3$ \\
MaxPool-3  & MaxPooling3D  & $(N,\ 4,\ 6,\ 2,\ 64)$      & $2\times2\times1$ \\
Conv3D-4   & Conv3D (ReLU) & $(N,\ 4,\ 6,\ 2,\ 128)$     & $3\times3\times3$ \\
MaxPool-4  & MaxPooling3D  & $(N,\ 2,\ 6,\ 2,\ 128)$     & $2\times1\times1$ \\
\hline
\multicolumn{4}{l}{\textit{Regression Head --- Fully Connected Block}} \\
\hline
Flatten    & Flatten       & $(N,\ 3072)$                 & ---  \\
Dense-1    & Dense (ReLU)  & $(N,\ 500)$                  & ---  \\
Dropout    & Dropout       & $(N,\ 500)$                  & ---  \\
Output   & Dense (ReLU)  & $(N,\ 283)$                  & ---  \\
\hline\hline
\end{tabular}
\end{table}
The first baseline is based on limited pre-processing of the data, which allows the neural network to use all the information contained in the spectroscopic time series at the cost of large inputs to the model.

% The first step of the baseline is the 1D-CNN used to extract the mean value of the transit depth of each planetary system.
% Then, a 3D-CNN is meant to retrieve the variations of the spectra around this mean value, corresponding to the atmospheric features.
% The steps are the following:
We outlined our treatment for the inputs, targets and architecture:
\begin{enumerate}
    \item \textit{Input}: the images are the direct product of the star spectrum normalisation, as shown in \autoref{fig:img_3DCNN}.
\begin{figure}
    \centering
    \includegraphics[width=\linewidth]{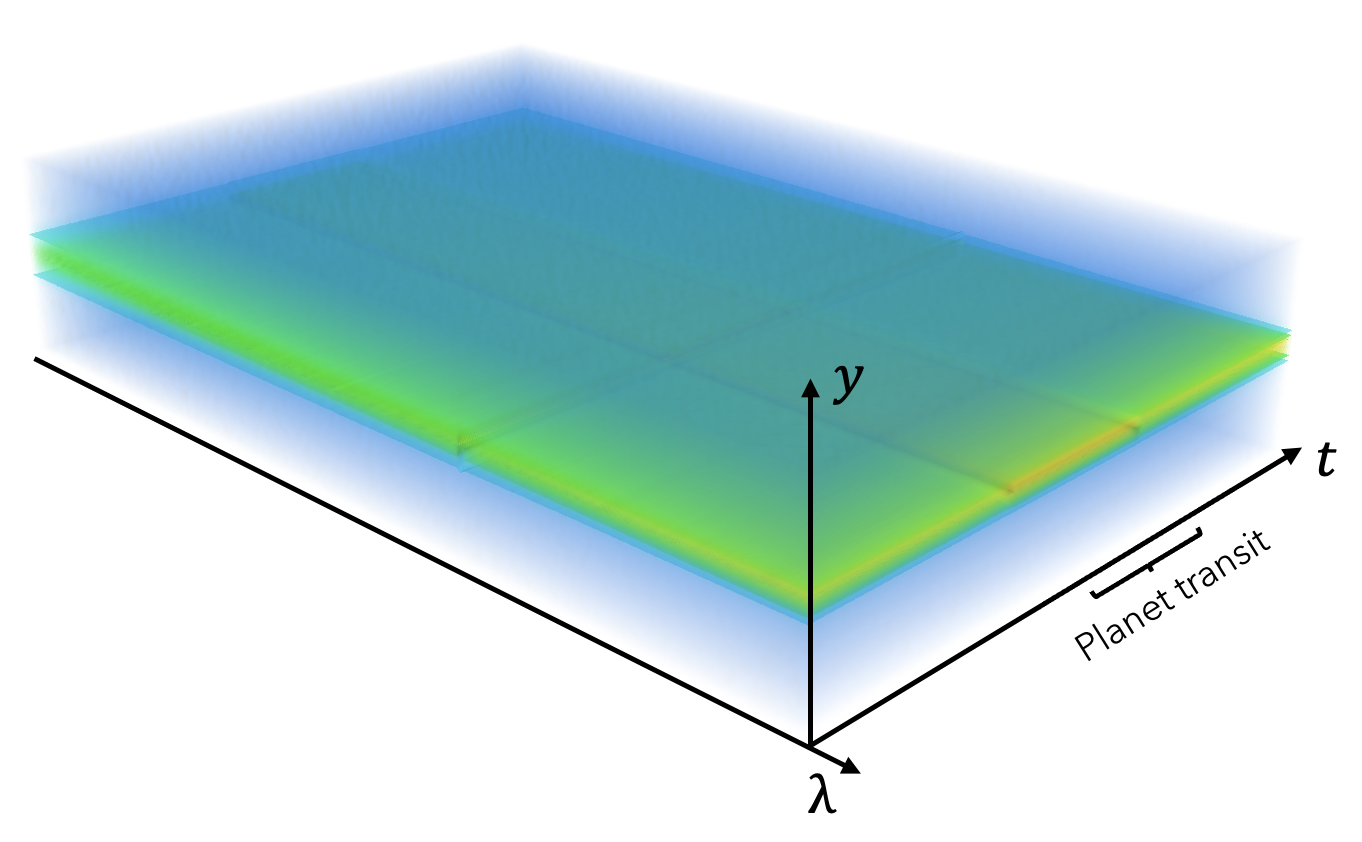}
    \caption{Example of a datacube used as an input for the first baseline with the 3D-CNN. The transit depth is exaggerated for clarity and better visualisation.}
    \label{fig:img_3DCNN}
\end{figure}
    \item \textit{Targets}: we remove the mean to fit atmospheric fluctuations only, and normalize the targets between 0 and 1, using the global maximum and minimum of all the examples, as shown in \autoref{fig:targets_3DCNN}.
\begin{figure}
    \centering
    \includegraphics[width=\linewidth]{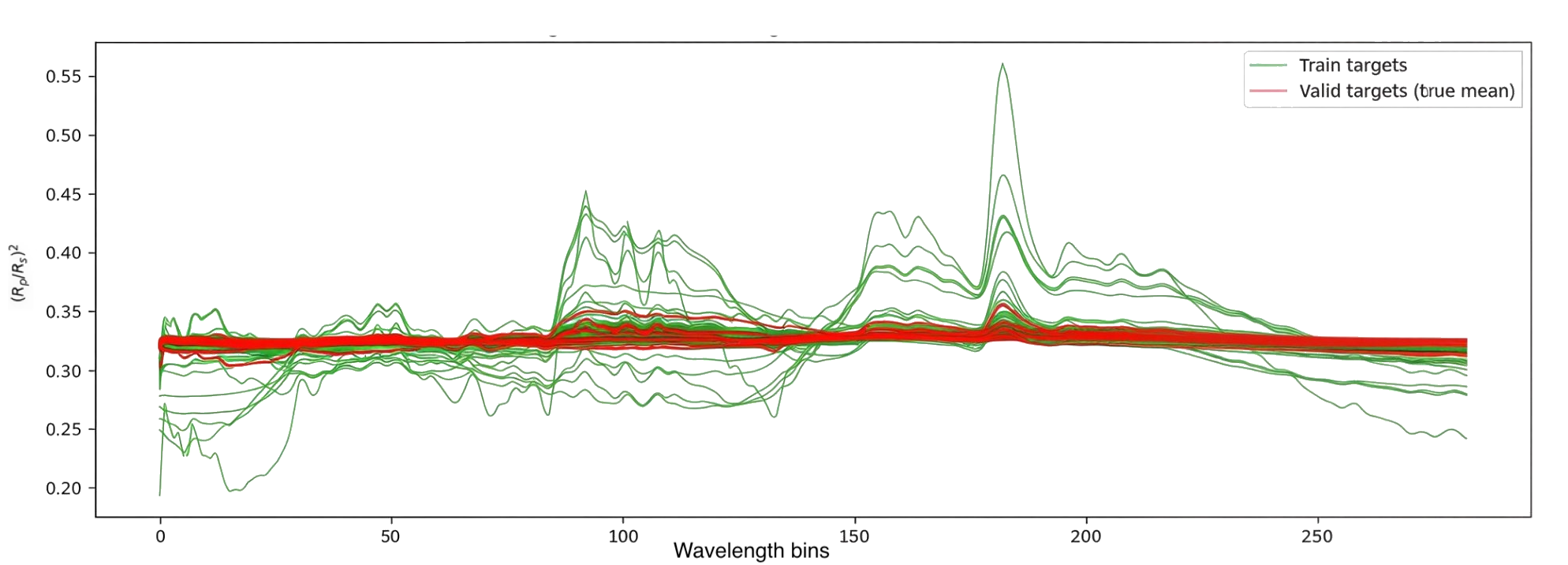}
    \caption{Normalised transmission spectra (Rp/Rs)² across wavelength bins for the 3D-CNN baseline. Green: training targets; red: validation true means.}
    \label{fig:targets_3DCNN}
\end{figure}

    \item \textit{Architecture}: The 3D CNN takes in the spatial-temporal cube and outputs the atmospheric features. Detailed architecture table in \ref{tab:3dcnn_fluctuation}

As illustrated in Figure \ref{fig:img_3DCNN_results}, the 3D-CNN model predicts an average spectrum, in black, that performs well across some targets but doesn't capture the actual atmospheric features. It adjusts only the mean value of this spectrum to match each observation but always predict the same variation pattern.

For estimating our model uncertainties and average spectra, we use Monte Carlo dropout \citep{gal2016} fixing at $1000$ and $30$ the number of instances for the first (1D-CNN) and second (3D-CNN) CNNs respectively.

%The score of this method remains correct because the uncertainties are sufficiently large, even without accounting for the accuracy of atmospheric variations.

\begin{figure}
    \centering
    \includegraphics[width=\linewidth]{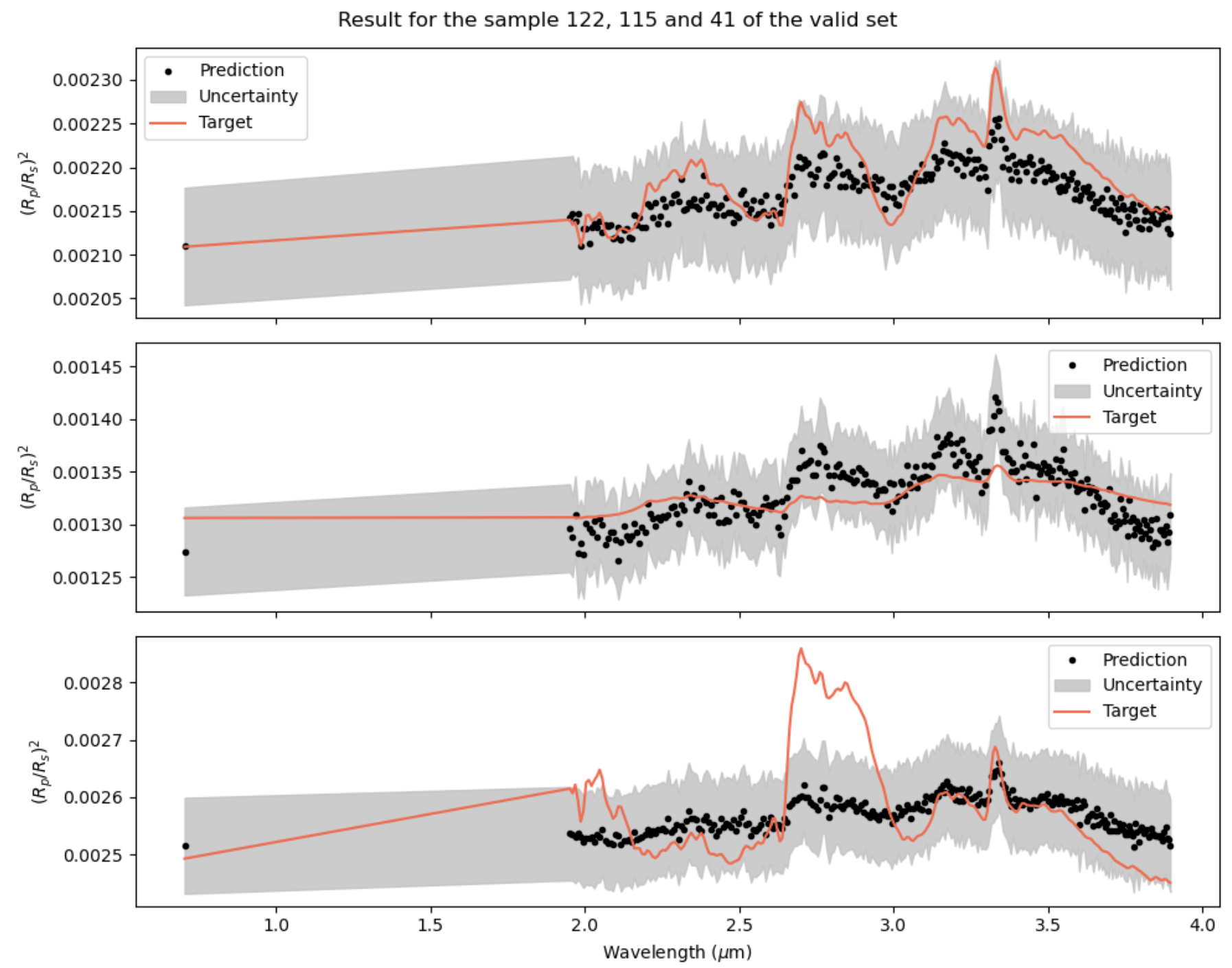}
    \caption{Examples of predictions obtained using the 3D-CNN on the validation set.}
    \label{fig:img_3DCNN_results}
\end{figure}

\end{enumerate}

\subsection{Second baseline}
\begin{table}[ht]
\centering
\caption{Architecture of the 2D convolutional neural network (CNN) used as a second
         baseline to model spectral light curve fluctuations. The input shape corresponds to the 40 in-transit time steps across
         283 wavelength bins. The network outputs a per-wavelength fluctuation
         vector of length 283.}
\label{tab:2dcnn_fluctuation}
\renewcommand{\arraystretch}{1.3}
\begin{tabular}{llll}
\hline\hline
\textbf{Layer} & \textbf{Type} & \textbf{Output Shape} & \textbf{Kernel Size} \\
\hline
\multicolumn{4}{l}{\textit{Block I --- Temporal Feature Extraction (time axis)}} \\
\hline
Input       & InputLayer         & $(N,\ 40,\ 283,\ 1)$    & ---               \\
Conv2D-1    & Conv2D (ReLU)      & $(N,\ 40,\ 283,\ 32)$   & $1\times3$        \\
MaxPool-1   & MaxPooling2D       & $(N,\ 20,\ 283,\ 32)$   & $2\times1$        \\
BatchNorm-1 & BatchNormalisation & $(N,\ 20,\ 283,\ 32)$   & ---               \\
Conv2D-2    & Conv2D (ReLU)      & $(N,\ 20,\ 283,\ 64)$   & $1\times3$        \\
MaxPool-2   & MaxPooling2D       & $(N,\ 10,\ 283,\ 64)$   & $2\times1$        \\
Conv2D-3    & Conv2D (ReLU)      & $(N,\ 10,\ 283,\ 128)$  & $1\times3$        \\
MaxPool-3   & MaxPooling2D       & $(N,\ 5,\ 283,\ 128)$   & $2\times1$        \\
Conv2D-4    & Conv2D (ReLU)      & $(N,\ 5,\ 283,\ 256)$   & $1\times3$        \\
Conv2D-5    & Conv2D (ReLU)      & $(N,\ 5,\ 283,\ 32)$    & $1\times3$        \\
\hline
\multicolumn{4}{l}{\textit{Block II --- Spectral Feature Extraction (wavelength axis)}} \\
\hline
MaxPool-4   & MaxPooling2D       & $(N,\ 5,\ 141,\ 32)$    & $1\times2$        \\
BatchNorm-2 & BatchNormalisation & $(N,\ 5,\ 141,\ 32)$    & ---               \\
Conv2D-6    & Conv2D (ReLU)      & $(N,\ 5,\ 141,\ 64)$    & $1\times3$        \\
MaxPool-5   & MaxPooling2D       & $(N,\ 5,\ 70,\ 64)$     & $1\times2$        \\
Conv2D-7    & Conv2D (ReLU)      & $(N,\ 5,\ 70,\ 128)$    & $1\times3$        \\
MaxPool-6   & MaxPooling2D       & $(N,\ 5,\ 35,\ 128)$    & $1\times2$        \\
Conv2D-8    & Conv2D (ReLU)      & $(N,\ 5,\ 35,\ 256)$    & $1\times3$        \\
MaxPool-7   & MaxPooling2D       & $(N,\ 5,\ 17,\ 256)$    & $1\times2$        \\
\hline
\multicolumn{4}{l}{\textit{Regression Head --- Fully Connected Block}} \\
\hline
Flatten     & Flatten            & $(N,\ 21760)$            & ---               \\
Dense-1     & Dense (ReLU)       & $(N,\ 700)$              & ---               \\
Dropout     & Dropout            & $(N,\ 700)$              & ---               \\
Output      & Dense (Linear)     & $(N,\ 283)$              & ---               \\
\hline\hline
\end{tabular}
\end{table}
% The 1D CNN searches for the average transit depth of each planetary system, $(Rp/Rs)^2$.

The second baseline fits the tiny fluctuations around the mean of the spectra to search for atmospheric features. This model is also used as a baseline for NeurIPS - Ariel Data Challenge 2024 \citep{yip2024ariel} and are publicly available online\footnote{\url{https://www.kaggle.com/code/gordonyip/host-starter-solution}}. Here we propose a method to enhance the signal variations between wavelengths, while keeping the amplitude ratios between the planetary systems, implementing the following treatment to the input:
\begin{enumerate}
    \item \textit{Time range}: For the second-stage 2D-CNN input, we retain only the in-transit portion of the time series, using the smallest transit duration in the dataset to define a consistent temporal window across all examples. Note that the first-stage 1D-CNN (Section~\ref{sec:ml_strategy}) uses the full white light curve including out-of-transit baseline to predict the mean transit depth; the 2D-CNN subsequently predicts only the spectral fluctuations around this mean value. This approach reduces the 2D-CNN input size while preserving the transit signal morphology.
    \item \textit{Pixel y-axis}: we sum up the fluxes along the pixel y-axis, for each wavelength, resulting in 2D images of dimension (N\_times, N\_wavelengths).
    \item \textit{Subtracting the mean and data normalization}: for each planetary system, we first subtract the mean signal across all wavelengths and time instants to obtain relative input data. This preprocessing step ensures coherence with our relative target values. We then normalize these relative signals to the range [-1, 1] using the maximum amplitude of the complete dataset as a reference, see \autoref{fig:data_norm_2DCNN}.
    \item \textit{Architecture}: We implement a 2D CNN to fit the atmospheric features (see Table \ref{tab:2dcnn_fluctuation} for more details)
\begin{figure}
    \centering
    \includegraphics[width=\linewidth]{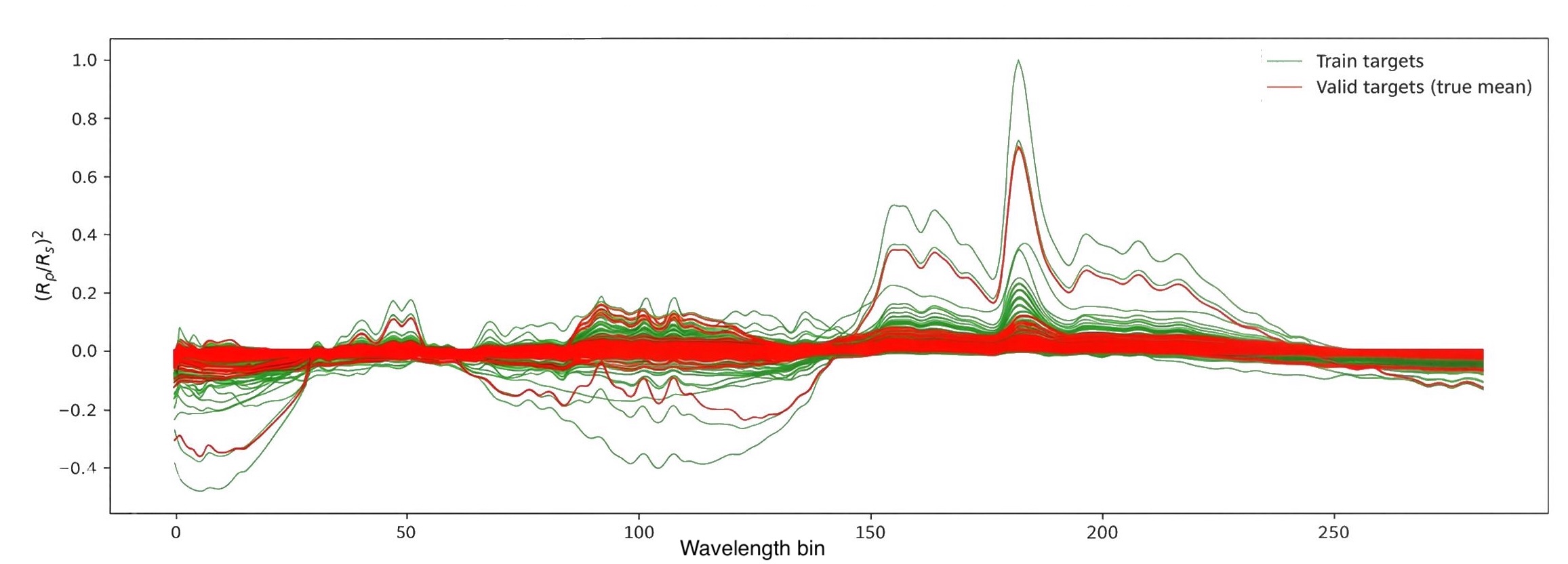}
    \caption{Normalised transmission spectra (Rp/Rs)² across wavelength bins for the 2D-CNN baseline. Green: training targets; red: validation true means.}
    \label{fig:targets_2DCNN}
\end{figure}
\begin{figure}
    \centering
    \includegraphics[width=\linewidth]{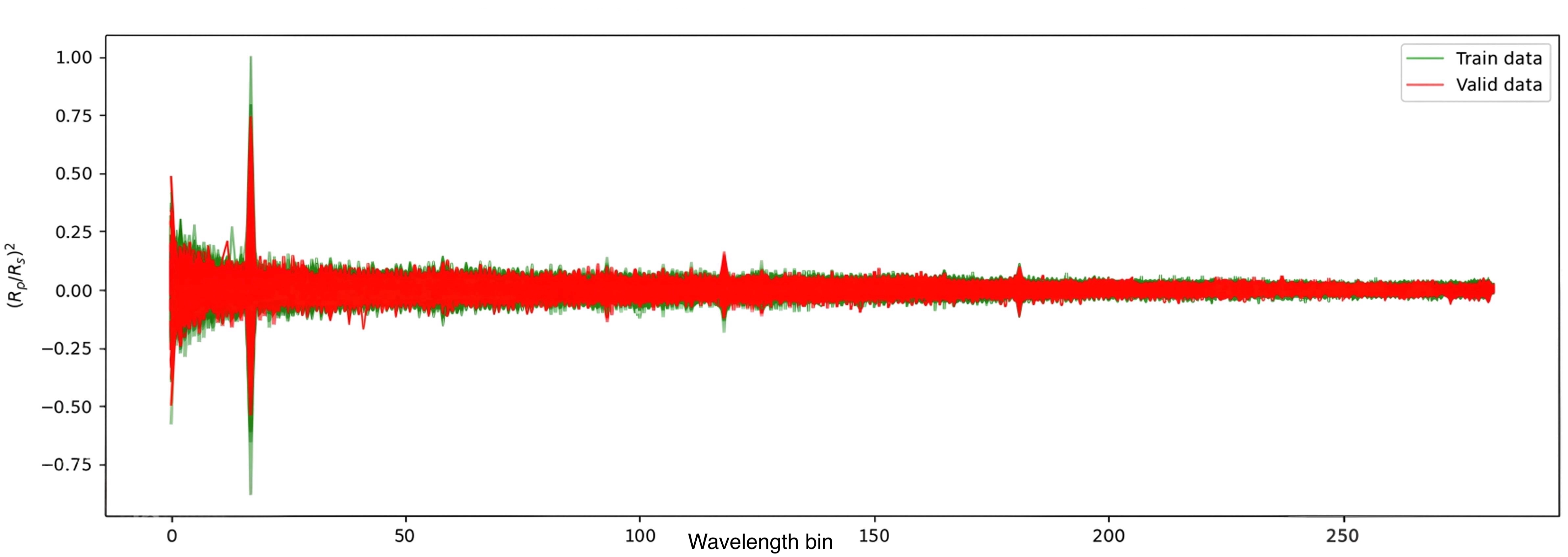}
    \caption{Normalization of input data (2D images) for the second baseline 2D CNN, shown here for a given planet at a given time.}
    \label{fig:data_norm_2DCNN}
\end{figure}

% \begin{figure}
%     \centering
%     \includegraphics[width=\linewidth]{original/valid_pred.png}
%     \caption{Second baseline: good feature predictions of the 2D CNN (validation set). Predicted spectra in black with uncertainties in grey and targets in red. The prediction of the mean value is not taken into account in these results. Combined with the mean value predictions, the final fits would show bigger error bars from the 1D CNN contribution and could be shifted from the GT spectra. }
%     \label{fig:valid_2DCNN}
% \end{figure}
\end{enumerate}

This approach does not use all the information contained in the datacube, e.g. the jitter effect is now accessible within the time domain only, but this implementation enables to perform fast calculations and to predict atmospheric spectra with different shapes contrary to the first baseline for which the actual implementation does not allow to produce variations in the spectral shape. For estimating our model uncertainties and average spectra, we again used the same configuration as the first approach use Monte Carlo dropout \citep{gal2016} fixing at $1000$ and $30$ the number of instances for the first and second CNNs respectively.
% \begin{table}[]
% \centering
% \resizebox{0.5\textwidth}{!}{%
% \begin{tabular}{lllllll}
% Pipeline & Entire Data & Test Data & Test case 1 & Test case 2 & Test case 3 & Test case 4 \\
% AB       &             & N/A       &             &             &             &
% \end{tabular}%
% }
% \caption{Example performance table}
% \label{tab:my-table}
% \end{table}

\subsection{Evaluation Metric}
We evaluate the quality of the predicted spectra, $(\tilde{x}_{\lambda})$, and their predicted uncertainties at different wavelengths, $(\tilde{\sigma})_{\lambda})$, $\tilde{\mathcal{S}}=(\tilde{x}_\lambda, \tilde{\sigma}_\lambda)$ against the ground truth spectrum, $x_\lambda$ using two metrics, 1.) Mean Squared Error (MSE) and 2.) Gaussian Log-Likelihood to incorporate the estimated uncertainties from the network.

The Mean Squared Error (MSE) is calculated by taking the squared difference between the ground truth spectra and the predicted spectra for each example. We then compute the mean of these squared differences for each pair of wavelengths across all examples.

The Gaussian Log-Likelihood (GLL) based scoring function, we will calculate the Gaussian Log-Likelihood (GLL) value for each pair of $x$ in $\mathcal{S}_i$ and $\tilde{\mathcal{S}}_i$, i.e.
\begin{align}
    GLL \Big( x_\lambda, \tilde{x}_\lambda, \tilde{\sigma}_\lambda \Big) = -\frac{1}{2}\log({2\pi\tilde{\sigma}^2_\lambda}) - \frac{(\tilde{x}_\lambda - x_\lambda)^2}{2 \tilde{\sigma}^2_\lambda}
\end{align}
The GLL values from each point will then be summed across all wavelengths and across the entire test set via $ \mathcal{L} = \sum_i \sum_\lambda GLL\Big(x_{i,\lambda}, \tilde{x}_{i,\lambda}, \sigma_{i,\lambda}\Big)$. A score will be generated for a given prediction set $\tilde{\mathcal{S}}_i$ via the following formulation:
\begin{align}
    \text{score} = \min \Big[ 0, \, \frac
    {\mathcal{L} - \mathcal{L}_{\text{naive}}}
    {\mathcal{L}_{\text{ideal}} - \mathcal{L}_{\text{naive}}} \Big]
\end{align}
Here we denote $\mathcal{L}_{\text{ideal}}$ as the optimum prediction set, for which $(\tilde{x}_\lambda) = (x_\lambda)$, and with $\tilde{\sigma}= 10\text{ppm}$. This uncertainty is lower than the minimum achievable theoretical uncertainty.

The score will be returned as a decimal in the interval $[0,1]$, with higher scores corresponding to better performing models.$\mathcal{L}_{\text{naive}}$ represents the reference score when the model always outputs the overall mean and standard deviation of the training set as its prediction.

\section{Results from ML baseline}
\label{sec:results}
% \subsection{Scores}

The resulting scores for the two approaches are given in \autoref{tab:scores} for the validation and test sets. We evaluate the models with both metric (GLL metric and MSE). For the mean computation, the SGD optimizer gives more permissive error bar on the GLL loss. Adam optimizer is used for predicting the fluctuations.

The results in \autoref{tab:scores} were obtained by keeping the best out of 4 different splits. These splits are identical across the two baselines, both sharing the first 1D CNN. We trained the models using the TRex computer from CNES, with a GPU device (V100). The computing times were 1478s and 183s for the first (3D-CNN) and second (2D-CNN) baselines respectively. Test scores are similar for the two models, whereas the validation score is usually better for the second baseline.

\begin{figure}
    \centering
    \includegraphics[width=\linewidth]{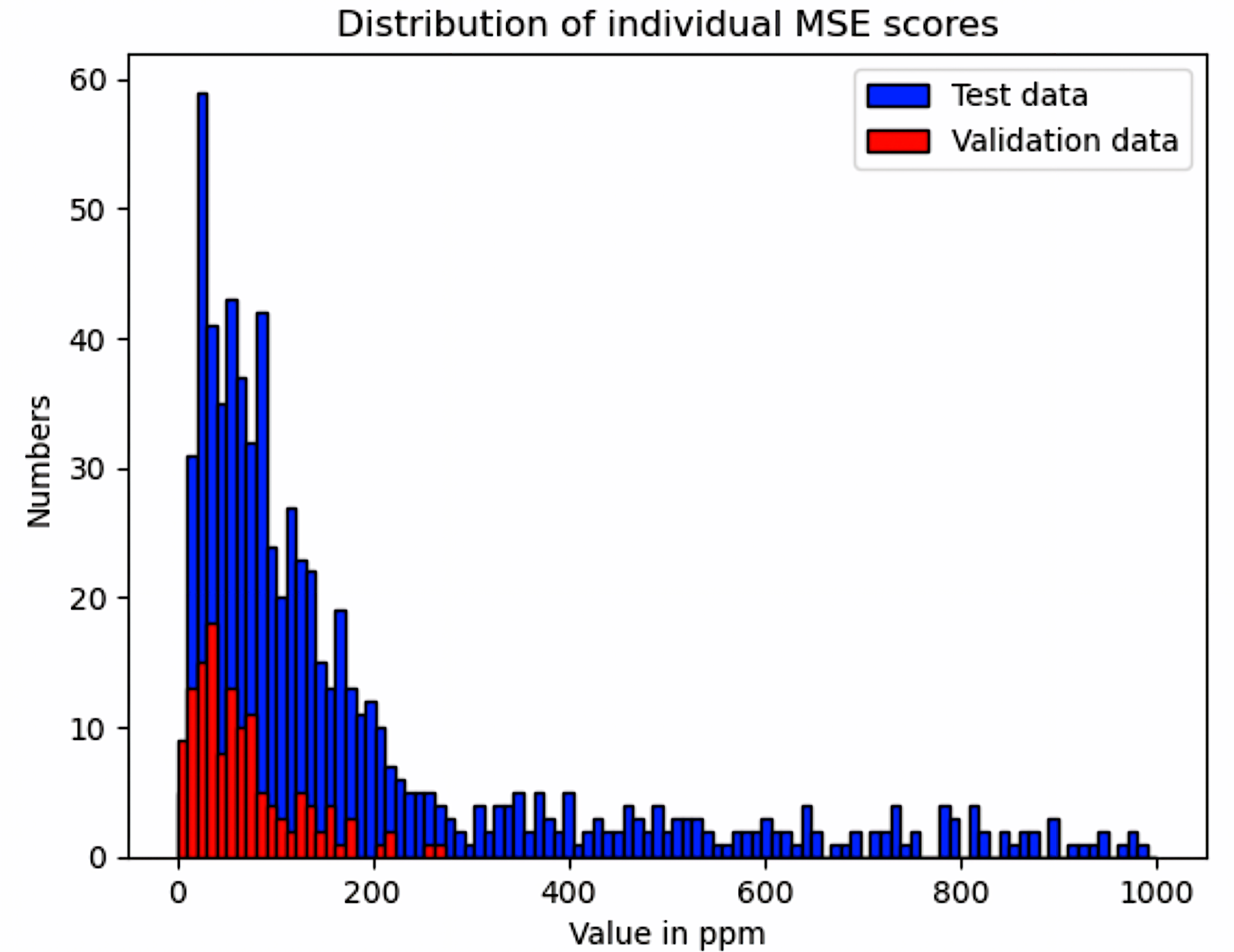}
    \caption{MSE distribution for validation and test data (second baseline). 35 test MSE higher than $1000$ are not captured in this plot, reaching a maximum of $3440$.
}
    \label{fig:mse_distrib}
\end{figure}

\begin{figure*}
    \centering
    \includegraphics[width=\linewidth]{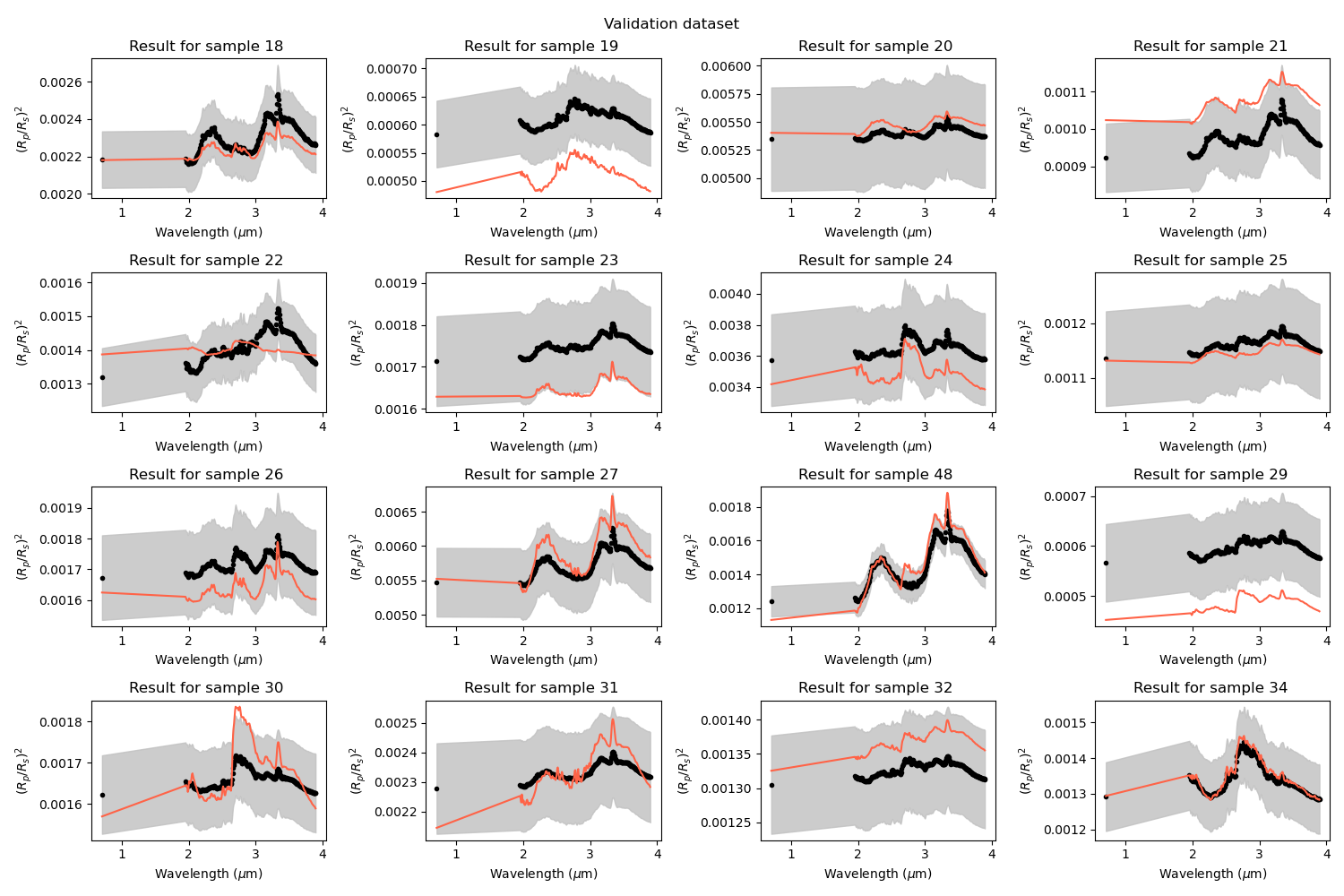}
    \caption{Model predictions from the second baseline across 16 randomly selected validation targets. Black dots/line: predicted mean spectrum; grey band: predicted uncertainty envelope; red line: true target spectrum.
}
    \label{fig:random_valid_results}
\end{figure*}

\begin{figure*}
    \centering
    \includegraphics[width=\linewidth]{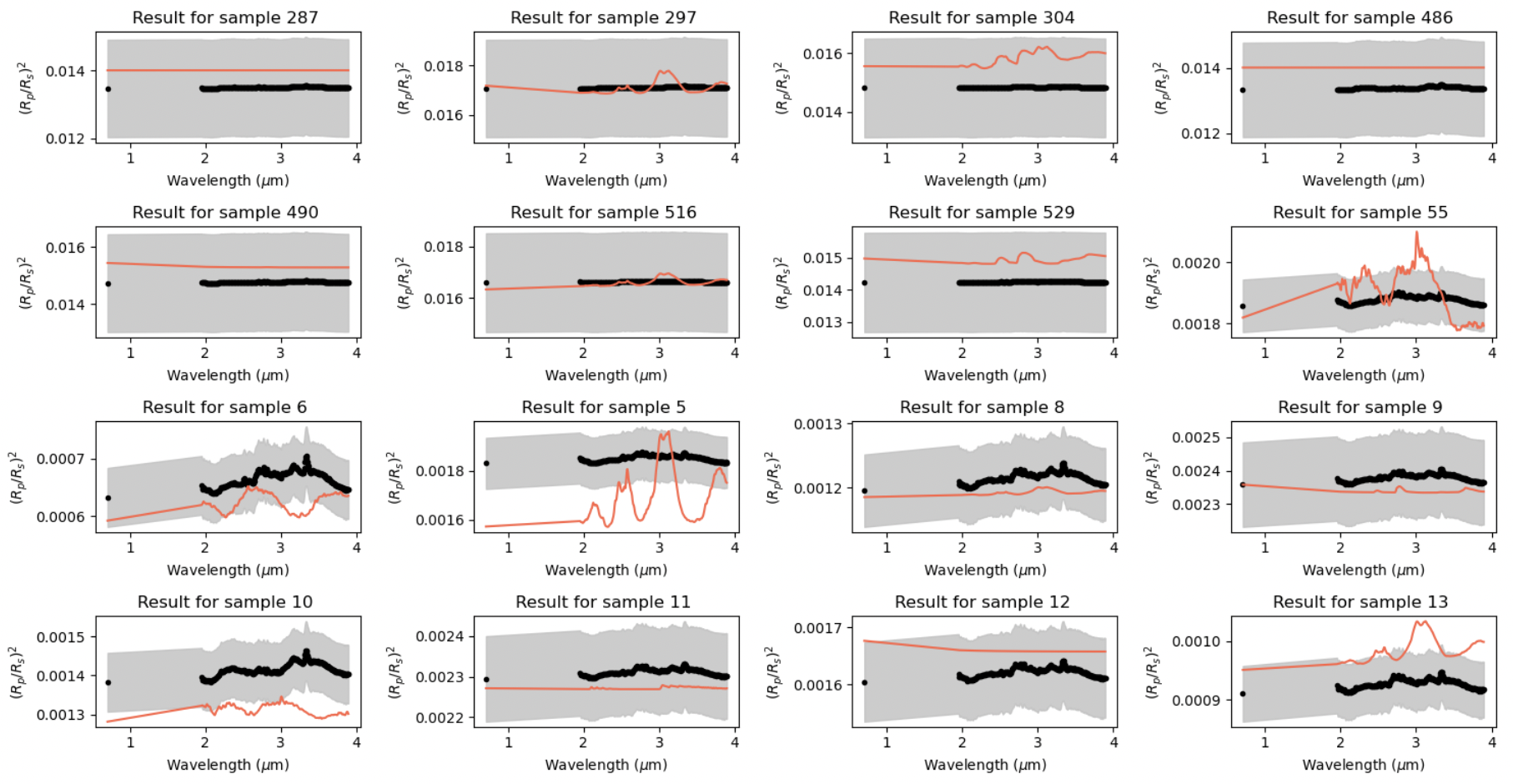}
    \caption{Model predictions from the second baseline across 16 randomly selected test targets. Black dots/line: predicted mean spectrum; grey band: predicted uncertainty envelope; red line: true target spectrum.
}
    \label{fig:random_test_results}
\end{figure*}

The first baseline model tend to converge to a single spectral shape that represents a compromise across the entire dataset. In contrast, the second baseline model predicts diverse shapes for the validation set. Neither model, however, successfully generalizes to the test set. The distribution of MSE scores for the second baseline model across both validation and test populations is shown in \autoref{fig:mse_distrib}. Random examples of predicted spectra compared to ground truth values are presented in \autoref{fig:random_valid_results} and \autoref{fig:random_test_results} for the validation and test sets, respectively.

Model performance is primarily determined by two factors. First, the ability to fit the mean spectral value significantly influences the scores. Second, the atmospheric extent relative to $(R_p/R_s)^2$ affects the magnitude of spectral fluctuations, and therefore on the overall score, e.g. thinner atmospheres result in reduced modulation to the flat line. Additionally, the error bars substantially influence the GLL based metric, with larger error bars generally yielding better scores up to a threshold value.

\begin{table}
\centering
\resizebox{0.5\textwidth}{!}{%
\begin{tabular}{lcccc}
\hline
& Valid GLL score & Valid MSE (ppm) & Test GLL score & Test MSE (ppm)  \\
\hline
3D-CNN Baseline & 0.4681  & 128 & 0.4232  & 544  \\
2D-CNN Baseline & 0.5020  & 86  & 0.4246  & 542 \\
\hline
\end{tabular}%
}
\caption{Results for the two baselines on validation and test sets on GLL based metric and MSE metric.}
\label{tab:scores}
\end{table}

\section{Discussion}
\label{sec:discussiom}

We have presented a comprehensive public dataset of simulated Ariel observations and demonstrated baseline machine learning approaches for data reduction. While our Convolutional Neural Networks successfully extract mean transit depths, they also reveal fundamental limitations when applied to test scenarios that differ systematically from training data—a challenge central to exoplanet survey science where target properties are inherently unknown.

This section examines the factors contributing to performance degradation between validation and test sets, explores mitigation strategies, and discusses the implications for developing robust data reduction pipelines for space-based transit spectroscopy missions.

\subsection{Performance degradation}
From the result section we can see a stark performance degradation from training data (as well as validation data) to test data. This is an expected outcome stemming from the way the dataset is constructed (see \autoref{tab:scens} for more details on each scenarios). To reiterate, we would like to explore the generalisation ability of the trained algorithm in unknown scenarios. In the following paragraphs we will look into possible causes that could contribute to the degradation.
\subsubsection{Distribution shifts from training to test data}
\begin{figure*}
    \centering
    \includegraphics[width=\linewidth]{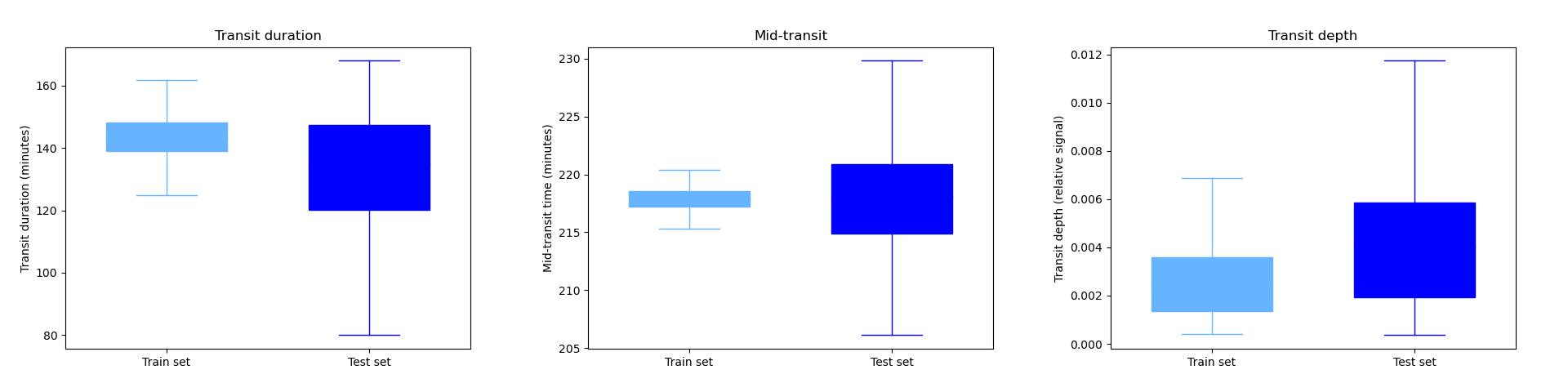}
    \caption{Distribution of transit duration, mid-transit time and transit depth, for train and test sets, represented by shades of blue. The solid rectangle is centered on the mean value, and its height is defined by the standard deviation. The whiskers extend to the minimum and maximum values, representing the range of the data. Time indices are converted in minutes (see conversion in section 4.1).
}
    \label{fig:train_test_transit}
\end{figure*}

\begin{figure*}
    \centering
    \includegraphics[width=0.8\linewidth]{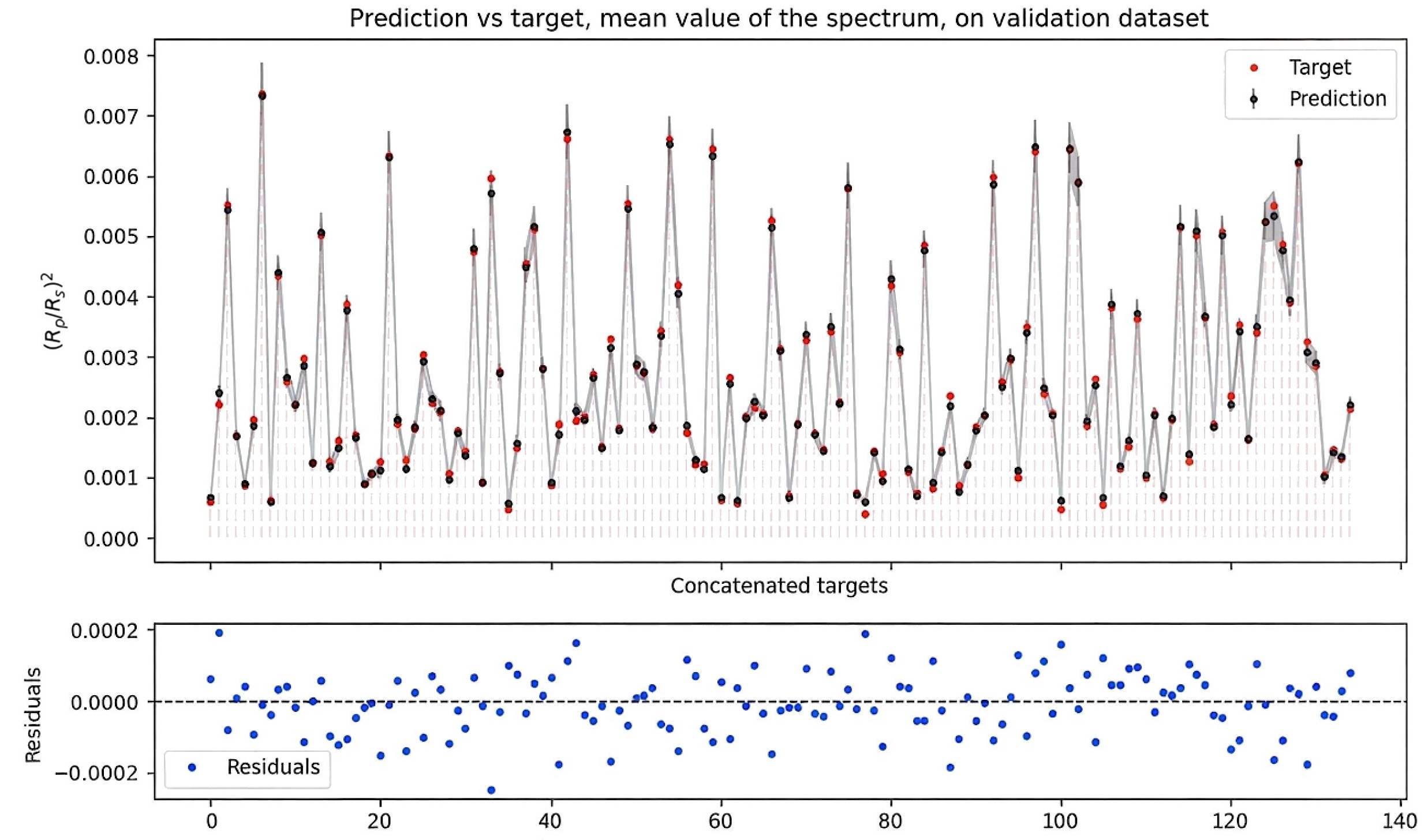}
    \includegraphics[width=0.8\linewidth]{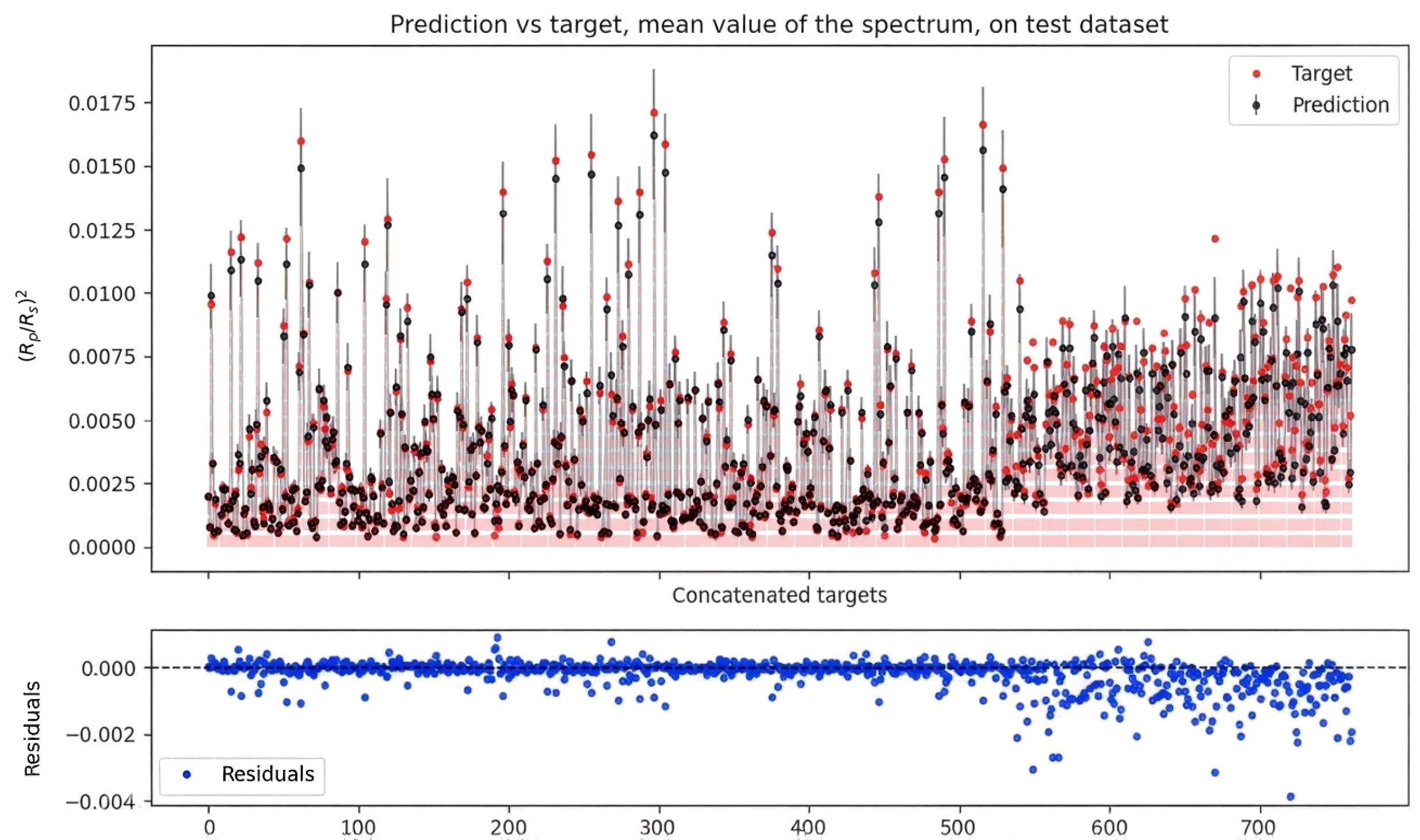}
    \caption{Predictions of spectra mean values for validation (top) and test (bottom) data, with error bars (black dots). Ground Truth mean value are represented by red dots and residuals are shown in the lower part of each figure (blue points).
}
    \label{fig:mean_pred}
\end{figure*}

% The baseline models presented above perform better at predicting the mean transit depth and atmospheric features for the training data than for the test data (lower MSE). This is explained by the fact that the test data are different from the training data.
\paragraph{Mean values}
Both the training and test datasets show variations in transit duration, mid-transit time and transit depth within their respective population. \autoref{fig:train_test_transit} displays these parameters distribution for training (cyan) and test data (blue), revealing that even for three parameters, the distribution of the test data is broader and exhibits more extreme minimum and maximum values. Within the test data, about $35$ planetary systems have extremely deep transits compared to the training population, resulting in a $5-10\%$ error or more in our predicted mean values, as shown in \autoref{fig:mean_pred}. Likewise, the predictions are degraded when the transits are narrower or shifted in time compared to the training set range. Given this shift in transit and atmospheric characteristics between training and test sets, uncertainties of the mean values for the test set are often underestimated while they could be overestimated for the validation set.

\begin{figure*}
    \centering
    \includegraphics[width=\linewidth]{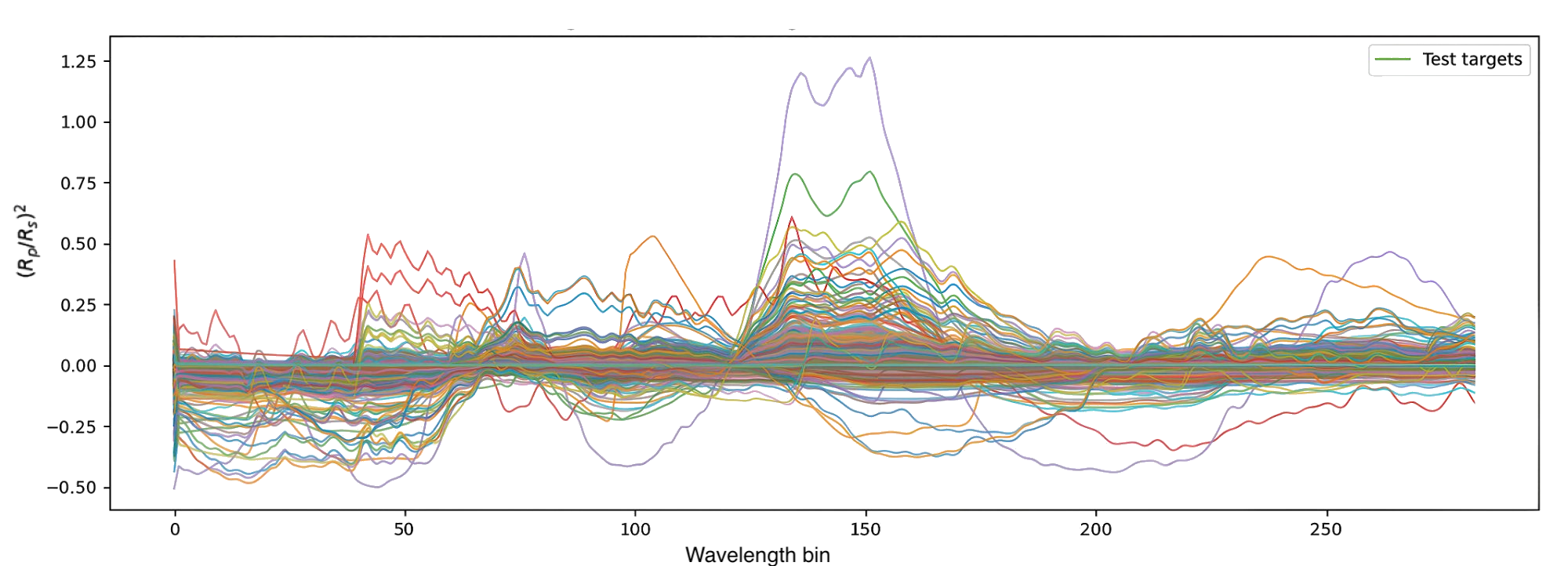}
    \caption{Test set transmission spectra after subtracting from the population median, highlighting relative atmospheric features across the wavelength range.}
    \label{fig:test_spectra}
\end{figure*}

\paragraph{Atmospheric features}
As presented in \autoref{tab:scens}, the atmospheric composition of planets that belong to the test data is different from the training data. This results in different atmospheric features in their spectra. \autoref{fig:test_spectra} shows the normalised fluctuations of the test set spectra that can be compared to the ones exhibited in \autoref{fig:data_norm_2DCNN} of the train and validation sets. The significant shift between the atmospheric features in the training and test datasets poses a significant challenge to the model's generalisation capabilities, as it encounters a markedly different distribution during evaluation. Such distribution is never presented in the training set means that the model cannot simply perform well by learning to memorise the pattern and recognise it during evaluation. Instead, it is forced to perform inference based on the incoming data alone. This distribution shift acts as a fundamental constraint on the model's ability to perform well on unseen data.

\subsubsection{Small training set}
The dataset design aligns with conservative estimates of Ariel's science yield, but its size (1,453 samples) presents a significant challenge for deep learning applications. Deep learning models derive their power from approximating arbitrary functions through substantial numbers of tunable parameters. Given the dense information content and noise within each sample, the data dimensionality necessitates a large parameter space. For example, our CNN architectures require 15M parameters — orders of magnitude larger than the number of training examples. This parameter-to-sample ratio impedes network convergence to generalizable solutions and increases susceptibility to overfitting.

\subsubsection{Uncertainty Quantification}
\label{sec:uq}
Reliable uncertainty quantification in neural networks remains a significant challenge in deep learning research. The high dimensionality of neural network parameter spaces complicates error propagation tracking, yet reliable confidence estimates are crucial for scientific applications. Current solutions typically rely on post-hoc approaches, such as the MC Dropout method implemented in this study.

Our results reveal two fundamental limitations of this approach. First, comparing \autoref{fig:random_valid_results} and \autoref{fig:random_test_results} shows minimal variation in error bar magnitude between validation and test sets, suggesting the model learned to replicate the training set's error distribution without adapting to the test distribution. Second, \autoref{fig:mean_pred} reveals that predicted uncertainties often exceed actual residuals for many validation examples while simultaneously failing to scale appropriately for out-of-distribution test examples with large prediction errors. In both regimes—whether predictions are accurate or inaccurate—the uncertainties remain similarly sized, indicating the method cannot distinguish between familiar and unfamiliar inputs.

This asymmetric behavior indicates that Monte Carlo (MC) Dropout learns the uncertainty distribution characteristic of the training data but cannot reliably extrapolate to novel scenarios. The method effectively memorizes the typical scatter in the training distribution rather than providing true epistemic uncertainty estimates that would appropriately increase for unfamiliar inputs. Ideally, the model should produce larger error bars for unseen examples, reflecting increased uncertainty. Instead, it applies learned uncertainty patterns to novel cases, adversely affecting the Gaussian Log Likelihood based score (GLL-score). This limitation is particularly problematic for survey missions like Ariel, where observations will inevitably encounter planetary systems outside the parameter space of any training set.

\subsection{Mitigation}
In this section we will discuss possible mitigations to the observed degradation in the previous section.

\subsubsection{Ad-hoc scaling factor}
Performance degradation in the test set can be attributed to underestimated error bars, resulting in severe scoring penalties. Applying differential scaling factors ($f=0.7$ for validation and $f=1.4$ for test sets) improved the second baseline scores to $0.5192$ and $0.4570$, respectively. While this adjustment improves model alignment with the test set, such dataset-specific modifications are not recommended as it generally does not scale well beyond the provided dataset. Notably, the model underestimates uncertainties even within its training distribution, indicating a fundamental limitation in uncertainty estimation methods that warrants future research.

\subsubsection{Choice of Optimisers}
Our investigation revealed a significant performance difference between optimisation algorithms. The Adam optimiser, despite its popularity for fast convergence and training stability, failed to achieve positive GLL-based metrics on the test set. Switching to SGD\footnote{Stochastic Gradient Descent (SGD) is an iterative optimisation algorithm that updates neural network parameters by computing gradients on small random subsets (mini-batches) of the training data, rather than the entire dataset. This introduces noise into the optimisation process, which can help escape local minima and improve generalisation.} substantially improved generalisation, achieving a test score of $\sim$0.45. This behaviour aligns with existing literature \citep{hardt2016train,wilson2017marginal}, supporting evidence that SGD's stochastic nature provides better generalisation performance on unseen data despite slower convergence.

\subsection{Supplementing test data to the training set}
We investigated the impact of incorporating limited test data into the training set, simulating early mission phases where initial observations become available. Table~\ref{tab:adding_test_in_train} presents a quantitative analysis of performance across different training–test mixtures, progressively adding random test samples (50, 100, 150, 400, and 700 out of 762 total test examples) to the original 576-sample training set.

Performance improvements show two distinct regimes. The MSE metric exhibits rapid initial improvement: adding just 50 test samples (8.6\% of available test data) reduces test MSE from 542 to 271 ppm, nearly a factor of two. This dramatic improvement suggests that even limited exposure to the test distribution helps the model learn critical features of the shifted domain. However, further additions show diminishing returns: increasing from 50 to 400 samples provides only incremental MSE reduction (271 to 185 ppm), and the curve appears to plateau unless the majority of test data is included.

GLL scores show a different pattern. While MSE improvements are front-loaded (large gains from first 50 samples), GLL improvements are more gradual and persist across the full range of added samples. Even with 700 test samples added, the GLL score (0.4945) remains below the validation set performance (0.5020 in the original split). This divergence indicates that \textit{accuracy} (MSE) improves faster than \textit{calibration} (GLL).

Implications for uncertainty quantification: The MSE–GLL divergence reveals that the network learns to predict spectral values more accurately but fails to appropriately adjust its uncertainty estimates. As prediction errors decrease with added test data, the model's error bars should correspondingly narrow, but they remain approximately constant, leading to overconfident predictions. This results in good MSE performance but continued GLL penalties, reinforcing our findings in Section~\ref{sec:uq} that Monte Carlo Dropout provides poorly calibrated uncertainties under distribution shift.

These results suggest that adaptive learning strategies incorporating even small amounts of mission data could substantially improve pipeline performance, but improved uncertainty quantification methods will be necessary to achieve well-calibrated confidence estimates.

\begin{table*}
\centering
\resizebox{1\textwidth}{!}{%
\begin{tabular}{lcccccccccccc}
\hline
& \multicolumn{2}{|c|}{Original scores} & \multicolumn{2}{|c|}{With 50 test data} & \multicolumn{2}{|c|}{With 100 test data} & \multicolumn{2}{|c|}{With 150 test data} & \multicolumn{2}{|c|}{With 400 test data} & \multicolumn{2}{|c|}{With 700 test data}  \\
\hline
& Valid & Test & Valid & Test & Valid & Test & Valid & Test & Valid & Test & Valid & Test \\
\hline
MSE mean values (in ppm) & 79  & 530 & 97  & 254 & 137 & 259 & 117 & 171 & 145 & 175 & 127 & 140 \\
MSE fluctuations (in ppm) & 40 & 110 & 49 &  95 & 49 & 86 & 43 & 67 & 45 & 61 & 50 & 55 \\
MSE total (in ppm) & 88 & 542 & 108 &  271 & 145 & 273 & 125 & 184 & 151 & 185 & 136 & 150 \\
GLL score & 5016 & 4246 & 4560 & 4912 & 4438 & 4702 & 4346 & 4860 & 4592 & 4880 & 4674 & 4945 \\

\hline
\end{tabular}%
}
\caption{Scores when adding some test data in the training set.}
\label{tab:adding_test_in_train}

\end{table*}

% \comment{[I feel like this results section lacks the hard-earned insights you have gained after training the models, for instances, we can comment on things like Rationale behind the two-step training approach, Analysis of performance deterioration in test scores - perhaps related to challenge design?
% Comparison of MSE between validation and test sets - Note the 7x increase in test MSE compared to validation MSE}

% \comment{Impact of Dataset Size and Composition: Virginie's exploration on moving test set into the training set, and the corresponding increase in the performance (if it is still there?) Given the small amount of data we have, it will be useful to see how small amount of test data could help to improve the score, and the implication of it in real cases. }

% \comment{Discussion on Model Behavior and possible Future directions? }

\subsection{Limitations}

While the dataset and methodology presented in this work represent a significant step forward in the preparation for Ariel’s data reduction challenges, a number of limitations must be acknowledged to frame the scope and applicability of the results.

Firstly, the simulated dataset, although based on updated payload configurations and realistic noise sources, including those derived from JWST calibration products, inevitably relies on a set of assumptions and simplifications. For instance, the pointing jitter is derived from a single realisation of a timeline, which is randomly shifted across simulations but may not fully encapsulate the range and variability expected during the mission’s operational lifetime. This constraint could limit the generality of conclusions drawn from jitter-related analyses.

Moreover, the atmospheric models used to generate the transit spectra adopt a simplified, isothermal, 1D structure with constant chemical abundances and a fixed set of trace gases. These models are intentionally not physically self-consistent in order to challenge the robustness of data reduction pipelines under diverse scenarios. However, this choice also means that certain effects, such as vertical chemical gradients, disequilibrium chemistry, or cloud heterogeneity, are not represented, reducing the astrophysical realism of the simulations.

Another important limitation concerns the stellar sample used in this exercise. The dataset includes only four host stars, selected for their brightness and compatibility with the imposed observing strategy. While this choice ensures consistency and allows for short integration times suitable for jitter analysis, it significantly restricts the diversity of stellar types and spectral energy distributions present in the dataset. Moreover, the sample omits both limb‑darkening effects and the presence of stellar spots—phenomena that significantly alter transit morphology. As a result, the generalisability of machine learning models trained on this dataset to a broader population of stars remains untested and should be approached with caution.

The observational setup is also standardised across the entire dataset, with fixed integration times, transit durations, and observing windows. While this uniformity supports the implementation and training of machine learning pipelines, it sacrifices the variability inherent in actual observations, where exposure times and transit coverage are tailored to individual targets. Consequently, the dataset may not fully capture the complexity of real-time scheduling and its implications for noise and signal extraction.

In terms of calibration, the decision to perturb the calibration products uniformly by 5$\%$  simulates uncertainties in a controlled fashion, but does not reproduce the full complexity and temporal evolution of real calibration errors, such as detector degradation, flat field instability, or long-term non-linear effects.

Finally, while the baseline neural network models presented in this study offer a proof-of-concept for learning-based reduction strategies, they are not optimised for scientific use. The training dataset remains limited in terms of astrophysical diversity and instrumental conditions, and the neural network architectures do not exploit advanced strategies such as self-supervised learning, transfer learning, or domain adaptation. Moreover, the risk associated with domain shift, that is, the discrepancy between the training and deployment data distributions, remains an open challenge and is only partially addressed through the proposed data partitioning strategy.

Despite these limitations, the dataset provides a valuable and well-characterised platform for benchmarking data reduction pipelines. It has been designed with modularity and extensibility in mind, enabling future developments that may address the limitations outlined above.

\section{Conclusion}
\label{sec:conclusion}

This work presents a comprehensive public dataset of simulated Ariel mission observations, addressing a critical need in the exoplanetary atmospheric characterisation community. By combining ExoSim2 and TauREx3 to create realistic synthetic data, we have developed one of the most extensive benchmark datasets based on the current payload design of the ESA Ariel Space Mission. The dataset includes 1435 simulated observations across diverse stellar and planetary configurations, featuring realistic noise models, calibration products from JWST NIRSpec observations, and systematic effects based on current space-based missions.

We provide, as an initial attempt, the use of deep learning models for a potentially scalable data reduction pipeline. We implemented and evaluated two baseline deep learning approaches demonstrating both the potential and challenges of using an end to end ML-based data reduction pipelines.

% These attempts reveal critical weaknesses in ML-based models in small data regime, and highlight issues when models experience shifts in the (incoming) data distribution as compared to its training data.

While these models proved to be quite successfully in extracting the average transit depths from noisy data, they also highlight critical issues with non-parametric models such as the tendency to overfit on the training datasets, unable to mitigate distribution shifts between training and test data and unreliable uncertainty quantification. These findings highlights the importance of developing robust, adaptable algorithms capable of handling diverse observational scenarios.

We have also investigated the impact of dataset composition on the ability for the model to generalise. Our analysis reveals that even small additions of test data to the training set can significantly improve model performance, hinting that adaptive learning strategies, or continuous learning strategies may be useful to optimise the reliability of these algorithms. The dataset has already proven its value through the Ariel Data Challenge 2024, providing a standardised and open platform for algorithm development and comparison.

Moving forward, this resource will support the broader scientific community in developing scalable, efficient, and reliable data reduction techniques essential for maximising the scientific return of the Ariel mission and future exoplanet characterisation efforts. As we approach the launch of the Ariel Space Telescope in the coming future, continued refinement of these methods will be crucial for achieving the mission's ambitious goals in comparative planetology and atmospheric science.

\subsection{Going beyond the data challenge}
Although the dataset was originally released in the context of the Ariel Data Challenge 2024, its structure, complexity, and realism render it suitable for a wide range of applications beyond the scope of the competition. Designed to replicate the full observational process, including time, wavelength and spatially-dependent noise sources and calibration uncertainties, the dataset can serve as a valuable benchmark for the community.

Firstly, the dataset can be used as a benchmark for the development and testing of data reduction pipelines. Its design incorporates a broad range of realistic noise sources, such as gain drifts, pointing jitter, pixel non-linearity, and detector-level systematics, making it a valuable resource for evaluating the performance and robustness of classical, Bayesian, and machine learning-based approaches under conditions representative of actual space-based observations.

A key feature of the dataset is the inclusion of the ground truth atmospheric spectra. This enables rigorous quantification of systematic biases introduced during the data reduction process. By comparing extracted transmission spectra to the known input values, users can evaluate the absolute accuracy of their methods and identify wavelength-dependent residuals or distortions, providing critical feedback for pipeline optimisation.

The dataset also provides an ideal testbed for methodological studies in machine learning, such as domain generalisation, transfer learning and anomaly detection. The structured shift between training and test distributions has been explicitly designed to reflect realistic deployment scenarios, offering an opportunity to assess model generalisability in the presence of non‑stationary noise. Ultimately, the extraction of planetary spectra from photometric and spectroscopic light curves for each Ariel target will balance target‑specific adjustments with a uniformly applied reduction framework across the entire sample. In this context, machine‑learning methods will play a key role in automatically detecting anomalous cases—such as those impacted by stellar contamination, thus flagging instances that may require more bespoke treatment.

Finally, given its accessibility and comprehensive documentation, the public dataset is well-suited for training and educational purposes. It can be employed to demonstrate the complexities of exoplanet transit spectroscopy, the role of calibration products, and the importance of error propagation in atmospheric retrievals.

Overall, the dataset aims to support the development of scalable, accurate, and reproducible methodologies for exoplanetary atmospheric characterisation, and to foster community readiness for the analysis of Ariel data and other forthcoming space-based surveys.

\section*{Acknowledgements}
The work presented in this paper was partially supported by UKSA,
grant ST/X002616/1, ST/W002507/1, and by the Italian Space Agency (ASI) with the Ariel grant n. 2021.5.HH.0.
This work used the DiRAC Data Intensive service (CSD3) at the University of Cambridge, managed by the University of Cambridge University Information Services on behalf of the STFC DiRAC HPC Facility (www.dirac.ac.uk). The DiRAC component of CSD3 at Cambridge was funded by BEIS, UKRI and STFC capital funding and STFC operations grants. DiRAC is part of the UKRI Digital Research Infrastructure.

The authors also wish to thank the Ariel S2MD Working Group for the technical support and input on the noise simulations.

%%%%%%%%%%%%%%%%%%%%%%%%%%%%%%%%%%%%%%%%%%%%%%%%%%
\section*{Data Availability}
The train data generated for this manuscript are available at \url{https://www.kaggle.com/competitions/ariel-data-challenge-2024/data}
%%%%%%%%%%%%%%%%%%%% REFERENCES %%%%%%%%%%%%%%%%%%

% The best way to enter references is to use BibTeX:

\bibliographystyle{rasti}
\bibliography{main} % if your bibtex file is called example.bib

@ARTICLE{Espinoza2025ApJ,
       author = {{Espinoza}, N{\'e}stor and {Allen}, Natalie H. and {Glidden}, Ana and {Lewis}, Nikole K. and {Seager}, Sara and {Ca{\~n}as}, Caleb I. and {Grant}, David and {Gressier}, Am{\'e}lie and {Courreges}, Shelby and {Stevenson}, Kevin B. and et al.},
        title = "{JWST-TST DREAMS: NIRSpec/PRISM Transmission Spectroscopy of the Habitable Zone Planet TRAPPIST-1 e}",
      journal = {\apjl},
         year = 2025,
        month = sep,
       volume = {990},
       number = {2},
          eid = {L52},
        pages = {L52},
          doi = {10.3847/2041-8213/adf42e},
archivePrefix = {arXiv},
       eprint = {2509.05414},
 primaryClass = {astro-ph.EP},
       adsurl = {https://ui.adsabs.harvard.edu/abs/2025ApJ...990L..52E}
}

@ARTICLE{Taylor2025MNRAS,
       author = {{Taylor}, Jake and {Radica}, Michael and {Chatterjee}, Richard D. and {Hammond}, Mark and {Meier}, Tobias and {Aigrain}, Suzanne and {MacDonald}, Ryan J. and {Albert}, Loic and {Benneke}, Bj{\"o}rn and {Coulombe}, Louis-Philippe and et al.},
        title = "{JWST NIRISS transmission spectroscopy of the super-Earth GJ 357b, a favourable target for atmospheric retention}",
      journal = {\mnras},
         year = 2025,
        month = jul,
       volume = {540},
       number = {4},
        pages = {3677-3692},
          doi = {10.1093/mnras/staf894},
archivePrefix = {arXiv},
       eprint = {2505.24462},
 primaryClass = {astro-ph.EP},
       adsurl = {https://ui.adsabs.harvard.edu/abs/2025MNRAS.540.3677T}
}

@ARTICLE{Carter2024NatAs,
       author = {{Carter}, A.~L. and {May}, E.~M. and {Espinoza}, N. and {Welbanks}, L. and {Ahrer}, E. and {Alderson}, L. and {Brahm}, R. and {Feinstein}, A.~D. and {Grant}, D. and {Line}, M. and et al.},
        title = "{A benchmark JWST near-infrared spectrum for the exoplanet WASP-39 b}",
      journal = {Nature Astronomy},
         year = 2024,
        month = aug,
       volume = {8},
        pages = {1008-1019},
          doi = {10.1038/s41550-024-02292-x},
archivePrefix = {arXiv},
       eprint = {2407.13893},
 primaryClass = {astro-ph.EP},
       adsurl = {https://ui.adsabs.harvard.edu/abs/2024NatAs...8.1008C}
}

@ARTICLE{Radica2023MNRAS,
       author = {{Radica}, Michael and {Welbanks}, Luis and {Espinoza}, N{\'e}stor and {Taylor}, Jake and {Coulombe}, Louis-Philippe and {Feinstein}, Adina D. and {Goyal}, Jayesh and {Scarsdale}, Nicholas and {Albert}, Lo{\"\i}c and {Baghel}, Priyanka and et al.},
        title = "{Awesome SOSS: transmission spectroscopy of WASP-96b with NIRISS/SOSS}",
      journal = {\mnras},
         year = 2023,
        month = sep,
       volume = {524},
       number = {1},
        pages = {835-856},
          doi = {10.1093/mnras/stad1762},
archivePrefix = {arXiv},
       eprint = {2305.17001},
 primaryClass = {astro-ph.EP},
       adsurl = {https://ui.adsabs.harvard.edu/abs/2023MNRAS.524..835R}
}

@ARTICLE{Challener2025NatAs,
       author = {{Challener}, Ryan C. and {Weiner Mansfield}, Megan and {Cubillos}, Patricio E. and {Piette}, Anjali A.~A. and {Coulombe}, Louis-Philippe and {Beltz}, Hayley and {Blecic}, Jasmina and {Rauscher}, Emily and {Bean}, Jacob L. and {Benneke}, Bj{\"o}rn and et al.},
        title = "{Horizontal and vertical exoplanet thermal structure from a JWST spectroscopic eclipse map}",
      journal = {Nature Astronomy},
         year = 2025,
        month = dec,
       volume = {9},
        pages = {1821-1832},
          doi = {10.1038/s41550-025-02666-9},
archivePrefix = {arXiv},
       eprint = {2510.24708},
 primaryClass = {astro-ph.EP},
       adsurl = {https://ui.adsabs.harvard.edu/abs/2025NatAs...9.1821C}
}

@ARTICLE{Tinetti2018ExA,
       author = {{Tinetti}, Giovanna and {Drossart}, Pierre and {Eccleston}, Paul and {Hartogh}, Paul and {Heske}, Astrid and {Leconte}, J{\'e}r{\'e}my and {Micela}, Giusi and {Ollivier}, Marc and {Pilbratt}, G{\"o}ran and {Puig}, Ludovic and et al.},
        title = "{A chemical survey of exoplanets with ARIEL}",
      journal = {Experimental Astronomy},
         year = 2018,
        month = nov,
       volume = {46},
       number = {1},
        pages = {135-209},
          doi = {10.1007/s10686-018-9598-x},
       adsurl = {https://ui.adsabs.harvard.edu/abs/2018ExA....46..135T}
}

@ARTICLE{Greene2016ApJ,
       author = {{Greene}, Thomas P. and {Line}, Michael R. and {Montero}, Cezar and {Fortney}, Jonathan J. and {Lustig-Yaeger}, Jacob and {Luther}, Kyle},
        title = "{Characterizing Transiting Exoplanet Atmospheres with JWST}",
      journal = {\apj},
         year = 2016,
        month = jan,
       volume = {817},
       number = {1},
          eid = {17},
        pages = {17},
          doi = {10.3847/0004-637X/817/1/17},
archivePrefix = {arXiv},
       eprint = {1511.05528},
 primaryClass = {astro-ph.EP},
       adsurl = {https://ui.adsabs.harvard.edu/abs/2016ApJ...817...17G}
}

@ARTICLE{Yip2020,
       author = {{Yip}, K.~H. and {Tsiaras}, A. and {Waldmann}, I.~P. and {Tinetti}, G.},
        title = "{Integrating Light Curve and Atmospheric Modeling of Transiting Exoplanets}",
      journal = {\aj},
         year = 2020,
        month = oct,
       volume = {160},
       number = {4},
          eid = {171},
        pages = {171},
          doi = {10.3847/1538-3881/abaabc},
archivePrefix = {arXiv},
       eprint = {1811.04686},
 primaryClass = {astro-ph.EP},
       adsurl = {https://ui.adsabs.harvard.edu/abs/2020AJ....160..171Y}
}

@ARTICLE{Changeat_2024,
       author = {{Changeat}, Q. and {Ito}, Y. and {Al-Refaie}, A.~F. and {Yip}, K.~H. and {Lueftinger}, T.},
        title = "{Toward Atmospheric Retrievals of Panchromatic Light Curves: EXPLOR-ing Generalized Inversion Techniques for Transiting Exoplanets with JWST and Ariel}",
      journal = {\aj},
         year = 2024,
        month = may,
       volume = {167},
       number = {5},
          eid = {195},
        pages = {195},
          doi = {10.3847/1538-3881/ad3032},
archivePrefix = {arXiv},
       eprint = {2403.02244},
 primaryClass = {astro-ph.EP},
       adsurl = {https://ui.adsabs.harvard.edu/abs/2024AJ....167..195C}
}

@ARTICLE{Yurchenko_AlH,
       author = {{Yurchenko}, Sergei N. and {Szajna}, Wojciech and {Hakalla}, Rafa{\l} and {Semenov}, Mikhail and {Sokolov}, Andrei and {Tennyson}, Jonathan and {Gamache}, Robert R. and {Pavlenko}, Yakiv and {Schmidt}, Mirek R.},
        title = "{ExoMol line lists - LIV. Empirical line lists for AlH and AlD and experimental emission spectroscopy of AlD in A$^{1}${\ensuremath{\Pi}} (v = 0, 1, 2)}",
      journal = {\mnras},
         year = 2024,
        month = feb,
       volume = {527},
       number = {4},
        pages = {9736-9756},
          doi = {10.1093/mnras/stad3802},
archivePrefix = {arXiv},
       eprint = {2312.05958},
 primaryClass = {astro-ph.SR},
       adsurl = {https://ui.adsabs.harvard.edu/abs/2024MNRAS.527.9736Y}
}

@ARTICLE{SiO,
       author = {{Buldyreva}, Jeanna and {Yurchenko}, Sergei N. and {Tennyson}, Jonathan},
        title = "{Simple semiclassical model of pressure-broadened infrared/microwave linewidths in the temperature range 200-3000 K}",
      journal = {RAS Techniques and Instruments},
         year = 2022,
        month = apr,
       volume = {1},
       number = {1},
        pages = {43-47},
          doi = {10.1093/rasti/rzac004},
       adsurl = {https://ui.adsabs.harvard.edu/abs/2022RASTI...1...43B}
}

@ARTICLE{Underwood2016,
       author = {{Underwood}, Daniel S. and {Tennyson}, Jonathan and {Yurchenko}, Sergei N. and {Huang}, Xinchuan and {Schwenke}, David W. and {Lee}, Timothy J. and {Clausen}, S{\o}nnik and {Fateev}, Alexander},
        title = "{ExoMol molecular line lists - XIV. The rotation-vibration spectrum of hot SO$_{2}$}",
      journal = {\mnras},
         year = 2016,
        month = jul,
       volume = {459},
       number = {4},
        pages = {3890-3899},
          doi = {10.1093/mnras/stw849},
archivePrefix = {arXiv},
       eprint = {1603.04065},
 primaryClass = {astro-ph.EP},
       adsurl = {https://ui.adsabs.harvard.edu/abs/2016MNRAS.459.3890U}
}

@ARTICLE{Bocchieri2025,
       author = {{Bocchieri}, Andrea and {Mugnai}, Lorenzo V. and {Pascale}, Enzo and {Papageorgiou}, Andreas and {Syty}, Ang{\`e}le and {Tsiaras}, Angelos and {Eccleston}, Paul and {Savini}, Giorgio and {Tinetti}, Giovanna and {Broquet}, Renaud and {Chapman}, Patrick and {Sechi}, Gianfranco},
        title = "{De-jittering Ariel: An optimized algorithm}",
      journal = {Experimental Astronomy},
         year = 2025,
        month = jun,
       volume = {59},
       number = {3},
          eid = {31},
        pages = {31},
          doi = {10.1007/s10686-025-09999-3},
archivePrefix = {arXiv},
       eprint = {2504.12907},
 primaryClass = {astro-ph.EP},
       adsurl = {https://ui.adsabs.harvard.edu/abs/2025ExA....59...31B}
}

@ARTICLE{Tinetti2013,
       author = {{Tinetti}, Giovanna and {Encrenaz}, Th{\'e}r{\`e}se and {Coustenis}, Athena},
        title = "{Spectroscopy of planetary atmospheres in our Galaxy}",
      journal = {\aapr},
         year = 2013,
        month = oct,
       volume = {21},
          eid = {63},
        pages = {63},
          doi = {10.1007/s00159-013-0063-6},
       adsurl = {https://ui.adsabs.harvard.edu/abs/2013A&ARv..21...63T}
}

@ARTICLE{Vinooja2025,
       author = {{Thurairethinam}, Vinooja and {Bocchieri}, Andrea and {Savini}, Giorgio and {Mugnai}, Lorenzo V. and {Pascale}, Enzo},
        title = "{Modeling phase variations introduced by extreme broadband dichroics for astronomical photometry}",
      journal = {Journal of Astronomical Telescopes, Instruments, and Systems},
         year = 2025,
        month = jan,
       volume = {11},
          eid = {014003},
        pages = {014003},
          doi = {10.1117/1.JATIS.11.1.014003},
       adsurl = {https://ui.adsabs.harvard.edu/abs/2025JATIS..11a4003T}
}

@article{EXOSIM,
  author = {Mugnai, Lorenzo V. and Bocchieri, Andrea and Pascale, Enzo and Lorenzani, Andrea and Papageorgiou, Andreas},
  title = {ExoSim 2: the new exoplanet observation simulator applied to the Ariel space mission},
  journal = {Experimental Astronomy},
  year = {2025},
  volume = {59},
  number = {1},
  pages = {9},
  doi = {10.1007/s10686-024-09976-2},
  url = {https://doi.org/10.1007/s10686-024-09976-2},
  issn = {1572-9508}
}

@article{wilson2017marginal,
  title={The marginal value of adaptive gradient methods in machine learning},
  author={Wilson, Ashia C and Roelofs, Rebecca and Stern, Mitchell and Srebro, Nati and Recht, Benjamin},
  journal={Advances in neural information processing systems},
  volume={30},
  year={2017}
}

@inproceedings{hardt2016train,
  title={Train faster, generalize better: Stability of stochastic gradient descent},
  author={Hardt, Moritz and Recht, Ben and Singer, Yoram},
  booktitle={International conference on machine learning},
  pages={1225--1234},
  year={2016},
  organization={PMLR}
}

@ARTICLE{2024MNRAS.531.2731S,
       author = {{Sarkar}, Subhajit and {Madhusudhan}, Nikku and {Constantinou}, Savvas and {Holmberg}, M{\r{a}}ns},
        title = "{Exoplanet transit spectroscopy with JWST NIRSpec: diagnostics and homogeneous case study of WASP-39 b}",
      journal = {\mnras},
         year = 2024,
        month = jun,
       volume = {531},
       number = {2},
        pages = {2731-2756},
          doi = {10.1093/mnras/stae1230},
archivePrefix = {arXiv},
       eprint = {2405.06737},
 primaryClass = {astro-ph.EP},
       adsurl = {https://ui.adsabs.harvard.edu/abs/2024MNRAS.531.2731S}
}

@ARTICLE{2023NIRCam,
       author = {{Ahrer}, Eva-Maria and {Stevenson}, Kevin B. and {Mansfield}, Megan and {Moran}, Sarah E. and {Brande}, Jonathan and {Morello}, Giuseppe and {Murray}, Catriona A. and {Nikolov}, Nikolay K. and {Petit dit de la Roche}, Dominique J.~M. and {Schlawin}, Everett and {Wheatley}, Peter J. and {Zieba}, Sebastian and {Batalha}, Natasha E. and {Damiano}, Mario and {Goyal}, Jayesh M. and {Lendl}, Monika and {Lothringer}, Joshua D. and {Mukherjee}, Sagnick and {Ohno}, Kazumasa and {Batalha}, Natalie M. and {Battley}, Matthew P. and {Bean}, Jacob L. and {Beatty}, Thomas G. and {Benneke}, Bj{\"o}rn and {Berta-Thompson}, Zachory K. and {Carter}, Aarynn L. and {Cubillos}, Patricio E. and {Daylan}, Tansu and {Espinoza}, N{\'e}stor and {Gao}, Peter and {Gibson}, Neale P. and {Gill}, Samuel and {Harrington}, Joseph and {Hu}, Renyu and {Kreidberg}, Laura and {Lewis}, Nikole K. and {Line}, Michael R. and {L{\'o}pez-Morales}, Mercedes and {Parmentier}, Vivien and {Powell}, Diana K. and {Sing}, David K. and {Tsai}, Shang-Min and {Wakeford}, Hannah R. and {Welbanks}, Luis and {Alam}, Munazza K. and {Alderson}, Lili and {Allen}, Natalie H. and {Anderson}, David R. and {Barstow}, Joanna K. and {Bayliss}, Daniel and {Bell}, Taylor J. and {Blecic}, Jasmina and {Bryant}, Edward M. and {Burleigh}, Matthew R. and {Carone}, Ludmila and {Casewell}, S.~L. and {Changeat}, Quentin and {Chubb}, Katy L. and {Crossfield}, Ian J.~M. and {Crouzet}, Nicolas and {Decin}, Leen and {D{\'e}sert}, Jean-Michel and {Feinstein}, Adina D. and {Flagg}, Laura and {Fortney}, Jonathan J. and {Gizis}, John E. and {Heng}, Kevin and {Iro}, Nicolas and {Kempton}, Eliza M. -R. and {Kendrew}, Sarah and {Kirk}, James and {Knutson}, Heather A. and {Komacek}, Thaddeus D. and {Lagage}, Pierre-Olivier and {Leconte}, J{\'e}r{\'e}my and {Lustig-Yaeger}, Jacob and {MacDonald}, Ryan J. and {Mancini}, Luigi and {May}, E.~M. and {Mayne}, N.~J. and {Miguel}, Yamila and {Mikal-Evans}, Thomas and {Molaverdikhani}, Karan and {Palle}, Enric and {Piaulet}, Caroline and {Rackham}, Benjamin V. and {Redfield}, Seth and {Rogers}, Laura K. and {Roy}, Pierre-Alexis and {Rustamkulov}, Zafar and {Shkolnik}, Evgenya L. and {Sotzen}, Kristin S. and {Taylor}, Jake and {Tremblin}, P. and {Tucker}, Gregory S. and {Turner}, Jake D. and {de Val-Borro}, Miguel and {Venot}, Olivia and {Zhang}, Xi},
        title = "{Early Release Science of the exoplanet WASP-39b with JWST NIRCam}",
      journal = {\nat},
         year = 2023,
        month = feb,
       volume = {614},
       number = {7949},
        pages = {653-658},
          doi = {10.1038/s41586-022-05590-4},
archivePrefix = {arXiv},
       eprint = {2211.10489},
 primaryClass = {astro-ph.EP},
       adsurl = {https://ui.adsabs.harvard.edu/abs/2023Natur.614..653A}
}

@ARTICLE{Polyansky2018,
       author = {{Polyansky}, Oleg L. and {Kyuberis}, Aleksandra A. and {Zobov}, Nikolai F. and et al.},
        title = "{ExoMol molecular line lists XXX: a complete high-accuracy line list for water}",
      journal = {\mnras},
         year = 2018,
        month = oct,
       volume = {480},
       number = {2},
        pages = {2597-2608},
          doi = {10.1093/mnras/sty1877},
archivePrefix = {arXiv},
       eprint = {1807.04529},
 primaryClass = {astro-ph.EP},
       adsurl = {https://ui.adsabs.harvard.edu/abs/2018MNRAS.480.2597P}
}

@ARTICLE{Rothman2010,
       author = {{Rothman}, L.~S. and {Gordon}, I.~E. and {Barber}, R.~J. and et al.},
        title = "{HITEMP, the high-temperature molecular spectroscopic database}",
      journal = {\jqsrt},
         year = 2010,
        month = oct,
       volume = {111},
        pages = {2139-2150},
          doi = {10.1016/j.jqsrt.2010.05.001},
       adsurl = {https://ui.adsabs.harvard.edu/abs/2010JQSRT.111.2139R}
}

@ARTICLE{Yurchenko2017,
       author = {{Yurchenko}, Sergei N. and {Amundsen}, David S. and {Tennyson}, Jonathan and et al.},
        title = "{A hybrid line list for CH$_{4}$ and hot methane continuum}",
      journal = {\aap},
         year = 2017,
        month = sep,
       volume = {605},
          eid = {A95},
        pages = {A95},
          doi = {10.1051/0004-6361/201731026},
archivePrefix = {arXiv},
       eprint = {1706.05724},
 primaryClass = {astro-ph.EP},
       adsurl = {https://ui.adsabs.harvard.edu/abs/2017A&A...605A..95Y}
}

@ARTICLE{Coles2019,
       author = {{Coles}, Phillip A. and {Yurchenko}, Sergei N. and {Tennyson}, Jonathan},
        title = "{ExoMol molecular line lists - XXXV. A rotation-vibration line list for hot ammonia}",
      journal = {\mnras},
         year = 2019,
        month = dec,
       volume = {490},
       number = {4},
        pages = {4638-4647},
          doi = {10.1093/mnras/stz2778},
archivePrefix = {arXiv},
       eprint = {1911.10369},
 primaryClass = {astro-ph.SR},
       adsurl = {https://ui.adsabs.harvard.edu/abs/2019MNRAS.490.4638C}
}

@ARTICLE{Chubb2020,
       author = {{Chubb}, Katy L. and {Tennyson}, Jonathan and {Yurchenko}, Sergei N.},
        title = "{ExoMol molecular line lists - XXXVII. Spectra of acetylene}",
      journal = {\mnras},
         year = 2020,
        month = apr,
       volume = {493},
       number = {2},
        pages = {1531-1545},
          doi = {10.1093/mnras/staa229},
archivePrefix = {arXiv},
       eprint = {2001.04550},
 primaryClass = {astro-ph.SR},
       adsurl = {https://ui.adsabs.harvard.edu/abs/2020MNRAS.493.1531C}
}

@ARTICLE{Barber2014,
       author = {{Barber}, R.~J. and {Strange}, J.~K. and {Hill}, C. and et al.},
        title = "{ExoMol line lists - III. An improved hot rotation-vibration line list for HCN and HNC}",
      journal = {\mnras},
         year = 2014,
        month = jan,
       volume = {437},
       number = {2},
        pages = {1828-1835},
          doi = {10.1093/mnras/stt2011},
archivePrefix = {arXiv},
       eprint = {1311.1328},
 primaryClass = {astro-ph.SR},
       adsurl = {https://ui.adsabs.harvard.edu/abs/2014MNRAS.437.1828B}
}

@ARTICLE{Li2015,
       author = {{Li}, Gang and {Gordon}, Iouli E. and {Rothman}, Laurence S. and et al.},
        title = "{Rovibrational Line Lists for Nine Isotopologues of the CO Molecule in the X $^{1}${\ensuremath{\Sigma}}$^{+}$ Ground Electronic State}",
      journal = {\apjs},
         year = 2015,
        month = jan,
       volume = {216},
       number = {1},
          eid = {15},
        pages = {15},
          doi = {10.1088/0067-0049/216/1/15},
       adsurl = {https://ui.adsabs.harvard.edu/abs/2015ApJS..216...15L}
}

@article{Azzam2016,
    author = {Azzam, Ala'a A. A. and Tennyson, Jonathan and Yurchenko, Sergei N. and Naumenko, Olga V.},
    title = "{ExoMol molecular line lists – XVI. The rotation–vibration spectrum of hot H2S}",
    journal = {Monthly Notices of the Royal Astronomical Society},
    volume = {460},
    number = {4},
    pages = {4063-4074},
    year = {2016},
    month = {05},
    issn = {0035-8711},
    doi = {10.1093/mnras/stw1133},
    url = {https://doi.org/10.1093/mnras/stw1133},
    eprint = {https://academic.oup.com/mnras/article-pdf/460/4/4063/13773124/stw1133.pdf},
}

@ARTICLE{2023PRISM,
       author = {{Rustamkulov}, Z. and {Sing}, D.~K. and {Mukherjee}, S. and {May}, E.~M. and {Kirk}, J. and {Schlawin}, E. and {Line}, M.~R. and {Piaulet}, C. and {Carter}, A.~L. and {Batalha}, N.~E. and {Goyal}, J.~M. and {L{\'o}pez-Morales}, M. and {Lothringer}, J.~D. and {MacDonald}, R.~J. and {Moran}, S.~E. and {Stevenson}, K.~B. and {Wakeford}, H.~R. and {Espinoza}, N. and {Bean}, J.~L. and {Batalha}, N.~M. and {Benneke}, B. and {Berta-Thompson}, Z.~K. and {Crossfield}, I.~J.~M. and {Gao}, P. and {Kreidberg}, L. and {Powell}, D.~K. and {Cubillos}, P.~E. and {Gibson}, N.~P. and {Leconte}, J. and {Molaverdikhani}, K. and {Nikolov}, N.~K. and {Parmentier}, V. and {Roy}, P. and {Taylor}, J. and {Turner}, J.~D. and {Wheatley}, P.~J. and {Aggarwal}, K. and {Ahrer}, E. and {Alam}, M.~K. and {Alderson}, L. and {Allen}, N.~H. and {Banerjee}, A. and {Barat}, S. and {Barrado}, D. and {Barstow}, J.~K. and {Bell}, T.~J. and {Blecic}, J. and {Brande}, J. and {Casewell}, S. and {Changeat}, Q. and {Chubb}, K.~L. and {Crouzet}, N. and {Daylan}, T. and {Decin}, L. and {D{\'e}sert}, J. and {Mikal-Evans}, T. and {Feinstein}, A.~D. and {Flagg}, L. and {Fortney}, J.~J. and {Harrington}, J. and {Heng}, K. and {Hong}, Y. and {Hu}, R. and {Iro}, N. and {Kataria}, T. and {Kempton}, E.~M. -R. and {Krick}, J. and {Lendl}, M. and {Lillo-Box}, J. and {Louca}, A. and {Lustig-Yaeger}, J. and {Mancini}, L. and {Mansfield}, M. and {Mayne}, N.~J. and {Miguel}, Y. and {Morello}, G. and {Ohno}, K. and {Palle}, E. and {Petit dit de la Roche}, D.~J.~M. and {Rackham}, B.~V. and {Radica}, M. and {Ramos-Rosado}, L. and {Redfield}, S. and {Rogers}, L.~K. and {Shkolnik}, E.~L. and {Southworth}, J. and {Teske}, J. and {Tremblin}, P. and {Tucker}, G.~S. and {Venot}, O. and {Waalkes}, W.~C. and {Welbanks}, L. and {Zhang}, X. and {Zieba}, S.},
        title = "{Early Release Science of the exoplanet WASP-39b with JWST NIRSpec PRISM}",
      journal = {\nat},
         year = 2023,
        month = feb,
       volume = {614},
       number = {7949},
        pages = {659-663},
          doi = {10.1038/s41586-022-05677-y},
archivePrefix = {arXiv},
       eprint = {2211.10487},
 primaryClass = {astro-ph.EP},
       adsurl = {https://ui.adsabs.harvard.edu/abs/2023Natur.614..659R}
}

@ARTICLE{2023G395H,
       author = {{Alderson}, Lili and {Wakeford}, Hannah R. and {Alam}, Munazza K. and {Batalha}, Natasha E. and {Lothringer}, Joshua D. and {Adams Redai}, Jea and {Barat}, Saugata and {Brande}, Jonathan and {Damiano}, Mario and {Daylan}, Tansu and {Espinoza}, N{\'e}stor and {Flagg}, Laura and {Goyal}, Jayesh M. and {Grant}, David and {Hu}, Renyu and {Inglis}, Julie and {Lee}, Elspeth K.~H. and {Mikal-Evans}, Thomas and {Ramos-Rosado}, Lakeisha and {Roy}, Pierre-Alexis and {Wallack}, Nicole L. and {Batalha}, Natalie M. and {Bean}, Jacob L. and {Benneke}, Bj{\"o}rn and {Berta-Thompson}, Zachory K. and {Carter}, Aarynn L. and {Changeat}, Quentin and {Col{\'o}n}, Knicole D. and {Crossfield}, Ian J.~M. and {D{\'e}sert}, Jean-Michel and {Foreman-Mackey}, Daniel and {Gibson}, Neale P. and {Kreidberg}, Laura and {Line}, Michael R. and {L{\'o}pez-Morales}, Mercedes and {Molaverdikhani}, Karan and {Moran}, Sarah E. and {Morello}, Giuseppe and {Moses}, Julianne I. and {Mukherjee}, Sagnick and {Schlawin}, Everett and {Sing}, David K. and {Stevenson}, Kevin B. and {Taylor}, Jake and {Aggarwal}, Keshav and {Ahrer}, Eva-Maria and {Allen}, Natalie H. and {Barstow}, Joanna K. and {Bell}, Taylor J. and {Blecic}, Jasmina and {Casewell}, Sarah L. and {Chubb}, Katy L. and {Crouzet}, Nicolas and {Cubillos}, Patricio E. and {Decin}, Leen and {Feinstein}, Adina D. and {Fortney}, Joanthan J. and {Harrington}, Joseph and {Heng}, Kevin and {Iro}, Nicolas and {Kempton}, Eliza M. -R. and {Kirk}, James and {Knutson}, Heather A. and {Krick}, Jessica and {Leconte}, J{\'e}r{\'e}my and {Lendl}, Monika and {MacDonald}, Ryan J. and {Mancini}, Luigi and {Mansfield}, Megan and {May}, Erin M. and {Mayne}, Nathan J. and {Miguel}, Yamila and {Nikolov}, Nikolay K. and {Ohno}, Kazumasa and {Palle}, Enric and {Parmentier}, Vivien and {Petit dit de la Roche}, Dominique J.~M. and {Piaulet}, Caroline and {Powell}, Diana and {Rackham}, Benjamin V. and {Redfield}, Seth and {Rogers}, Laura K. and {Rustamkulov}, Zafar and {Tan}, Xianyu and {Tremblin}, P. and {Tsai}, Shang-Min and {Turner}, Jake D. and {de Val-Borro}, Miguel and {Venot}, Olivia and {Welbanks}, Luis and {Wheatley}, Peter J. and {Zhang}, Xi},
        title = "{Early Release Science of the exoplanet WASP-39b with JWST NIRSpec G395H}",
      journal = {\nat},
         year = 2023,
        month = feb,
       volume = {614},
       number = {7949},
        pages = {664-669},
          doi = {10.1038/s41586-022-05591-3},
archivePrefix = {arXiv},
       eprint = {2211.10488},
 primaryClass = {astro-ph.EP},
       adsurl = {https://ui.adsabs.harvard.edu/abs/2023Natur.614..664A}
}

@ARTICLE{2023NIRISS,
       author = {{Feinstein}, Adina D. and {Radica}, Michael and {Welbanks}, Luis and {Murray}, Catriona Anne and {Ohno}, Kazumasa and {Coulombe}, Louis-Philippe and {Espinoza}, N{\'e}stor and {Bean}, Jacob L. and {Teske}, Johanna K. and {Benneke}, Bj{\"o}rn and {Line}, Michael R. and {Rustamkulov}, Zafar and {Saba}, Arianna and {Tsiaras}, Angelos and {Barstow}, Joanna K. and {Fortney}, Jonathan J. and {Gao}, Peter and {Knutson}, Heather A. and {MacDonald}, Ryan J. and {Mikal-Evans}, Thomas and {Rackham}, Benjamin V. and {Taylor}, Jake and {Parmentier}, Vivien and {Batalha}, Natalie M. and {Berta-Thompson}, Zachory K. and {Carter}, Aarynn L. and {Changeat}, Quentin and {dos Santos}, Leonardo A. and {Gibson}, Neale P. and {Goyal}, Jayesh M. and {Kreidberg}, Laura and {L{\'o}pez-Morales}, Mercedes and {Lothringer}, Joshua D. and {Miguel}, Yamila and {Molaverdikhani}, Karan and {Moran}, Sarah E. and {Morello}, Giuseppe and {Mukherjee}, Sagnick and {Sing}, David K. and {Stevenson}, Kevin B. and {Wakeford}, Hannah R. and {Ahrer}, Eva-Maria and {Alam}, Munazza K. and {Alderson}, Lili and {Allen}, Natalie H. and {Batalha}, Natasha E. and {Bell}, Taylor J. and {Blecic}, Jasmina and {Brande}, Jonathan and {Caceres}, Claudio and {Casewell}, S.~L. and {Chubb}, Katy L. and {Crossfield}, Ian J.~M. and {Crouzet}, Nicolas and {Cubillos}, Patricio E. and {Decin}, Leen and {D{\'e}sert}, Jean-Michel and {Harrington}, Joseph and {Heng}, Kevin and {Henning}, Thomas and {Iro}, Nicolas and {Kempton}, Eliza M. -R. and {Kendrew}, Sarah and {Kirk}, James and {Krick}, Jessica and {Lagage}, Pierre-Olivier and {Lendl}, Monika and {Mancini}, Luigi and {Mansfield}, Megan and {May}, E.~M. and {Mayne}, N.~J. and {Nikolov}, Nikolay K. and {Palle}, Enric and {Petit dit de la Roche}, Dominique J.~M. and {Piaulet}, Caroline and {Powell}, Diana and {Redfield}, Seth and {Rogers}, Laura K. and {Roman}, Michael T. and {Roy}, Pierre-Alexis and {Nixon}, Matthew C. and {Schlawin}, Everett and {Tan}, Xianyu and {Tremblin}, P. and {Turner}, Jake D. and {Venot}, Olivia and {Waalkes}, William C. and {Wheatley}, Peter J. and {Zhang}, Xi},
        title = "{Early Release Science of the exoplanet WASP-39b with JWST NIRISS}",
      journal = {\nat},
         year = 2023,
        month = feb,
       volume = {614},
       number = {7949},
        pages = {670-675},
          doi = {10.1038/s41586-022-05674-1},
archivePrefix = {arXiv},
       eprint = {2211.10493},
 primaryClass = {astro-ph.EP},
       adsurl = {https://ui.adsabs.harvard.edu/abs/2023Natur.614..670F}
}

@ARTICLE{2023MNRAS.524..377H,
       author = {{Holmberg}, M{\r{a}}ns and {Madhusudhan}, Nikku},
        title = "{Exoplanet spectroscopy with JWST NIRISS: diagnostics and case studies}",
      journal = {\mnras},
         year = 2023,
        month = sep,
       volume = {524},
       number = {1},
        pages = {377-402},
          doi = {10.1093/mnras/stad1580},
archivePrefix = {arXiv},
       eprint = {2306.04676},
 primaryClass = {astro-ph.EP},
       adsurl = {https://ui.adsabs.harvard.edu/abs/2023MNRAS.524..377H}
}

@inproceedings{
yip2024ariel,
title={Ariel Data Challenge 2024: Extracting exoplanetary signals from the Ariel Space Telescope},
author={Kai Hou Yip and Lorenzo V. Mugnai and Andrea Bocchieri and Andreas Papageorgiou and Orph{\'e}e Faucoz and Tara Tahseen and Batista Virginie and Ang{\`e}le Syty and Enzo Pascale and Quentin Changeat and Billy Edwards and Paul Eccleston and Clare Jenner and Ryan King and Theresa Lueftinger and Nikolaos Nikolaou and Pascale Danto and Sudeshna Boro Saikia and Lu{\'\i}s F. Sim{\~o}es and Giovanna Tinetti and Ingo P. Waldmann},
booktitle={NeurIPS 2024 Competition Track},
year={2024},
url={https://openreview.net/forum?id=1mGm9tHrFT}
}

@article{Baraffe2015,
	author = {Baraffe, Isabelle and Homeier, Derek and Allard, France and Chabrier, Gilles},
	doi = {10.1051/0004-6361/201425481},
	issn = {0004-6361},
	journal = {Astronomy {\&} Astrophysics},
	month = {may},
	pages = {A42},
	title = {New evolutionary models for pre-main sequence and main sequence low-mass stars down to the hydrogen-burning limit},
	volume = {577},
	year = {2015}
}

@ARTICLE{Arielrad,
 author = {{Mugnai}, Lorenzo V. and {Pascale}, Enzo and {Edwards}, Billy and
         {Papageorgiou}, Andreas and {Sarkar}, Subhajit},
        title = "{ArielRad: the Ariel radiometric model}",
      journal = {Experimental Astronomy},
         year = 2020,
        month = oct,
       volume = {50},
       number = {2-3},
        pages = {303-328},
          doi = {10.1007/s10686-020-09676-7},
archivePrefix = {arXiv},
       eprint = {2009.07824},
 primaryClass = {astro-ph.IM},
       adsurl = {https://ui.adsabs.harvard.edu/abs/2020ExA....50..303M}
}

@ARTICLE{Taurex,
       author = {{Al-Refaie}, A.~F. and {Changeat}, Q. and {Waldmann}, I.~P. and {Tinetti}, G.},
        title = "{TauREx 3: A Fast, Dynamic, and Extendable Framework for Retrievals}",
      journal = {\apj},
         year = 2021,
        month = aug,
       volume = {917},
       number = {1},
          eid = {37},
        pages = {37},
          doi = {10.3847/1538-4357/ac0252},
archivePrefix = {arXiv},
       eprint = {1912.07759},
 primaryClass = {astro-ph.IM},
       adsurl = {https://ui.adsabs.harvard.edu/abs/2021ApJ...917...37A}
}

@inproceedings{PAOS,
author = {Andrea Bocchieri and Lorenzo V. Mugnai and Enzo Pascale},
title = {{PAOS: a fast, modern, and reliable Python package for physical optics studies}},
volume = {13092},
booktitle = {Space Telescopes and Instrumentation 2024: Optical, Infrared, and Millimeter Wave},
editor = {Laura E. Coyle and Shuji Matsuura and Marshall D. Perrin},
organization = {International Society for Optics and Photonics},
publisher = {SPIE},
pages = {130924K},
year = {2024},
doi = {10.1117/12.3018333},
URL = {https://doi.org/10.1117/12.3018333}
}

@ARTICLE{Sing2018,
       author = {{Sing}, David K.},
        title = "{Observational Techniques With Transiting Exoplanetary Atmospheres}",
      journal = {arXiv e-prints},
         year = 2018,
        month = apr,
          eid = {arXiv:1804.07357},
        pages = {arXiv:1804.07357},
          doi = {10.48550/arXiv.1804.07357},
archivePrefix = {arXiv},
       eprint = {1804.07357},
 primaryClass = {astro-ph.EP},
       adsurl = {https://ui.adsabs.harvard.edu/abs/2018arXiv180407357S}
}

@ARTICLE{Barron2007,
       author = {{Barron}, N. and {Borysow}, M. and {Beyerlein}, K. and {Brown}, M. and {Lorenzon}, W. and {Schubnell}, M. and {Tarl{\'e}}, G. and {Tomasch}, A. and {Weaverdyck}, C.},
        title = "{Subpixel Response Measurement of Near-Infrared Detectors}",
      journal = {\pasp},
         year = 2007,
        month = apr,
       volume = {119},
       number = {854},
        pages = {466-475},
          doi = {10.1086/517620},
archivePrefix = {arXiv},
       eprint = {astro-ph/0611339},
 primaryClass = {astro-ph},
       adsurl = {https://ui.adsabs.harvard.edu/abs/2007PASP..119..466B}
}

@INPROCEEDINGS{edwards2019,
       author = {{Edwards}, Billy and {Mugnai}, Lorenzo and {Tinetti}, Giovanna and {Pascale}, Enzo and {Sarkar}, Subhajit},
        title = "{An Updated Study of Potential Targets for Ariel}",
    booktitle = {EPSC-DPS Joint Meeting 2019},
         year = 2019,
       volume = {2019},
        month = sep,
          eid = {EPSC-DPS2019-603},
        pages = {EPSC-DPS2019-603},
       adsurl = {https://ui.adsabs.harvard.edu/abs/2019EPSC...13..603E}
}

@article{elad2023image,
  title={Image denoising: The deep learning revolution and beyond—a survey paper},
  author={Elad, Michael and Kawar, Bahjat and Vaksman, Gregory},
  journal={SIAM Journal on Imaging Sciences},
  volume={16},
  number={3},
  pages={1594--1654},
  year={2023},
  publisher={SIAM}
}

@article{su2022survey,
  title={A survey of deep learning approaches to image restoration},
  author={Su, Jingwen and Xu, Boyan and Yin, Hujun},
  journal={Neurocomputing},
  volume={487},
  pages={46--65},
  year={2022},
  publisher={Elsevier}
}

@article{bie2023renaissance,
  title={RenAIssance: A Survey into AI Text-to-Image Generation in the Era of Large Model},
  author={Bie, Fengxiang and Yang, Yibo and Zhou, Zhongzhu and Ghanem, Adam and Zhang, Minjia and Yao, Zhewei and Wu, Xiaoxia and Holmes, Connor and Golnari, Pareesa and Clifton, David A and others},
  journal={arXiv preprint arXiv:2309.00810},
  year={2023}
}

@article{dubey2024transformer,
  title={Transformer-based generative adversarial networks in computer vision: A comprehensive survey},
  author={Dubey, Shiv Ram and Singh, Satish Kumar},
  journal={IEEE Transactions on Artificial Intelligence},
  year={2024},
  publisher={IEEE}
}

@article{zhang2023text,
  title={Text-to-image diffusion model in generative ai: A survey},
  author={Zhang, Chenshuang and Zhang, Chaoning and Zhang, Mengchun and Kweon, In So},
  journal={arXiv preprint arXiv:2303.07909},
  year={2023}
}

@ARTICLE{Vojtekova21,
       author = {{Vojtekova}, Antonia and {Lieu}, Maggie and {Valtchanov}, Ivan and {Altieri}, Bruno and {Old}, Lyndsay and {Chen}, Qifeng and {Hroch}, Filip},
        title = "{Learning to denoise astronomical images with U-nets}",
      journal = {\mnras},
         year = 2021,
        month = may,
       volume = {503},
       number = {3},
        pages = {3204-3215},
          doi = {10.1093/mnras/staa3567},
archivePrefix = {arXiv},
       eprint = {2011.07002},
 primaryClass = {astro-ph.IM},
       adsurl = {https://ui.adsabs.harvard.edu/abs/2021MNRAS.503.3204V}
}

@ARTICLE{sweere22,
       author = {{Sweere}, Sam F. and {Valtchanov}, Ivan and {Lieu}, Maggie and {Vojtekova}, Antonia and {Verdugo}, Eva and {Santos-Lleo}, Maria and {Pacaud}, Florian and {Briassouli}, Alexia and {C{\'a}mpora P{\'e}rez}, Daniel},
        title = "{Deep learning-based super-resolution and de-noising for XMM-newton images}",
      journal = {\mnras},
         year = 2022,
        month = dec,
       volume = {517},
       number = {3},
        pages = {4054-4069},
          doi = {10.1093/mnras/stac2437},
archivePrefix = {arXiv},
       eprint = {2205.01152},
 primaryClass = {astro-ph.IM},
       adsurl = {https://ui.adsabs.harvard.edu/abs/2022MNRAS.517.4054S}
}

@ARTICLE{Park24,
       author = {{Park}, Hyosun and {Jo}, Yongsik and {Kang}, Seokun and {Kim}, Taehwan and {Jee}, M. James},
        title = "{Deeper, Sharper, Faster: Application of Efficient Transformer to Galaxy Image Restoration}",
      journal = {arXiv e-prints},
         year = 2024,
        month = mar,
          eid = {arXiv:2404.00102},
        pages = {arXiv:2404.00102},
          doi = {10.48550/arXiv.2404.00102},
archivePrefix = {arXiv},
       eprint = {2404.00102},
 primaryClass = {astro-ph.IM},
       adsurl = {https://ui.adsabs.harvard.edu/abs/2024arXiv240400102P}
}

@ARTICLE{Krick2020,
       author = {{Krick}, Jessica E. and {Fraine}, Jonathan and {Ingalls}, Jim and {Deger}, Sinan},
        title = "{Random Forests Applied to High-precision Photometry Analysis with Spitzer IRAC}",
      journal = {The Astronomical Journal},
         year = 2020,
        month = sep,
       volume = {160},
       number = {3},
          eid = {99},
        pages = {99},
          doi = {10.3847/1538-3881/aba11f},
archivePrefix = {arXiv},
       eprint = {2006.14095},
 primaryClass = {astro-ph.IM},
       adsurl = {https://ui.adsabs.harvard.edu/abs/2020AJ....160...99K}
}

@ARTICLE{Ingalls2016,
       author = {{Ingalls}, James G. and {Krick}, J.~E. and {Carey}, S.~J. and {Stauffer}, John R. and {Lowrance}, Patrick J. and {Grillmair}, Carl J. and {Buzasi}, Derek and {Deming}, Drake and {Diamond-Lowe}, Hannah and {Evans}, Thomas M. and {Morello}, G. and {Stevenson}, Kevin B. and {Wong}, Ian and {Capak}, Peter and {Glaccum}, William and {Laine}, Seppo and {Surace}, Jason and {Storrie-Lombardi}, Lisa},
        title = "{Repeatability and Accuracy of Exoplanet Eclipse Depths Measured with Post-cryogenic Spitzer}",
      journal = {The Astronomical Journal},
         year = 2016,
        month = aug,
       volume = {152},
       number = {2},
          eid = {44},
        pages = {44},
          doi = {10.3847/0004-6256/152/2/44},
archivePrefix = {arXiv},
       eprint = {1601.05101},
 primaryClass = {astro-ph.IM},
       adsurl = {https://ui.adsabs.harvard.edu/abs/2016AJ....152...44I}
}

@article{nikolaou2023lessons,
  title={Lessons learned from the 1st Ariel Machine Learning Challenge: Correcting transiting exoplanet light curves for stellar spots},
  author={Nikolaou, Nikolaos and Waldmann, Ingo P and Tsiaras, Angelos and Morvan, Mario and Edwards, Billy and Yip, Kai Hou and Thompson, Alexandra and Tinetti, Giovanna and Sarkar, Subhajit and Dawson, James M and others},
  journal={RAS Techniques and Instruments},
  volume={2},
  number={1},
  pages={695--709},
  year={2023},
  publisher={Oxford University Press}
}

@ARTICLE{Morvan2020,
       author = {{Morvan}, Mario and {Nikolaou}, Nikolaos and {Tsiaras}, Angelos and {Waldmann}, Ingo P.},
        title = "{Detrending Exoplanetary Transit Light Curves with Long Short-term Memory Networks}",
      journal = {The Astronomical Journal},
         year = 2020,
        month = mar,
       volume = {159},
       number = {3},
          eid = {109},
        pages = {109},
          doi = {10.3847/1538-3881/ab6aa7},
archivePrefix = {arXiv},
       eprint = {2001.03370},
 primaryClass = {astro-ph.EP},
       adsurl = {https://ui.adsabs.harvard.edu/abs/2020AJ....159..109M}
}

@INPROCEEDINGS{Morvan2022,
       author = {{Morvan}, Mario and {Nikolaou}, Nikolaos and {Yip}, Kai and {Waldmann}, Ingo},
        title = "{Don't Pay Attention to the Noise: Learning Self-supervised Representations of Light Curves with a Denoising Time Series Transformer}",
    booktitle = {Machine Learning for Astrophysics},
         year = 2022,
        month = jul,
          eid = {11},
        pages = {11},
          doi = {10.48550/arXiv.2207.02777},
archivePrefix = {arXiv},
       eprint = {2207.02777},
 primaryClass = {astro-ph.IM},
       adsurl = {https://ui.adsabs.harvard.edu/abs/2022mla..confE..11M}
}

@misc{gal2016,
      title={Dropout as a Bayesian Approximation: Representing Model Uncertainty in Deep Learning}, 
      author={Yarin Gal and Zoubin Ghahramani},
      year={2016},
      eprint={1506.02142},
      archivePrefix={arXiv},
      primaryClass={stat.ML},
      url={https://arxiv.org/abs/1506.02142}, 
}

@misc{yip_esa-ariel_2022,
    title = {{ESA}-{Ariel} {Data} {Challenge} {NeurIPS} 2022: {Inferring} {Physical} {Properties} of {Exoplanets} {From} {Next}-{Generation} {Telescopes}},
    copyright = {arXiv.org perpetual, non-exclusive license},
    shorttitle = {{ESA}-{Ariel} {Data} {Challenge} {NeurIPS} 2022},
    url = {https://arxiv.org/abs/2206.14642},
    doi = {10.48550/ARXIV.2206.14642},
    urldate = {2024-12-06},
    publisher = {arXiv},
    author = {Yip, Kai Hou and Waldmann, Ingo P. and Changeat, Quentin and Morvan, Mario and Al-Refaie, Ahmed F. and Edwards, Billy and Nikolaou, Nikolaos and Tsiaras, Angelos and de Oliveira, Catarina Alves and Lagage, Pierre-Olivier and Jenner, Clare and Cho, James Y-K. and Thiyagalingam, Jeyan and Tinetti, Giovanna},
    year = {2022},
    note = {Version Number: 1},
}

@article{changeat_esa-ariel_2023,
    title = {{ESA}-{Ariel} {Data} {Challenge} {NeurIPS} 2022: introduction to exo-atmospheric studies and presentation of the {Atmospheric} {Big} {Challenge} ({ABC}) {Database}},
    volume = {2},
    copyright = {https://creativecommons.org/licenses/by/4.0/},
    issn = {2752-8200},
    shorttitle = {{ESA}-{Ariel} {Data} {Challenge} {NeurIPS} 2022},
    url = {https://academic.oup.com/rasti/article/2/1/45/6998590},
    doi = {10.1093/rasti/rzad001},
    language = {en},
    number = {1},
    urldate = {2024-12-05},
    journal = {RAS Techniques and Instruments},
    author = {Changeat, Quentin and Yip, Kai Hou},
    month = jan,
    year = {2023},
    pages = {45--61},
}

@ARTICLE{Schmidt2025AJ,
       author = {{Schmidt}, Stephen P. and {MacDonald}, Ryan J. and {Tsai}, Shang-Min and {Radica}, Michael and {Wang}, Le-Chris and {Ahrer}, Eva-Maria and {Bell}, Taylor J. and {Fisher}, Chloe and {Thorngren}, Daniel P. and {Wogan}, Nicholas and {May}, Erin M. and {Ferrari}, Piero and {Bennett}, Katherine A. and {Rustamkulov}, Zafar and {L{\'o}pez-Morales}, Mercedes and {Sing}, David K.},
        title = "{A Comprehensive Reanalysis of K2-18 b's JWST NIRISS+NIRSpec Transmission Spectrum}",
      journal = {\aj},
         year = 2025,
        month = dec,
       volume = {170},
       number = {6},
          eid = {298},
        pages = {298},
          doi = {10.3847/1538-3881/ae019a},
archivePrefix = {arXiv},
       eprint = {2501.18477},
 primaryClass = {astro-ph.EP},
       adsurl = {https://ui.adsabs.harvard.edu/abs/2025AJ....170..298S}
}

@ARTICLE{Welbanks2025NatAs,
       author = {{Welbanks}, Luis and {Nixon}, Matthew C. and {McGill}, Peter and {Tilke}, Lana J. and {Wiser}, Lindsey S. and {Rotman}, Yoav and {Mukherjee}, Sagnick and {Feinstein}, Adina D. and {Line}, Michael R. and {Benneke}, Bj{\"o}rn and {Seager}, Sara and {Beatty}, Thomas G. and {Seligman}, Darryl Z. and {Parmentier}, Vivien and {Sing}, David K.},
        title = "{Challenges in the detection of gases in exoplanet atmospheres}",
      journal = {Nature Astronomy},
         year = 2025,
        month = dec,
          doi = {10.1038/s41550-025-02730-4},
archivePrefix = {arXiv},
       eprint = {2504.21788},
 primaryClass = {astro-ph.EP},
       adsurl = {https://ui.adsabs.harvard.edu/abs/2025NatAs.tmp..257W}
}

@ARTICLE{Stassun2017AJ,
       author = {{Stassun}, Keivan G. and {Collins}, Karen A. and {Gaudi}, B. Scott},
        title = "{Accurate Empirical Radii and Masses of Planets and Their Host Stars with Gaia Parallaxes}",
      journal = {\aj},
         year = 2017,
        month = mar,
       volume = {153},
       number = {3},
          eid = {136},
        pages = {136},
          doi = {10.3847/1538-3881/aa5df3},
archivePrefix = {arXiv},
       eprint = {1609.04389},
 primaryClass = {astro-ph.EP},
       adsurl = {https://ui.adsabs.harvard.edu/abs/2017AJ....153..136S}
}

%%%%%%%%%%%%%%%%%%%%%%%%%%%%%%%%%%%%%%%%%%%%%%%%%%

%%%%%%%%%%%%%%%%% APPENDICES %%%%%%%%%%%%%%%%%%%%%

\appendix
\section{Data calibration} \label{sec:calibration}

\begin{figure*}
    \centering
    \includegraphics[width=0.8\textwidth]{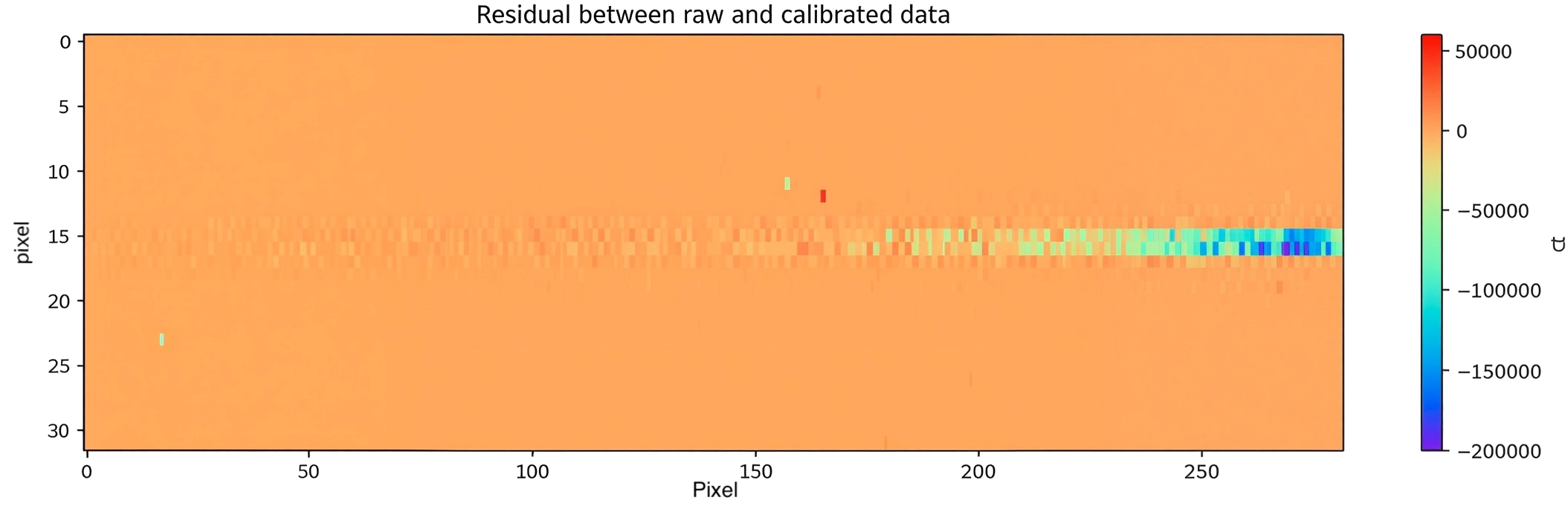}
    \caption{Difference between a raw image from the train set and the associated calibrated frame, applying flat field, dark and dead pixels' correction and non-linearity of the pixels' response correction.}
    \label{fig:calibration}
\end{figure*}

\autoref{fig:calibration} presents the difference between the original raw frame and the same image after calibration, including flat-field, dark current, dead pixel, and pixel non-linearity corrections. Outside the signal region, the residuals are minimal and remain close to zero, as indicated by the orange colour scale. In contrast, larger deviations are observed in regions of higher signal. This behaviour indicates that the dominant contribution to the calibration difference arises from the non-linearity correction, which becomes increasingly significant at higher count levels.

\bsp	% typesetting comment
\label{lastpage}
\end{document}